\documentclass[usenatbib]{mn2e}
\usepackage{graphicx}
\usepackage{amssymb}
\usepackage{times}
\usepackage{mn2e-breakabs}
\begin{document}

\title[Systematic Analysis of BOSS 3D Clustering] 
{The clustering of galaxies in the SDSS-III Baryon Oscillation Spectroscopic Survey: Analysis of potential systematics}

\author[A. J. Ross et al.]{\parbox{\textwidth}{Ashley J. Ross\thanks{Email: Ashley.Ross@port.ac.uk}$^{1}$,
Will J. Percival$^{1}$, 
Ariel G. S\'anchez$^{2}$, 
Lado Samushia$^{1,3}$, 
Shirley Ho$^{4,5}$,
 Eyal Kazin$^{6}$, 
 Marc Manera$^{1}$, 
 Beth Reid$^{4,7}$,
  Martin White$^{3,8,9}$, 
 Rita Tojeiro$^{1}$,
 Cameron K. McBride$^{10}$,
 Xiaoying Xu$^{11}$, 
 David A. Wake$^{12}$,
  Michael A. Strauss$^{13}$,
 Francesco Montesano$^{2}$,
  Molly E. C. Swanson$^{10}$,
  Stephen Bailey$^4$,
   Adam S. Bolton$^{14}$,
   Antonio Montero Dorta$^{15}$,
   Daniel J. Eisenstein$^{10}$, 
   Hong Guo$^{16}$, 
   Jean-Christophe Hamilton$^{17}$, 
   Robert C. Nichol$^1$,
    Nikhil Padmanabhan$^{18}$, 
    Francisco Prada$^{19,20,15}$,
     David J. Schlegel$^3$,     
       Mariana Vargas Maga\~na$^{17}$,
         Idit Zehavi$^{16}$,
         Michael Blanton$^{21}$,
         Dmitry Bizyaev$^{22}$
         Howard Brewington$^{22}$,
         Antonio J. Cuesta$^{17}$,
         Elena  Malanushenko$^{22}$,
         Viktor Malanushenko$^{20}$,
         Daniel Oravetz$^{22}$,
         John Parejko$^{18}$,
         Kaike Pan$^{22}$,
         Donald P. Schneider$^{23,24}$
         Alaina Shelden$^{22}$,
         Audrey Simmons$^{22}$,
         Stephanie Snedden$^{22}$, 
         Gong-bo Zhao$^1$
         }
 \vspace*{4pt}
 \\
 $^1$Institute of Cosmology \& Gravitation, Dennis Sciama Building, University of Portsmouth, Portsmouth, PO1 3FX, UK\\
 $^{2}$Max-Planck-Institut f\"ur extraterrestrische Physik, Postfach 1312, Giessenbachstr.,
85748 Garching, Germany\\
$^3$National Abastumani Astrophysical Observatory, Ilia State University, 2A Kazbegi Ave., GE-1060 Tbilisi, Georgia\\ 
$^4$Lawrence Berkeley National Laboratory, 1 Cyclotron Road, Berkeley, CA 94720, USA\\
$^{5}$Department of Physics, Carnegie Mellon University, 5000 Forbes Avenue, Pittsburgh, PA 15213, USA\\
$^{6}$ Centre for Astrophysics and Supercomputing, Swinburne University of Technology, P.O. Box 218, Hawthorn, Victoria 3122, Australia\\
$^7$Hubble Fellow\\
$^{8}$Department of Physics, University of California, 366 LeConte Hall, Berkeley, CA 94720, USA\\
$^{9}$Department of Astronomy, 601 Campbell Hall, University of California at Berkeley, Berkeley, CA 94720, USA\\
$^{10}$Harvard-Smithsonian Center for Astrophysics, 60 Garden St., MS \#20, Cambridge, MA 02138\\
$^{11}$Steward Observatory, University of Arizona, 933 North Cherry Ave., Tucson, AZ 85721, USA\\
$^{12}$Yale Center for Astronomy and Astrophysics, Yale University, New Haven, CT 06511, USA\\
$^{13}$Department of Astrophysical Sciences, Princeton University,  Ivy Lane, Princeton, NJ 08544, USA\\
$^{14}$Department Physics and Astronomy, University of Utah, UT 84112, USA\\
$^{15}$Instituto de Astrof\'isica de Andaluc\'ia (CSIC), E-18080 Granada, Spain\\
$^{16}$Department of Astronomy, Case Western Reserve University, Cleveland, Ohio 44106, USA\\
$^{17}$APC, Universit\'e Paris-Diderot-Paris 7, CNRS/IN2P3, CEA, Observatoire de Paris, 10, rue A. Domon \& L. Duquet, Paris, France\\
$^{18}$Department of Physics, Yale University, 260 Whitney Ave, New Haven, CT 06520, USA\\
$^{19}$Campus of International Excellence UAM+CSIC, Cantoblanco, E-28049 Madrid, Spain\\
$^{20}$Instituto de Fisica Teorica (UAM/CSIC), Universidad Autonoma de Madrid, Cantoblanco, E-28049 Madrid, Spain\\
$^{21}$Center for Cosmology and Particle Physics, New York University, New York, NY 10003, USA\\
$^{22}$Apache Point Observatory, P.O. Box 59, Sunspot, NM 88349-0059, USA\\
$^{23}$Department of Astronomy and Astrophysics, The Pennsylvania State University,
  University Park, PA 16802\\
$^{24}$Institute for Gravitation and the Cosmos, The Pennsylvania State University,
  University Park, PA 16802  
 }

\date{Accepted by MNRAS} 
\pagerange{\pageref{firstpage}--\pageref{lastpage}} \pubyear{2012}
\maketitle
\label{firstpage}

\begin{abstract}
We analyze the density field of galaxies observed by the Sloan Digital Sky Survey (SDSS)-III Baryon Oscillation Spectroscopic Survey (BOSS) included in the SDSS Data Release Nine (DR9). DR9 includes spectroscopic redshifts for over 400,000 galaxies spread over a footprint of 3,275 deg$^2$. We identify, characterize, and mitigate the impact of sources of systematic uncertainty on large-scale clustering measurements, both for angular moments of the redshift-space correlation function, $\xi_{\ell}(s)$ and the spherically averaged power spectrum, $P(k)$, in order to ensure that robust cosmological constraints will be obtained from these data. A correlation between the projected density of stars and the higher redshift ($0.43 < z < 0.7$) galaxy sample (the `CMASS' sample) due to imaging systematics imparts a systematic error that is larger than the statistical error of the clustering measurements at scales $s > 120h^{-1}$Mpc or $k < 0.01h$Mpc$^{-1}$. We find that these errors can be ameliorated by weighting galaxies based on their surface brightness and the local stellar density. The clustering of CMASS galaxies found in the Northern and Southern Galactic footprints of the survey generally agrees to within 2$\sigma$. We use mock galaxy catalogs that simulate the CMASS selection function to determine that randomly selecting galaxy redshifts in order to simulate the radial selection function of a random sample imparts the least systematic error on $\xi_{\ell}(s)$ measurements and that this systematic error is negligible for the spherically averaged correlation function, $\xi_0$. We find a peak in $\xi_0$ at $s\sim200h^{-1}$Mpc, with a corresponding feature with period $\sim0.03h$Mpc$^{-1}$ in $P(k)$, and find features at least as strong in 4.8\% of the mock galaxy catalogs, concluding this feature is likely to be a consequence of cosmic variance. The methods we recommend for the calculation of clustering measurements using the CMASS sample are adopted in companion papers that locate the position of the baryon acoustic oscillation feature \citep{alph}, constrain cosmological models using the full shape of $\xi_0$ \citep{SanchezCos12}, and measure the rate of structure growth \citep{ReidRSD12}.
\end{abstract}


\section{Introduction}

In the last decade, wide-field surveys such as the Two Degree Field Galaxy Redshift Survey (2dFGRS; \citealt{2df}), the Sloan Digital Sky Survey (SDSS; \citealt{SDSS}), and the WiggleZ Redshift Survey \citep{wiggz} have obtained accurate spectroscopic redshifts of well over one million galaxies, allowing maps of the 3-dimensional structure of the Universe to be constructed out to $z = 0.9$. These maps encode a wealth of information on cosmology (e.g., \citealt{Teg04,2dfcos,Eis05,Per09BAO,ReidDR7,BlakeKazinBAO,MontesanoPk}) and the nature of galaxies (e.g. \citealt{Norberg02,Gomez03,Swanson08,WakeLRGev,TojeiroDR7,ARoss11a,ZehaviHODfinal}).

The Baryon Oscillation Spectroscopic Survey (BOSS) is designed to obtain spectroscopy of 1.5 million galaxies to $z = 0.7$ over an imaging area of 10,000 deg$^2$ \citep{Eisenstein11}. \cite{White11BEDR} investigated an early sample from this survey, confirming the survey was making a high-quality map of massive galaxies with bias $\sim$ 2. We utilize spectroscopic redshifts for over 400,000 BOSS galaxies that will be released as part of the SDSS Data Release Nine (DR9). These galaxies cover close to 1/3 of the final (planned) footprint, and currently comprise the largest effective volume \citep{Teg98} of any spectroscopic galaxy catalog --- 2.2${\rm Gpc}^3$ (assuming a concordance $\Lambda$CDM model). These data should therefore allow the best-to-date statistical uncertainty on the measured power spectrum, $P(k)$, and thus the best-to-date cosmological measurements determined using a galaxy catalog. As such, discovery and elimination of systematic uncertainty is of vital importance to realizing the survey goals. Potential systematic effects on the observed density of galaxies must be robustly tested and ameliorated in an un-biased way.

The purpose of this study is to identify and minimize the impact of sources of systematic uncertainty in the large-scale clustering of BOSS galaxies, in order to ensure robust cosmological constraints are obtained. \cite{imsys} studied systematic effects on the projected density of BOSS galaxy targets, finding a strong relationship with stellar density and differences in the sample in occupying the Northern and Southern Galactic Cap (NGC and SGC from hereon). We follow up and extend this work by testing how these systematic variations effect spatial clustering measurements and by testing against systematic effects incurred when obtaining spectroscopic redshifts. We aim to answer the following questions:\begin{enumerate}  \item How do variations in photometric calibration, e.g., between the NGC and SGC footprints, affect the selection of BOSS galaxies?  \item How does the observed density of galaxies depend on observing conditions? \item What is the best way to simulate the radial selection function and how important are effects related to galaxy evolution? \item How do permutations of (i)-(iii) affect the clustering we measure?\end{enumerate}

Our results have already been used in studies of the clustering of BOSS DR9 galaxies. \cite{alph} localize the position of the baryon acoustic oscillation (BAO) feature to better than 2\% accuracy. \cite{ReidRSD12} and measure redshift-space distortions (RSD) and \cite{SamRSD12} thereby constrain dark energy and modified gravity models. See also \cite{Toj12RSD} for a complementary method of measuring structure growth using DR9 galaxies. \cite{Nuza12} found that the clustering of BOSS galaxies can be well approximated by using a sub-halo abundance matching method applied to a dark matter simulation. \cite{SanchezCos12} obtain cosmological constraints by fitting the full shape of the correlation function. We hope that future studies heed, and improve upon, our analysis, which we feel is the most careful analysis of observational systematics to date. 

The presentation of our analysis is organized as follows: In Section \ref{sec:data} we describe the BOSS DR9 sample of galaxies and its corresponding angular mask. In Section \ref{sec:calc}, we describe how we estimate clustering statistics, their covariance, and compare to models. For both the covariance and the models we utilize the mock catalogs of galaxies (hereafter `mocks') generated by \cite{Manera12mock}. In Section \ref{sec:NS}, we investigate and explain the differences we find in the densities of galaxies in the NGC and SGC, addressing question (i). In Section \ref{sec:angvar} we describe potential sources of systematic variation in the density of galaxies targeted for spectroscopy and the methods we employ to remove these variations in an unbiased way, addressing question (ii). In Section \ref{sec:zsys}, we investigate the radial distribution of our galaxy sample, using mock catalogs to determine the least biased way to simulate the radial selection function of BOSS galaxies and to check that the clustering we measure is robust when the galaxies are split into two samples above/below redshift 0.52, thus addressing question (iii). Throughout Sections \ref{sec:NS} through \ref{sec:zsys}, we address question (iv) using $\xi_{\ell}(s)$ measurements at $s < 150h^{-1}$Mpc. In Section \ref{sec:remain}, we adress consider the clustering at scales $s > 150h^{-1}$Mpc, also utilizing measurements of anisotropic clustering and the power spectrum. We conclude in Section \ref{sec:con}. Throughout, we assume a flat cosmology with $\Omega_{m} = 0.274, \Omega_{b}h^2=0.0224, h=0.70, n_s=0.95$, and $\sigma_8=0.8$ (identical to that used in \citealt{White11BEDR} and \citealt{alph}) unless otherwise noted.

\section{Data} 
 \label{sec:data}

The SDSS-III Baryon Oscillation Spectroscopic Survey (\citealt{Eisenstein11}) obtains targets using SDSS imaging data. In combination, the SDSS-I, SDSS-II, and SDSS-III surveys obtained wide-field CCD photometry (\citealt{C,Gunn06}) in five passbands ($u,g,r,i,z$; e.g., \citealt{F}), amassing a total footprint of 14,555 deg$^2$, internally calibrated using the `uber-calibration' process described in \cite{Pad08}, and with a 50\% completeness limit of point sources at $r = 22.5$ (\citealt{DR8}). After completing the imaging, BOSS has targeted 1.5 million massive galaxies, 150,000 quasars, and over 75,000 ancillary targets for spectroscopic observation over an area of 10,000 deg$^2$ \citep{Eisenstein11}. BOSS observations began in fall 2009, and the last data will be acquired in 2014. The BOSS spectrographs (R = 1300-3000) are fed by 1000 optical fibres in a single pointing, each with a 2$^{\prime\prime}$ aperture. Each observation is performed in a series of 15-minute exposures and integrated until a fiducial minimum signal-to-noise ratio, chosen to ensure a high redshift success rate, is reached. This ensures a sample with nearly isotropic redshift selection complete to 98\%. We test this isotropy in Section \ref{sec:zfail}. 

\begin{figure}
  \includegraphics[width=84mm]{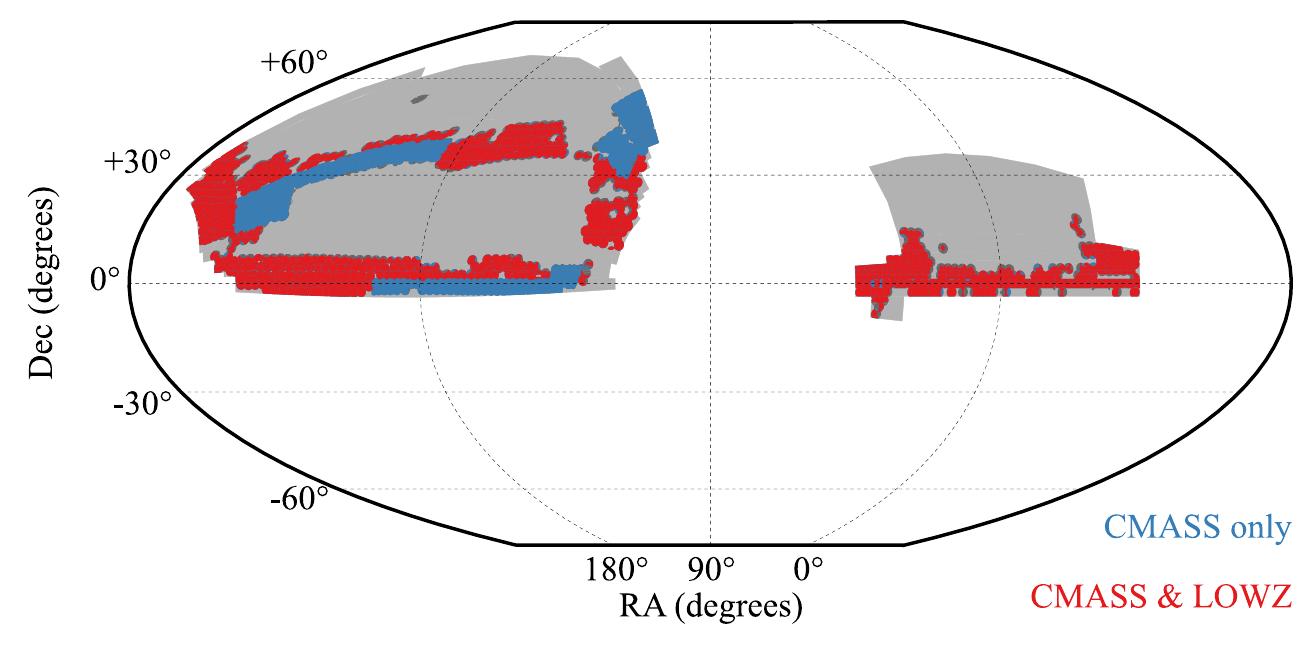}
  \caption{The footprint of BOSS DR9 galaxies, projected into two dimensions using the McBryde-Thomas Flat Polar Quartic projection, is shaded in blue and red. Areas with CMASS data only are shaded blue. The CMASS and LOWZ footprints cover 3275 and 2208 deg$^2$, respectively. The grey area represents the final (planned) BOSS footprint.}
  \label{fig:angular_distribution}
  \label{fig:footprint}
\end{figure}

\subsection{Target Selection}

BOSS targets two samples of galaxies. These are the `LOWZ' and `CMASS' samples, as described by \cite{Eisenstein11}. We are careful throughout this paper to distinguish between target objects and true galaxies --- the majority of LOWZ and CMASS targets are galaxies, but a small fraction are stars (3\% of CMASS) and high-redshift quasars (1\%, i.e., not objects sampling the intended density field). Anything we refer to as a galaxy has been spectroscopically confirmed as such. The SDSS measures magnitudes using both PSF-convolved fits to DeVaucouleurs (we denote these with a $dev$ subscript) and exponential profiles (we denote these with an $exp$ subscript), Each of these magnitudes are used to determine `model', which we denote using the subscript $_{mod}$, and `cmodel' magnitudes,  denoted using the subscript $_{cmod}$, which are used in target selection. Model magnitudes denote the best-fit of the two profiles in the $r$-band (see \citealt{EDR} for further information on model magnitudes), The cmodel magnitudes, first defined in \cite{DR2}, represent the best-fitting linear combination of the exponential and DeVaucouleurs model fluxes. We will also use PSF magnitudes, which we denote using the subscript $_{psf}$.

We select BOSS targets using the photometry of objects identified as galaxies by the SDSS pipeline. Most of the sample (100\% for LOWZ and 90.9\% for CMASS) was targeted using the SDSS DR8 photometry designated as `primary'. The remaining sample was targeted from images now designated in DR8 as secondary. This data was superseded by overlapping imaging runs of better quality but whose reductions were unavailable at the time of targeting. Photometric scatter across the multiple selection boundaries listed below implies that many objects targeted using primary photometry would not have been targeted using secondary photometry (and vice-versa). However, this result is simply due to the known statistical distribution of measured magnitudes, quantified by the magnitude error. This effect should not cause any additional systematic error beyond that potentially induced by targeting from a sample with magnitude errors that vary with angular position, assuming one always uses the photometry used {\it at the time of targeting} in an analysis. Indeed, we find restricting our analyses to data targeted using DR8 primary photometry results in no significant change in any clustering statistic we measure. 
\begin{figure*}
\begin{minipage}{7in}
 \centering
 \includegraphics[width=4.5in]{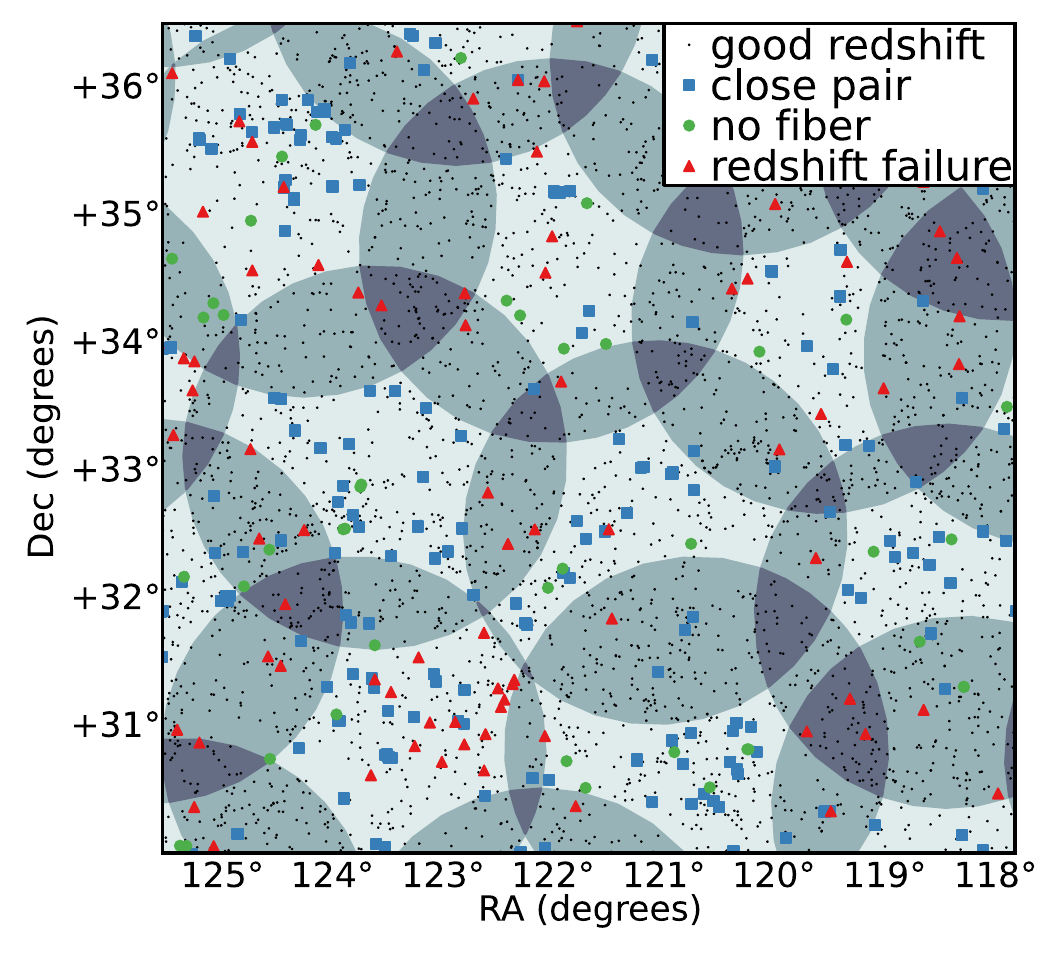}
  \caption{The distribution of CMASS targets for a selection of mask sectors. Circles represent the sky area covered by observing tiles and the number of overlapping tiles is indicated by the level of shading. Black dots indicate the positions of targets for which we obtained a `good' redshift, as defined in Section \ref{sec:zfail}. Blue squares denote targets for which we did not allocate a fiber for spectroscopic observation because the target is within 62$^{\prime\prime}$ of another CMASS target (`close pair'). Green circles denote targets for which we did allocate a fiber that are not close pairs. Red triangles denote targets for which we allocated a fiber, but did not obtain a good redshift.}
  \label{fig:maskcls}
\end{minipage}
\end{figure*}

\cite{Eisenstein11} define the selection criteria for BOSS galaxy targets. We repeat them here for completeness and ease of reference. The CMASS selection is defined by\footnote{In the early part of the survey, various super-sets of this selection were used, e.g., the fiber magnitude limit has changed from $i_{fib2} < 21.7$ to $i_{fib2} < 21.5$. We only use data satisfying the above selection cuts in our analysis and recommend the same for any cosmological analysis using BOSS galaxy data, as this provides a more isotropic selection and discards less than 3\% of the total available DR9 CMASS redshifts.}:
\begin{eqnarray}
 17.5 < i_{cmod}  < 19.9\\
r_{mod} - i_{mod}  < 2 \\
d_{\perp} > 0.55 \label{eq:hcut}\\
i_{fib2} < 21.5\\
i_{cmod}  < 19.86 + 1.6(d_{\perp} - 0.8) \label{eq:slide}
\end{eqnarray}
where all magnitudes are corrected for Galactic extinction (via the \citealt{SFD} dust maps), $i_{fib2}$ is the $i$-band magnitude within a $2^{\prime \prime}$ aperture, and 
\begin{equation}
d_{\perp} = r_{mod} - i_{mod} - (g_{mod} - r_{mod})/8.0 .
\label{eq:dp}
\end{equation}
These color cuts are designed to obtain a sample of galaxies with approximately constant stellar mass with $z \gtrsim 0.43$ and include many galaxies that would be considered `blue' by traditional SDSS (rest-frame) color cuts (see, e.g., \citealt{Strateva02}). Indeed, \cite{Masters11} find that 26\% of CMASS galaxies have a late-type (i.e., spiral disc) morphology. See \cite{Toj12} for a detailed description of the CMASS population of galaxies.

For CMASS targets, stars are further separated from galaxies by only keeping objects with
\begin{eqnarray}
i_{psf} - i_{mod} > 0.2 + 0.2(20.0-i_{mod})  \label{eq:sgsep1}\\
z_{psf}-z_{mod} > 9.125 -0.46z_{mod} \label{eq:sgsep2}
\end{eqnarray}

\noindent unless the object also passes the LOWZ cuts (only 0.5\% of objects passing the LOWZ selection cuts are stars), which are defined by
\begin{eqnarray}
r_{cmod} < 13.5 + c_{\parallel}/0.3 \label{eq:lzslide}\\
|c_{\perp}| < 0.2 \\
16 < r_{cmod} < 19.6 \\
r_{psf}-r_{mod} > 0.3 
\end{eqnarray}
where
\begin{equation}
c_{\parallel} = 0.7(g_{mod}-r_{mod})+1.2(r_{mod}-i_{mod}-0.18)
\end{equation}
and
\begin{equation}
c_{\perp} = r_{mod}-i_{mod}-(g_{mod}-r_{mod})/4.0 -0.18 .
\end{equation}

\noindent Some objects satisfy both the LOWZ and CMASS  selection criteria. We therefore apply a minimum (maximum) redshift of 0.43 to the CMASS (LOWZ) sample, after obtaining a redshift in order to have two mutually exclusive samples.

The earliest set of spectra obtained for LOWZ data used an overly restrictive algorithm designed to remove stellar contamination, which unfortunately removed a significant number of galaxies from the target sample. This algorithm was changed for later data and, to maximize the size of the sample with an isotropic selection algorithm, we reduce the area by excluding the regions observed with the restrictive algorithm. Thus, the coverage (after accounting for completeness, see Section \ref{sec:mask}) of the LOWZ sample we use (2208 deg$^2$) is smaller than that of the CMASS sample (3275 deg$^2$). Fig. \ref{fig:angular_distribution} displays the angular footprint of the LOWZ sample in red and the area that contains only CMASS data in blue. The footprint contains 327,349 CMASS targets and 132,060 LOWZ targets. All of the data in these catalogs will be publicly released in the SDSS DR9.

\subsection{Mask}
\label{sec:mask}
The BOSS DR9 geometry is constructed from a series of spectroscopic observations, each of which is a 3$^{\rm o}$ diameter circle on the sky, corresponding to one pointing of the telescope. Each of these circles\footnote{Each observation corresponds to a `plate'. Each set of targets has a unique tile, but multiple plates can observe the same tile (and thus repeat observation of the exact same set of targets).} contains a unique set of targets and its area represents a `tile' (see \citealt{Blanton03} and Dawson et al. in prep.). The total area covered by these tiles forms the basis of our angular mask. Some tiles are not fully covered by observation, usually due to a lack of imaging data in the targeted region, so these are additional boundaries that we include in the mask. We divide the total area into `sectors', which are defined as the areas covered by a unique set of tiles --- i.e. the regions where the spectroscopic observing conditions are the same. We use the software package Mangle\footnote{http://space.mit.edu/~molly/mangle/} (see \citealt{mangleH,mangleHT,mangle}) to use the tile positions to divide the area into the unique sectors that we use to define the mask. 

We also apply veto masks that exclude areas surrounding bright stars and imaging fields deemed not photometric, both of these processes are described in \cite{alph}. In these areas, we do not expect the observing conditions to allow uniform detection of BOSS galaxies. Additionally, we mask the 92$^{\prime\prime}$ diameter region at the center of every tile where fibers cannot be placed due to physical limitations. The veto mask is only applied to the data after targeting; that is, galaxies observed in non-photometric fields and near bright stars are excluded from our analysis. This provides a cleaner sample, albeit with a more complicated mask, than if we were to quantify the varying target selection due to these effects. These masks remove 4\% of the observed footprint.

The density of galaxy targets on the sky varies due to galaxy clustering. Therefore, given that there are a finite number of fibers for each tile, the percentage of targets receiving a fiber will vary. Additionally, fibers cannot be placed closer than 62$^{\prime\prime}$ due to the size of the cladding around each fiber. We denote a `close pair' as any object not assigned a fiber due to a collision with an object of the same target type (i.e., CMASS target with CMASS target), since collisions with objects of different types should show no spatial correlation. 

In each sector, we compile statistics using the same definitions as \cite{alph}. The angular completeness, $C_{\rm BOSS}$, and the redshift completeness, $C_{\rm red}$ are determined by first counting the number of objects in each sector that are:
\begin{enumerate}
\item spectroscopically confirmed stars ($N_{\rm star}$),
\item galaxies with redshifts from good BOSS spectra ($N_{\rm gal}$),
\item galaxies with redshifts from SDSS-II spectra ($N_{\rm known}$),
\item objects with BOSS spectra from which stellar classification or redshift determination failed ($N_{\rm fail}$),
\item objects with no spectra, in a close-pair ($N_{\rm cp}$),
 \item objects with no spectra, not in a close-pair ($N_{\rm missed}$).
\end{enumerate}
These definitions represent a complete accounting for the possible outcomes of BOSS targets. Objects contributing to $N_{\rm missed}$ will either be observed in the future or are fiber collisions with objects of a different target type. For each sector, \cite{alph} then define the following:
\begin{equation}
N_{\rm targ} = N_{\rm star}+N_{\rm gal}+N_{\rm fail}+
 N_{\rm cp}+N_{\rm missed}+N_{\rm known},
\end{equation}
where $N_{\rm targ}$ is the total number of target objects, and
\begin{equation}
N_{\rm obs} = N_{\rm star}+N_{\rm gal}+N_{\rm fail},
\end{equation}
where $N_{\rm obs}$ is the total number of objects within the sector with a BOSS spectrum. $C_{\rm BOSS}$ is then
\begin{equation}  \label{eq:comp}
 C_{\rm BOSS} = \frac{N_{\rm obs}+N_{\rm cp}}{N_{\rm targ}-N_{\rm known}},
\end{equation}
and finally $C_{red}$ is
\begin{equation}  \label{eq:red_comp}
 C_{\rm red} = \frac{N_{\rm gal}}{N_{\rm obs}-N_{\rm star}}.
\end{equation}
The $C_{\rm BOSS}$ completeness varies from sector to sector due to fiber collisions with objects with a different type and the fact that many objects will be observed in future observations. We subsample the known redshift sample (complete by definition) so that its completeness matches $C_{\rm BOSS}$. We discard from our analysis any sectors where $C_{\rm BOSS} < 0.7$ or $C_{\rm red} < 0.8$. These cuts remove 8\% of the total footprint covered by BOSS DR9, but only 3.5\% of galaxy redshifts. Making the completeness cuts more restrictive does not significantly affect any clustering statistic we measure. 

Fig. \ref{fig:maskcls} displays CMASS targets for a selected observed area. Areas covered by more than one tile are shaded such that the outline of each tile is clearly visible. These overlapping regions cover 41\% of the total DR9 footprint. Targets with good redshifts are plotted as small black points. Targets not allocated a fiber are green. Within a given sector, these should be random with respect to the position of other CMASS targets, and these are therefore accounted for by $C_{\rm BOSS}$. (Although they are more likely to occur in sectors where future observations are planned, they are still random {\it within these sectors}.). Targets not allocated a fiber due to close pair collisions are blue. This happens most frequently (but not exclusively) in regions covered by only one tile. Targets that were allocated a fiber but whose observation did not result in a good redshift measurement are red. In general, these occur more frequently near to the tile boundaries. We discuss these cases further in Section \ref{sec:zfail}. 

 We create random (unclustered) catalogs by isotropically populating the sky, then selecting only those positions lying inside sectors with $C_{\rm BOSS} > 0.7$ and $C_{\rm red} > 0.8$ and outside of the veto mask. We then cull the random positions in every sector based on their $C_{\rm BOSS}$ (i.e., if $C_{\rm BOSS} = 0.9$, we randomly remove 10\% of the random points). This process yields a random catalog that mimics the angular distribution of our galaxy catalogs, save for fiber collisions with galaxies of the same type, redshift completeness, and systematic effects in the imaging. We correct for these remaining effects using a series of weights, as described in Section \ref{sec:weights}. Our default approach is to assign to each random position the redshift of randomly selected galaxy and we test this approach in Section \ref{sec:ztest}.
\begin{figure}
\includegraphics[width=84mm]{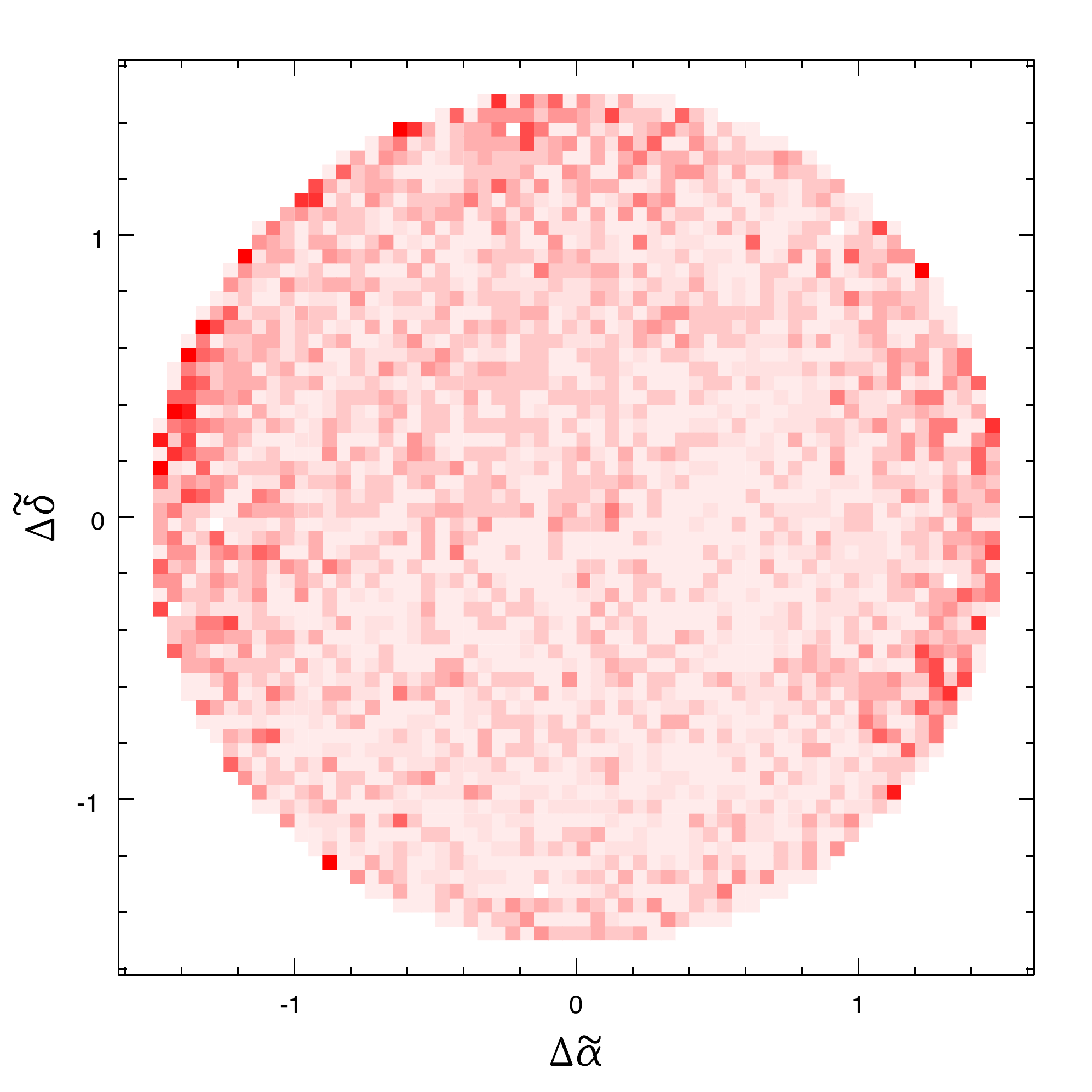}
  \caption{The percentage of failed CMASS redshifts as a function of the position on the tile, averaged over 817 DR9 tiles. The lightest regions are 0\%  and the darkest regions are 12\%. $\Delta \tilde{\alpha}$ is the distance along the right ascension direction and $\Delta \tilde{\delta}$ is the distance along the declination direction (both transformed so that the true angular separations are represented). }
  \label{fig:fracfail}
\end{figure}

\begin{figure}
\includegraphics[width=84mm]{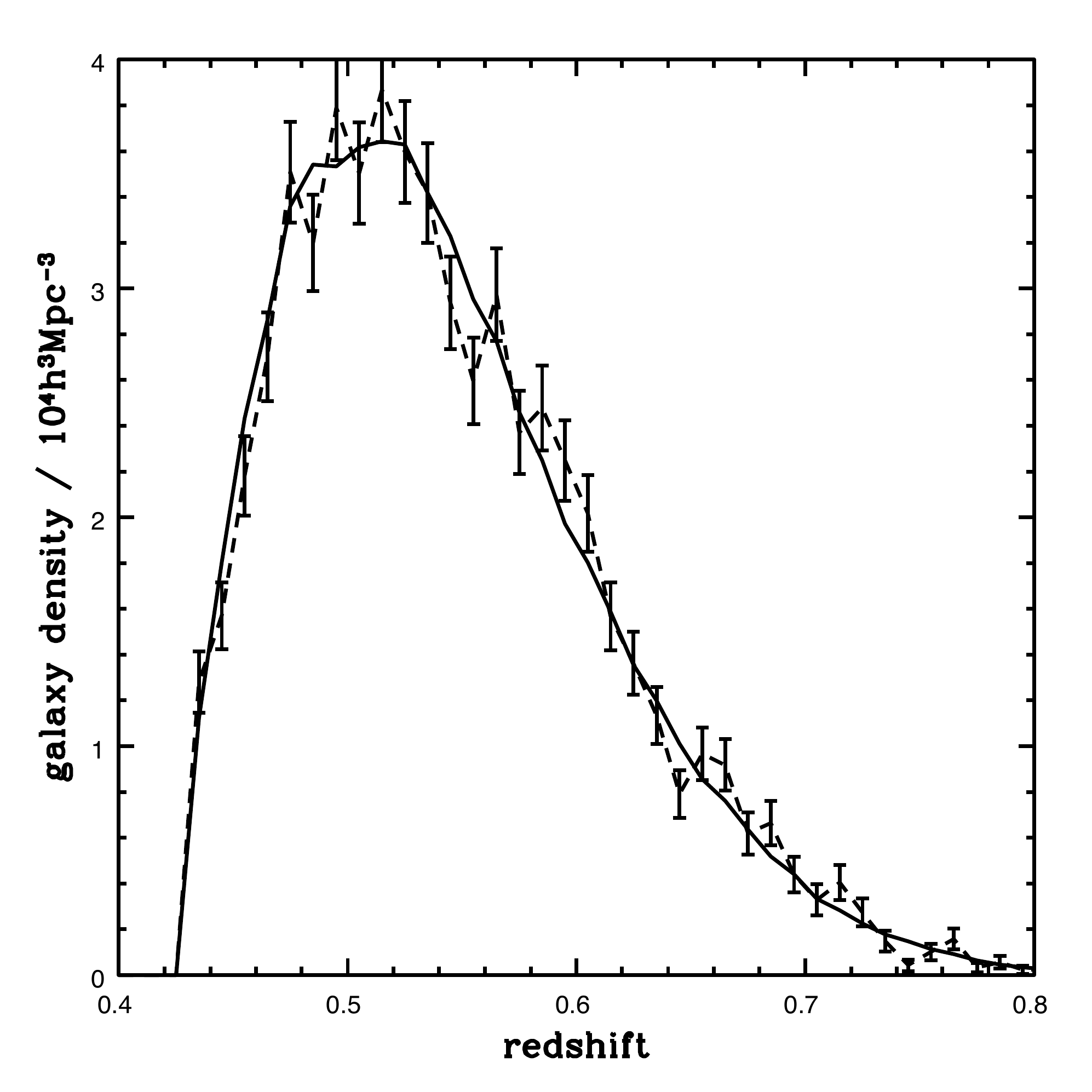}
  \caption{Galaxy spatial co-moving number density assuming a flat $\Lambda$CDM
    cosmology with $\Omega_m=0.274$, for CMASS galaxies. The solid line is
    calculated for all galaxies, while the dashed line only includes
    those galaxies nearest to a redshift failure, renormalised to
    match the total density of the full sample. The error-bars assume Poissonian distribution for the number counts in each bin.}
  \label{fig:number_density_vs_redshift}
\end{figure}

\
\subsection{Redshift Failures}
\label{sec:zfail}

We define a `good' redshift as any galaxy that does not have any `zWARNING' flags (as defined in \citealt{DR6}) determined by the spectroscopic pipeline. This flag indicates that the redshifts are unreliable typically because there are multiple acceptable redshift solutions (usually due to low signal to noise) or that the spectrum is defective. Analysis of repeat observations of BOSS targets and visual inspection reveal that galaxies with zWarning = 0 are reliable (accurate to $< 0.001$ in $\Delta z/(1+z)$) at the 99.7\% level whereas those with zWarning $> 0$  are reliable just 67\% of the time. For CMASS targets, good redshifts are obtained for 98.2\% of targets; for LOWZ, it is 99.6\%. Although this completeness is quite high, one may worry that the failures may relate to observational systematics or have a preferred location on an observing tile.

\begin{figure*}
\begin{minipage}{7in}
\centering
 \includegraphics[width=5.5in]{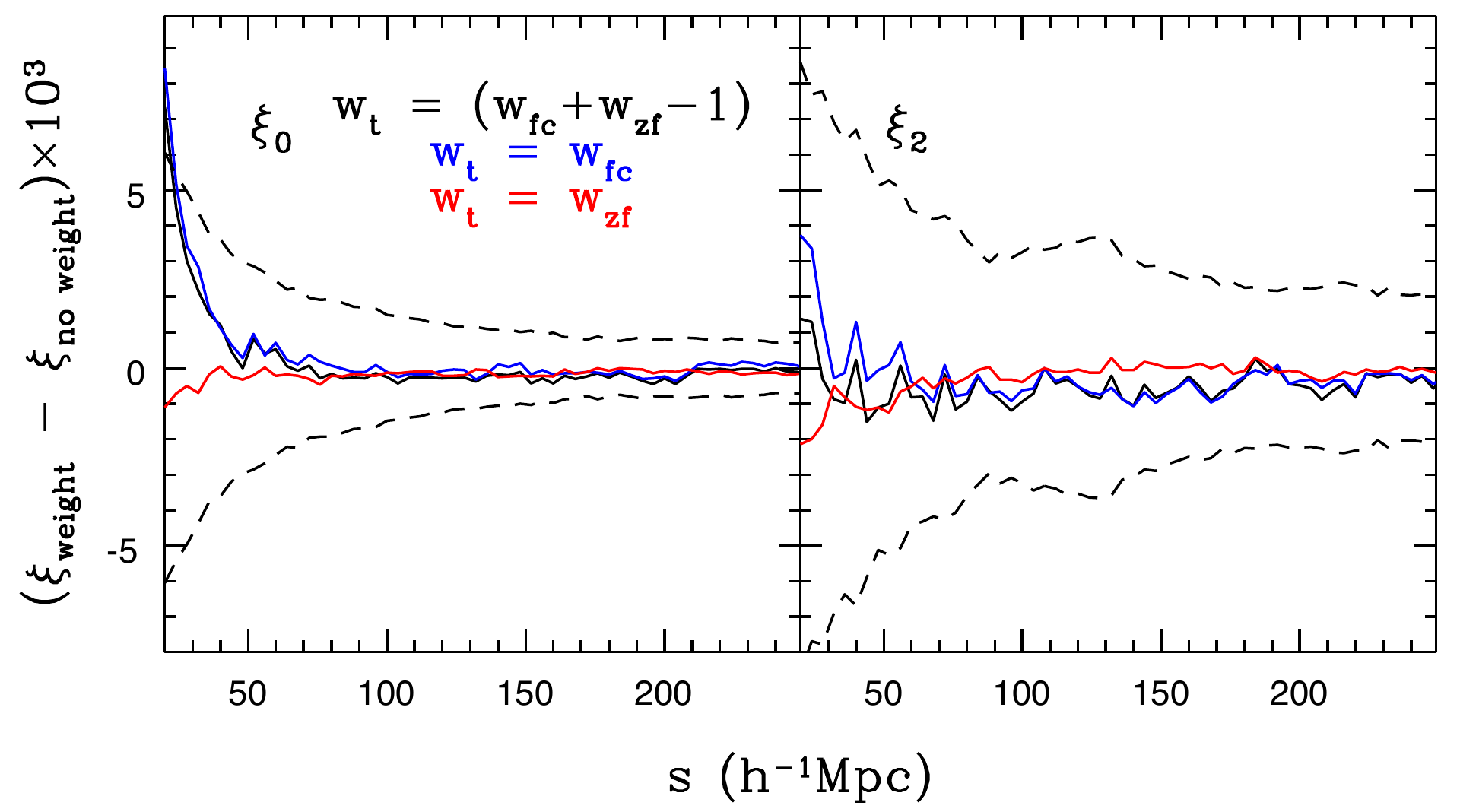}
 \caption{The effect of including weights for redshift failures ($w_{zf}$, red) and un-observed close-pairs due to fiber-collisions ($w_{fc}$, blue) and their combination (black) on $\xi_{0}$ and $\xi_{2}$. The dashed lines display the $1\sigma$ statistical uncertainty expected from mock catalogs.}
 \label{fig:cpzf}
 \end{minipage}
 \end{figure*}

BOSS fibers are numbered such that a given fiber corresponds to a particular position on the CCD. The spectrograph optics point spread function degrades near the edges of the CCDs, thus lowering the quality of the extracted spectra and reducing the likelihood of obtaining good quality redshifts from spectra near the CCD edges. (See Gunn et al. in prep. for more details on the performance of the BOSS spectrograph.) This correlation between redshift quality and fiber number translates into a spatial dependence on the sky, given that fibers are not assigned randomly. In order to test this effect, we translate all of the fiber positions of galaxies targeted by BOSS to positions relative to the center of the tile. This allows us to determine the redshift failure rate as a function of position on the tile (and thus whether redshift failures may impart angular fluctuations in the density of observed galaxies). The result of this test is displayed in Fig. \ref{fig:fracfail}. The failed redshifts are not only more likely to be on the edge of a tile, but appear concentrated near the minimum and maximum right ascension of each tile. We apply weights (see Section \ref{sec:weights}) to correct for this spatial dependence, but find there to be a negligible affect on the measured clustering (see Fig. \ref{fig:cpzf}).

Fig.~\ref{fig:number_density_vs_redshift} shows the galaxy spatial number
density for the CMASS sample, as a function of redshift. We
also plot the normalised (so that it has the same integral) number density against redshift for the
galaxies nearest to a redshift failure, $n_{nzf}(z)$. Were there a strong trend with redshift, for example that we were only missing redshifts for high-redshift galaxies, we should
expect that the nearest neighbours to the redshift failures (which
should be selected with similar properties, such as fiber ID, seeing
and extinction), should predominantly be at low redshift. In fact we
see no such trend --- if anything we find evidence to the contrary. We estimate the uncertainty on $n_{nzf}(z)$ in each $z$ bin by assuming a Poissonian distribution for the number counts in each bin and we determine the $\chi^2$ when the (normalized) $n_{nzf}(z)$ is compared to that of the full $n(z)$. We find $\chi^2 =$ 34.4 for $0.43 < z < 0.7$ (27 bins; 15\% of consistent samples would have a higher $\chi^2$) and $\chi^2 =$ 23.7 for $0.5 < z < 0.7$ (20 bins; 26\% of consistent sample would have a higher $\chi^2$). We therefore find no evidence that the spatially dependent component (which is the component that we are interested in, as it may create a spurious clustering signal) of the redshift-failure probability is dependent on redshift.

\section{Analysis Techniques}
\label{sec:calc}

\subsection{Clustering Estimators}
\label{sec:est}
We use the \cite{LZ} estimator to calculate the anisotropic redshift space correlation function, $\xi(s,\mu)$, where $s$ is the redshift-space separation in $h^{-1}$Mpc and $\mu$ is the angle to the line-of-sight.

\begin{equation}
\xi(s,\mu) = \frac{DD(s,\mu)-2DR(s,\mu)}{RR(s,\mu)}+1
\end{equation}

\noindent where $D$ represents the data sample (i.e., BOSS galaxies) and $R$ represents the random sample (occupying the angular footprint and with the same redshift distribution as the data sample) and the paircounts are normalised to the total number.

In linear theory, the first three moments of $\xi(s,\mu)$, expanded in Legendre polynomials, contain all of the information: 
\begin{equation}
\xi_{\ell}(s) =  \frac{(2\ell + 1)}{2}\int_{-1}^{1} {\rm d}\mu P_{\ell}(\mu)\xi(s,\mu).
\end{equation}
We therefore weight pairs by $P_{\ell}$ (yielding separate pair-counts for $\ell = 0,2,4$ for each of $DD$, $DR$, and $RR$). Labelling the $P_{\ell}$ weighted pair-counts with subscript $\ell$, we determine $\xi_{\ell}(s)$ via
\begin{equation}
\frac{2\xi_{\ell}(s)}{2\ell+1} =\frac{DD_{\ell}(s)-2DR_{\ell}(s)+RR_{\ell}(s)}{RR_0(s)}
\end{equation}
We count pairs in bin 7 $h^{-1}$ Mpc wide in $s$ and focus our efforts on understanding $\xi_0$ and $\xi_2$, (rather than the full $\xi(s,\mu)$) as these two measurements are expected to contain almost all of the information. In general, one must be careful, as our procedure implicitly assumes that pairs are isotropically selected in $\mu$, and this is not true for a general survey geometry. In this study, we do not need to be concerned, as we will always be comparing our results to those obtained via mock catalogs (which accurately match the survey geometry; see Section \ref{sec:mocks} and \citealt{Manera12mock}), but it must be accounted for in studies that compare more general models to the data. For a discussion of the procedures one may use, in general, when a survey does not provide an isotropically distributed sample of pairs, see \cite{Sam11}. 

We also use {\rm HEALPix} \citep{Gor05} maps, using Nside$=$256, to calculate projected auto/cross-correlation functions, $\xi_p(r_{eff})$ of galaxies and potential systematics (such as Galactic extinction). We split the sample into redshift shells of width $\Delta z = 0.01$ and define the overdensity, $\delta$, in redshift shell and pixel $i$ as
\begin{equation}
\delta_{i,z} = x_{i,z}/x_{ave,z} - 1,
\end{equation}
where $x_i$ is the value of the quantity in question in pixel $i$ and $x_{ave}$ is the average of the quantity over all pixels. We can thus calculate $\xi_p$, as a function of the effective scale, $r_{eff}$, using pixelized maps via 
\begin{equation}
\xi_p(r_{eff}) = \frac{\sum_{i,j,z1,z2} \delta^1_{i,z1}\delta^2_{j,z2} \Theta_{i,j,z1,z2}(r)N_1(z1)N_2(z2)w_iw_j}{\sum_{i,j,z1,z2}\Theta_{i,j,z1,z2}(r)N_1(z1)N_2(z2)w_iw_j}
\label{eq:xipix}
\end{equation}
where the indices $i,j$ represent the angular positions of pixels $i$ and $j$ and $z1,z2$ represent redshift slices, $\Theta_{i,j,z1,z2}(r)$ is 1 if the distance between the pixels (as determined by $i,j,z1,z2$) is within the bin defined by $r_{eff}\pm\delta r_{eff}$ and 0 otherwise, $N(z_1)$ is the number of galaxies in shell $z_1$, and $w_i$ is the weight of the pixel (see the following section), which we determine using the random catalog. It is straightforward to insert a purely angular map (e.g., the Galactic extinction) to determine how its angular cross-correlation with the galaxy field is translated to the physical scale $r_{eff}$. One simply holds its overdensity field constant with redshift and assigns it a flat $n(z)$ and proceeds through the sum defined above. 

We also calculate power spectra, $P(k)$, using the standard Fourier technique of \cite{FKP}, as described by \cite{ReidDR7}. Here the window function is accounted for as a convolution of the model. This implies power spectra can only be easily compared, without correcting for varying window functions, when they have the same selection and weighting. We therefore predominantly use the correlation function to present results, but we will show $P(k)$ measurements in Section \ref{sec:remain}, as these measurements isolate the largest wavelength density perturbations (separate bins in $s$ are highly covariant). 

\subsection{Weights}
\label{sec:weights}
We use weights to account for spatial variations in redshift failures, fiber collisions, and imaging systematics, i.e., those effects that are not quantified via the $C_{\rm BOSS}$ completeness as described above. Given the total weight, $w_{tot}$, for each galaxy,
\begin{equation}
DD(s,\mu) = \sum_i\sum_j w_{tot,i}w_{tot,j}\Theta_{ij}(s,\mu),
\end{equation}
where $\Theta_{ij}(s,\mu)$ is 1 if the separation between the two galaxies and the angle they make to the line of sight is within the particular bin, and 0 otherwise. These weights correct the galaxy densities to provide a more isotropic selection. They should therefore not be applied to a random catalog, if it is based on an isotropic selection.

We will find that there is systematic relationship between the density of targets (see Section \ref{sec:angvar}) and the density of stars, and we therefore require a weight, $w_{star}$. We describe how we determine $w_{star}$ in Section \ref{sec:angvar}. We account for systematics that affect the process of going from the target catalog to a redshift catalog with weights for both redshift failures, $w_{zf}$, and fiber collisions with targets of the same type (`close pairs'), $w_{fc}$. We start by assigning each $w_{zf}$ and $w_{fc}$ unit weight. We have found that probability of a redshift failure depends on position of the fiber on the tile center (see Fig. \ref{fig:fracfail}.) To account for this, we find the nearest neighbor on the sky to the object with the failed redshift, and increase its $w_{zf}$ by one. Such a weighting preserves the large-scale angular auto-correlation function. Further, this action should not bias the redshift-space correlation function, as we find no evidence of a redshift dependence on the probability of a redshift failure (see Fig. \ref{fig:number_density_vs_redshift}). Similarly, we determine $w_{fc}$, by up-weighting the colliding neighbor by one.  \cite{HGuo} have developed an optimized method to account for fiber collisions at all scales, but at large scales ($s\gtrsim10h^{-1}$Mpc), our method should produce identical results. We combine the redshift failure and fiber-collision weights into a single weight equal to $w_{fc} + w_{zf} -1$. Thus, the total weight we apply to each galaxy is
\begin{equation}
w_{tot} = w_{star}(w_{fc} + w_{zf} -1).
\label{eq:wtot}
\end{equation}
The weight for any galaxy can be arbitrarily large, e.g., a galaxy will be given a weight of three if it is nearest to a target with a redshift failure and also causes a fiber collision with another galaxy (and has $w_{star}=1$).

Fig. \ref{fig:cpzf} displays the difference we find in the measured $\xi_{0}$ and $\xi_{2}$ of CMASS galaxies when applying the redshift failure (red) and fiber-collision (blue) weights and their combination (black). At scales greater than 50 $h^{-1}$Mpc, the effect is negligible compared to the expected statistical uncertainty for the sample (displayed in dashed black lines). Redshift failures have no significant effect on the measured clustering.

The application of the fiber-collision weights increases the clustering amplitudes at scales less than $80h^{-1}$ Mpc. This effect is expected, as fiber collisions will be more likely to occur in highly clustered samples. At $20h^{-1}$Mpc, the difference is nearly 1$\sigma$. The weights for redshift failures impart a slight decrease in the measured amplitudes, at a level consistently less than 20\% of the statistical uncertainty determined using the mock galaxy catalogs (see Section \ref{sec:mocks}).  

We also want to optimally weight galaxies based on their number density, as suggested by \cite{FKP}. We refer to these weights as `FKP' weights. To do so, we use a constant $P_{FKP} = 20000 h^3$Mpc$^{-3}$ (as do \citealt{alph}; this is roughly the amplitude of the CMASS $P(k)$ at $k =0.1 h$Mpc$^{-1}$) and weight by
\begin{equation}
 w_{P} =1/(1+n(z)P_{FKP}),
 \label{eq:fkp}
 \end{equation}
 where $n(z)$ is the number density of galaxies at redshift $z$, determined using our assumed cosmology. The purpose of these weights is to optimally weight areas with different number densities, not to correct observed number densities for a systematic effect. Thus, they are applied to both the random objects and the real galaxies; the final pair counts can be expressed as:
\begin{equation}
DD_{\ell}(s) = \sum_{Di}\sum_{Dj} w_{P,Di}w_{tot,Di}w_{P,Dj}w_{tot,Dj}\Theta_{DiDj}(s)P_{\ell}(\mu),
\end{equation}
\begin{equation}
DR_{\ell}(s) = \sum_{Di}\sum_{Rj} w_{P,Di}w_{tot,Di}w_{P,Rj}\Theta_{DiRj}(s)P_{\ell}(\mu),
\end{equation}
\begin{equation}
RR_{\ell}(s) = \sum_{Ri}\sum_{Rj} w_{P,Ri}w_{P,Rj}\Theta_{RiRj}(s)P_{\ell}(\mu).
\end{equation}

\subsection{Covariance Matrices}
\label{sec:mocks}

We use mock galaxy catalogs with realistic clustering to determine covariance matrices for the distributions of galaxies and their clustering measurements. The mock catalogs have been produced with the method explained in \cite{Manera12mock}. This method is inspired by the Perturbation Theory Halos (PTHalos) paper of \cite{PTHalos}, and have been calibrated using 40 N-Body galaxy and halo mock catalogs generated using the LasDamas simulations\footnote{http://lss.phy.vanderbilt.edu/lasdamas/} (McBride et al in prep.). 

Using Second Order Perturbation Theory (2LPT), \cite{Manera12mock} generate 600 matter fields at redshift 0.55 drawn using periodic boxes of size L=2400 $h^{-1}$Mpc (one matter field from each box). Each uses a flat cosmology defined by $\Omega_{m}=0.274, \Omega_{b}h^2=0.0224, h=0.70, n_s=0.95$, and $\sigma_8=0.8$. This is the same cosmology as \cite{White11BEDR}  and is close to WMAP7 parameters \citep{larsonwmap}. Halos from 2LPT runs are identified using a friends-of-friends algorithm and are then mass-calibrated using the \cite{TinkerMF} mass function. These halos are then populated with galaxies using the halo-occupation-distribution (HOD) parameterization defined by \cite{ZCZ}. We determine the HOD parameters by fitting the CMASS $\xi_0(s)$ measurement using data in a range of $30<s<80 h^{-1}$Mpc. 

For each mock realization, the periodic boxes are reshaped to match the final BOSS geometries for the NGC or the SGC footprints (see Fig \ref{fig:footprint}; the box sizes are not large enough to accommodate both NGC and SGC simultaneously). Redshift distortions are then applied based on the 2LPT velocity field, combined with a model for the intra-halo velocity dispersion. The final DR9 angular masks are then applied, and galaxies are sampled from the full simulations to match the CMASS radial selection function (displayed in Fig \ref{fig:number_density_vs_redshift}), thus yielding 600 mock galaxy catalogs mimicking the clustering of BOSS CMASS galaxies. For further details of the methods, see \cite{Manera12mock}. 

\begin{figure}
  \includegraphics[width=84mm]{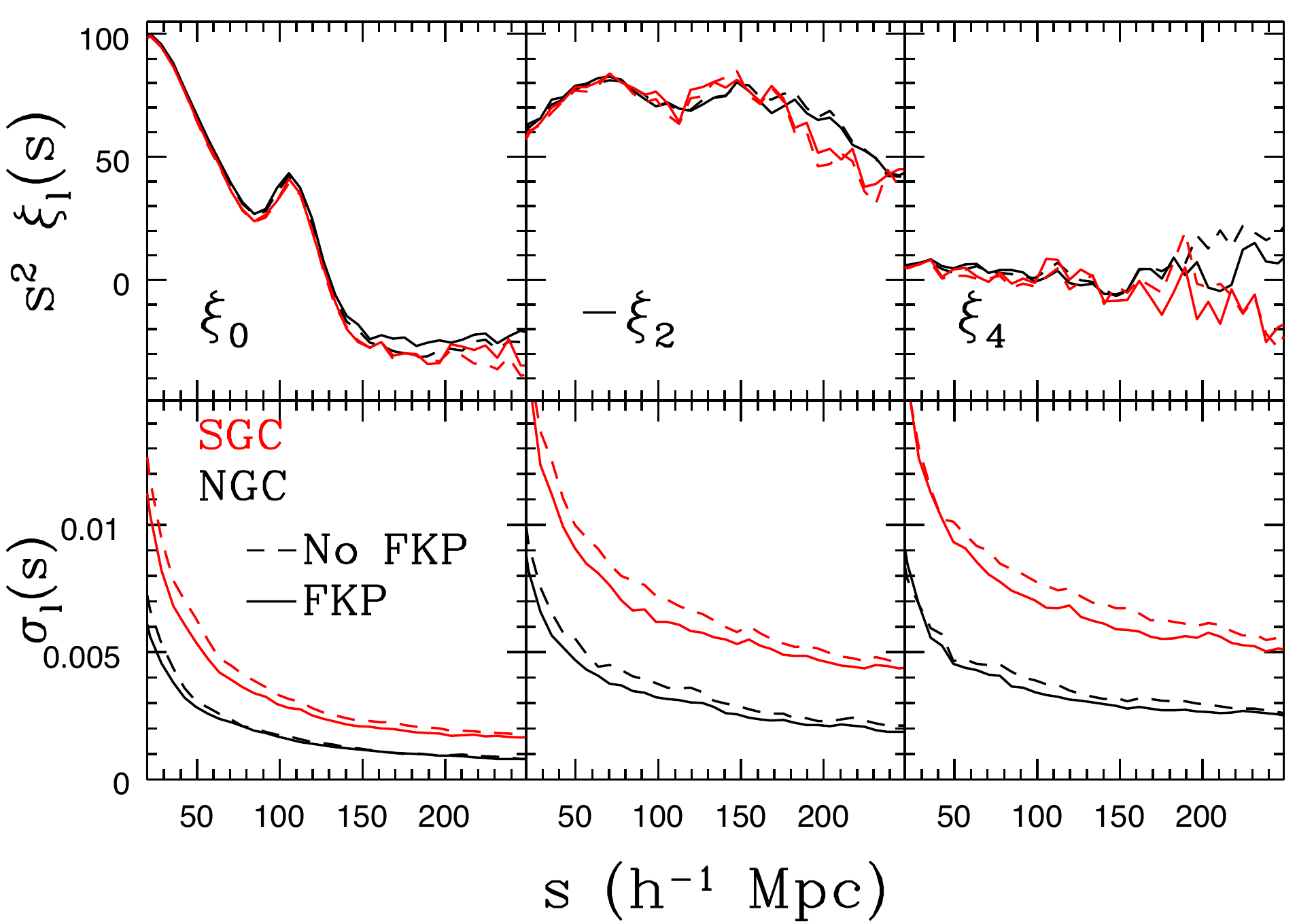}
  \caption{Top panels: The average of $\xi_{\ell}(s)$ we determine using 600 mocks, for the NGC (black) and SGC (red) footprints and with (solid lines) and without (dashed lines) FKP weights. The baryon acoustic oscillation peak can clearly be seen at $s\sim100h^{-1}$Mpc in $\xi_0$. Bottom panels: the standard deviation of the 600 mocks, using the same scheme as the top panels.}
  \label{fig:ximock}
\end{figure}

For each of the 600 mock galaxy catalogs, in both the NGC and SGC, we calculate $\xi_{\ell}$ with and without FKP weights (see Section \ref{sec:weights}). The mean results of these measurements (top panels) and their standard deviations (bottom panels) are displayed in Fig. \ref{fig:ximock}. The FKP weights decrease the standard deviation, typically by 10\%. We note that from here on we do not test $\xi_4(s)$ measurements, as, in \cite{ReidRSD12} the information added by incorporating $\xi_4(s)$ affords only marginal improvements. 

The covariance of $\xi_{\ell}(s)$ at separations $s_1,s_2$ and moments $\ell,\ell^{\prime}$ is determined using the standard mathematical definition, i.e., 
\begin{equation}
C_{\ell,\ell^{\prime}}(s_1,s_2) = \frac{1}{599}\sum_{i=1}^{600}\left(\xi^i_{\ell}(s_1)-\xi^{ave}_{\ell}(s_1)\right)\left(\xi^i_{\ell^{\prime}}(s_2)-\xi^{ave}_{\ell^{\prime}}(s_2)\right).
\end{equation}
To calculate the covariance matrix over the total footprint, we assume the two regions are independent and thus $C^{-1}_{total} = C^{-1}_{North}+C^{-1}_{South}$. We use the full covariance matrix to calculate $\chi^2$ in the usual manner and apply tests on how our treatment of the data affects derived parameters.  Any time we quote a $\chi^2$ value, it is calculated using the covariance matrix determined from the mocks. 

\subsection{Modelling}
\label{sec:mod}
We use the mean of clustering measurements determined using the mock catalogs to define the fiducial models we test. We derive parameters relating to the amplitude of the real-space galaxy density field, $\tilde{b}$, the amplitude of the galaxy velocity field, $\tilde{f}$, and the factor by which the distances assumed by our cosmological model are incorrect, $\tilde{\alpha}$ (as this relates to the position of the BAO feature, see \citealt{alph}). We expect that if we obtain robust results on these parameters, one should be able to derive robust constraints on any derived parameter, over the same range in $s$.

\begin{figure}
  \includegraphics[width=84mm]{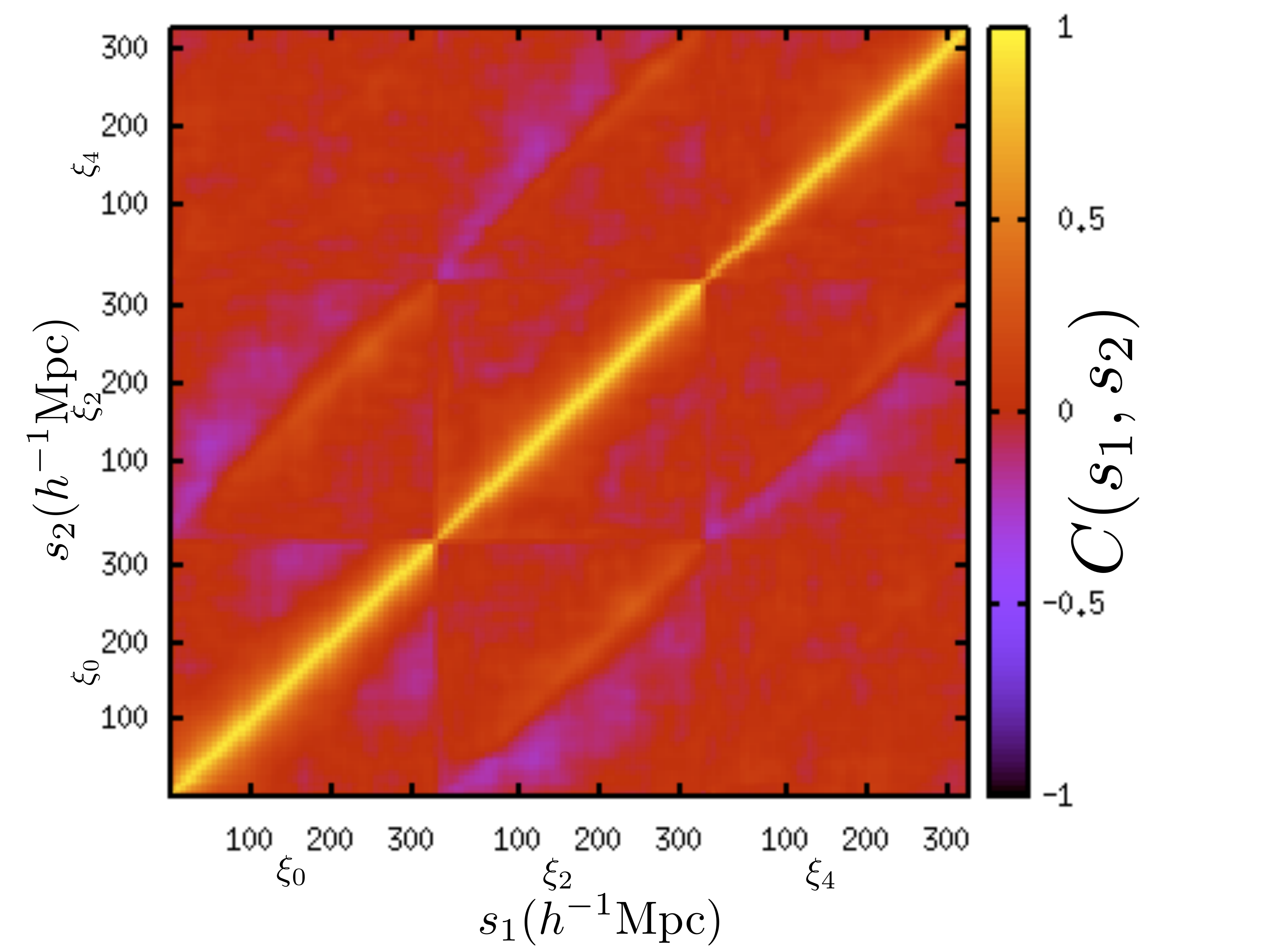}
  \caption{The normalized covariance matrix of and between $\xi_{\ell}$ determined using 600 mock galaxy catalogs simulating the BOSS DR9 selection function.}
  \label{fig:covmock}
\end{figure}

We assume a linear model for redshift-space clustering \citep{Kaiser} and assume linear biasing. In this model, given the real-space correlation function of matter, $\xi_{M}$,
\begin{eqnarray}
\xi_0(b,f) & = &\xi_{M}(b^2+2/3bf+0.2f^2)\label{eq:m0}\\
  \xi_2(b,f) &=& (\frac{4}{3}bf+\frac{4}{7}f^2)[\xi_{M}-\xi_{M}'], \label{eq:m2}\\
  \xi_4(f) &=& \frac{8}{35}f^2[\xi_{M}+\frac{5}{2}\xi_{M}'-\frac{7}{2}\xi_{M}''],
\label{eq:m4}
\end{eqnarray}
 and
\begin{eqnarray}
  \xi'\equiv3s^{-3}\int^s_0\xi(r')(r')^2dr' \\ 
  \xi''\equiv5s^{-5}\int^s_0\xi(r')(r')^4dr',
\end{eqnarray}
\citep{hamilton92} where $b$ is the real-space linear bias of the galaxy sample and $f$ is the rate of change of the linear growth rate: $f\equiv d\log D/d\log a$, where $D$ is the linear growth rate and $a$ is the scale factor of the Universe. We therefore determine our model $\xi_{\ell}^{mod}$ by re-scaling the average mock $\xi^{mock}_{\ell}$ according to:
\begin{eqnarray}
\xi^{mod}_0(\tilde{b},\tilde{f}) = \xi^{mock}_0(\tilde{b}^2+2/3\tilde{b}\tilde{f}+0.2\tilde{f}^2)/(4.657)\label{eq:mod0}\\
\xi^{mod}_2(\tilde{b},\tilde{f}) = \xi^{mock}_2(4/3\tilde{b}\tilde{f}+4/7\tilde{f}^2)/(2.188)\label{eq:mod2}\\
\xi^{mod}_4(\tilde{f}) = \xi^{mock}_4\tilde{f}^2/0.548,\label{eq:mod4}
\end{eqnarray}
The values in the denominators account for a bias of the mocks of $b=1.9$ (fits to the model of \citealt{RW11} suggest this is accurate to within 2\% and we fix it at 1.9 for simplicity) and the fact that $f(z=0.55) = 0.74$ for the cosmology used by the mocks (flat $\Omega_{m} = 0.274$; e.g., the denominator for Eq. \ref{eq:mod0} is given by $1.9^2 + 2/3\times1.9\times0.74+ 0.2\times0.74^2 = 4.657$). This model thus assumes the same scalings in amplitude as linear RSD theory, but uses the shape of the the mean mock galaxy $\xi_{\ell}(s)$, which include non-linear RSD features. We note that linear theory is not appropriate to obtain accurate estimates of $b$ and $f$ (see \citealt{RW11}); it is for this reason that we quote our measurements as, e.g., $\tilde{b}$ rather than $b$. However, we expect that if our treatment of the data yields robust estimates of $\tilde{b}$ and $\tilde{f}$, the measurements will also be robust when more accurate models are tested in \cite{SamRSD12} and \cite{ReidRSD12}. 

We also test the robustness of the data to a simple dilation in scale, using models where we vary a `stretch parameter', $\tilde{\alpha}$. In this case
\begin{equation}
\xi^{mod}_0(\tilde{b},\tilde{\alpha},s) = \xi^{mock}_0(\tilde{\alpha} s)(\tilde{b}^2+0.48\tilde{b}+0.1036)/(4.657),
\end{equation}
(where we are now fixing $\tilde{f}$). We determine $\xi^{mock}_0(\tilde{\alpha} s)$ via power-law interpolation of the fiducial mean mock result at scales $s < 80 h^{-1}$Mpc and linear interpolation at larger scales. We note that this stretch parameter will contain information on, at least, $\Omega_{m}h^2$ (from the overall peak of the power spectrum) and the position of the BAO peak. We therefore believe that if our treatment of the data yields robust $\tilde{\alpha}$ values, robust measurements of both the BAO position and $\Omega_{m}h^2$ will be obtained when more sophisticated models are applied, as in \cite{alph} and \cite{SanchezCos12}. We study the recovered values of $\tilde{b}$, $\tilde{f}$, and $\tilde{\alpha}$ according to Galactic hemisphere (Section \ref{sec:nsxi}), angular weights (Section \ref{sec:wxi}), and redshift (Section \ref{sec:xizspl}) and these results are summarized in Table \ref{tab:allpar}, found in Section \ref{sec:con}.

\subsection{Default Survey Window}
The following three sections define and justify our recommendations for how to treat the survey window in regard to the NGC and SGC footprints, photometric systematics, and the radial selection function. In each section, these recommendations are used, unless otherwise noted. The default is to: treat the NCG and SGC as having separate selection functions and optimally combining their individual pair-counts, apply weights to each galaxy based on linear relationships between $i_{fib2}$ magnitude and stellar density, and apply redshifts to the random sample by randomly selecting redshifts from the galaxy sample.

\section{Dependence on Galactic Hemisphere}
\label{sec:NS}
\begin{figure*}
\begin{minipage}{7in}
\centering
 \includegraphics[width=5.5in]{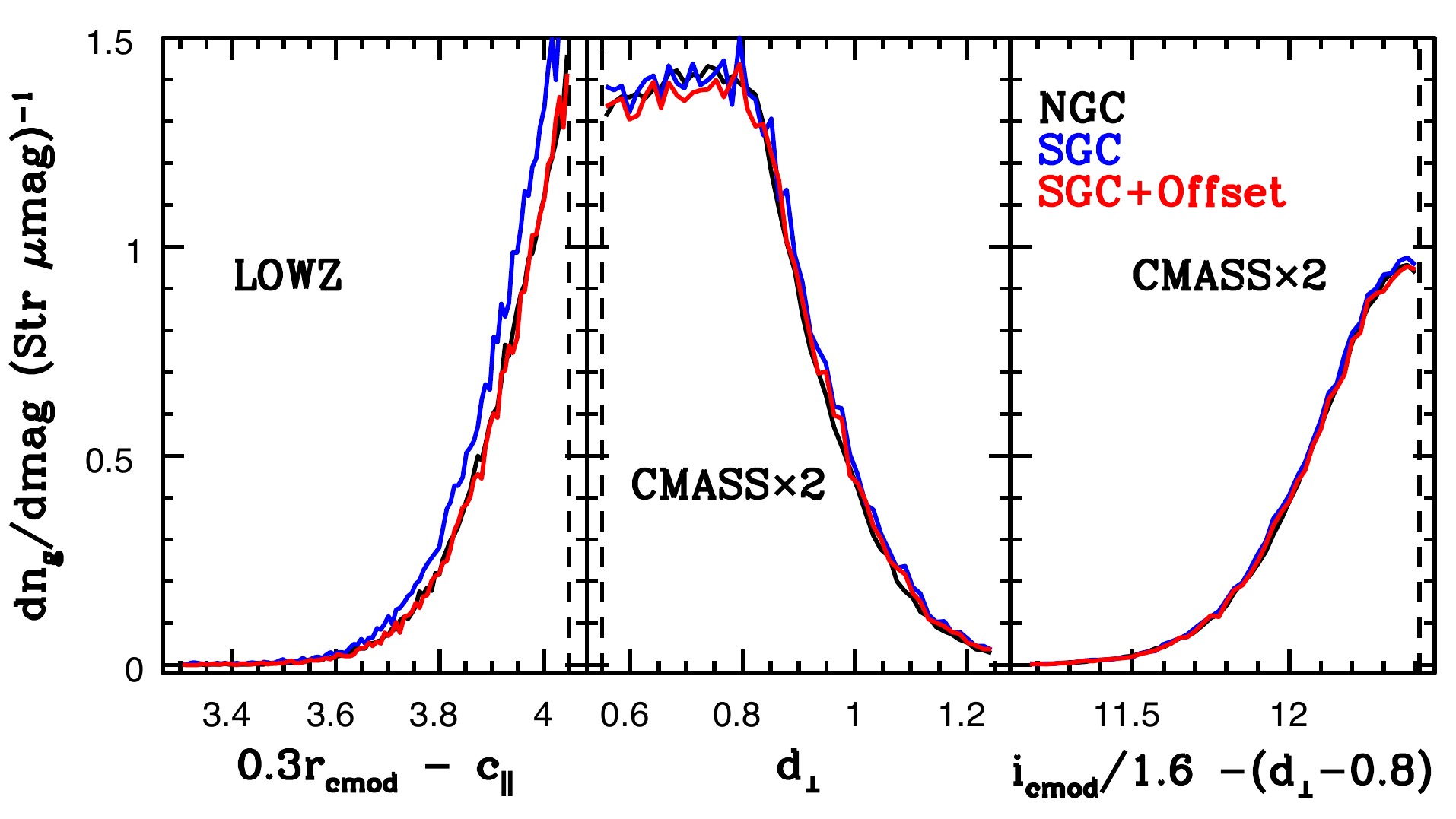}
 \caption{The distribution of DR9 spectroscopically-identified galaxies as a function of color combinations used in their selection, for objects in the Northern (black) and Southern (blue) Galactic Caps, and in Southern Galactic Cap after applying the Schlafly \& Finkbeiner (2010) offsets to the target selection (red). Dashed vertical lines display the location of the cut applied to BOSS targeting.}
 \label{fig:schcut}
 \end{minipage}
 \end{figure*}
 
The SDSS imaging was carried out in two large contiguous areas in the NGC and SGC (see Fig. \ref{fig:footprint}). The mean sky background and airmass are both higher for the SGC imaging, which results in larger uncertainties on the measured magnitudes of this data, as the mean uncertainty for $i$-band CMASS targets is 0.076 in the NGC and 0.101 in the SGC. However, we show in Sections \ref{sec:angvar} and \ref{sec:zsys} that the projected number density and redshift distributions of BOSS targets do not depend on either sky background or airmass and we therefore find no evidence that the observing conditions should produce systematic differences between the properties of targets selected in the SGC and NGC. Of greater concern is the fact that the two regions are tied together with relatively few scans to measure the relative photometric calibration \citep{Pad08}. This suggests the possibility of a significant photometric offset between these two regions. 

\cite{Sch10} and \cite{Sch10b} have found systematic variations in the colors of the population of SDSS stars as a function of their position. These offsets reflect a combination of variations in stellar populations across the Galaxy, calibration errors in the SDSS photometry (at the 1\% level), and errors in the corrections for Galactic extinction. In particular, they find that there is a systematic offset in the measured photometry between the SGC and NGC (the amplitude of this offset is within the expected 1\% rms of DR8 photometric calibration errors). The CMASS cut is sensitive to the $d_{\perp}$ color, both due to the hard cut (Eq. \ref{eq:hcut}) and the sliding cut (Eq. \ref{eq:slide}). The LOWZ sample is sensitive to the $c_{||}$ color, due to its sliding cut (Eq. \ref{eq:lzslide}).  \cite{Sch10b} find a 0.015 mag offset in $c_{||}$ and a 0.0064 mag offset in $d_{\perp}$ between the NGC and SGC (based on their `spectrum based' method; see their Table 6). \cite{imsys} found that the 2\% difference in the number density of CMASS targets between the NGC and SGC hemisphere was consistent with this offset in $d_{\perp}$. In what follows, we repeat and improve upon the analysis performed in \cite{imsys} using only spectroscopically confirmed galaxies (albeit over a footprint 1/3 the size). 

\begin{figure}
  \includegraphics[width=84mm]{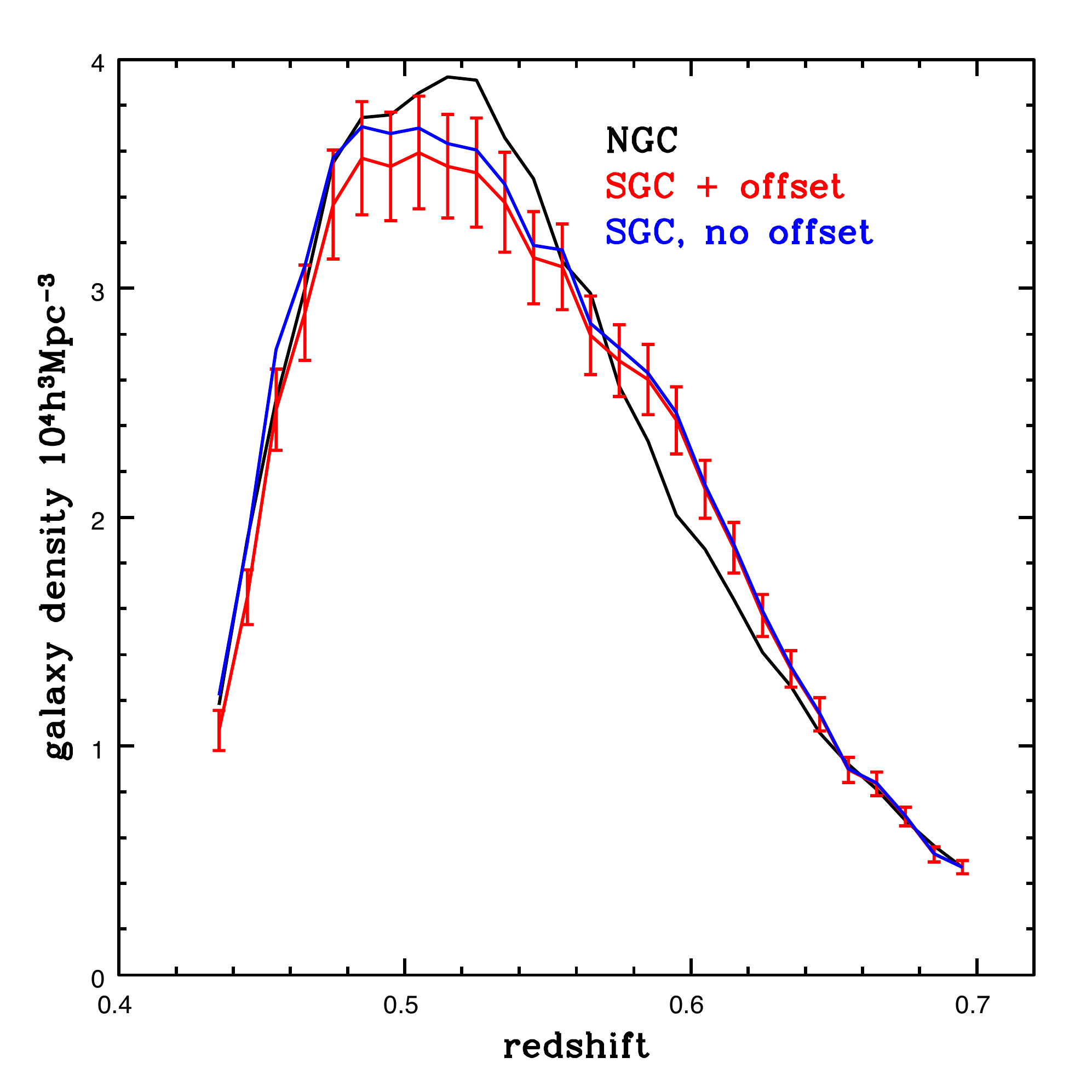}
  \caption{The galaxy spatial number density assuming a flat $\Lambda$CDM
    cosmology with $\Omega_m=0.274$ of CMASS objects in Northern (NGC) and Southern Galactic Caps (SGC). The red line displays the result when we apply the Schlafly \& Finkbeiner (2010) offsets to the target selection in the SGC. The error bars are determined using 600 mock catalogs cut to the angular footprint of the SGC.}
  \label{fig:nzNSwmerr}
\end{figure}

\subsection{Number densities}
Fig. \ref{fig:schcut} displays the distribution in galaxies vs. the color/magnitude information used to select them. We show the relations for spectroscopically identified galaxies and apply the redshift failure and close pair weights described in Section \ref{sec:weights} when determining number densities. The left panel shows the distribution of LOWZ galaxies against the value of the sliding cut, with the relationship for the NGC plotted in black and the relationship for the South plotted in red. At the cut (at 4.05 mag) the slope of the number density relationship is roughly 1.4$\times10^{6}$ galaxies per steradian per magnitude change in $c_{||}$. Thus, a 0.015 mag offset in $c_{||}$ (as implied by \citealt{Sch10b}) should cause a change of $\sim$ 2$\times 10^4$ galaxies per steradian. If we apply this offset, we get the distribution in blue in the left-hand panel of Fig. \ref{fig:schcut}. The curve appears much more consistent with the distribution in the NGC (black) than the fiducial SGC relationship (red).

In the DR9 CMASS sample, we find a 3.2\% higher number density in the SGC than in the NGC ($2.832\times 10^5$ str$^{-1}$ compared to $2.744\times 10^5$ str$^{-1}$). This is 1.2\% higher than found by \cite{imsys} due to a combination of the different footprints and the fact that we use spectroscopically identified galaxies. The middle panel displays the distribution of CMASS galaxies against their value of $d_{\perp}$, and the right-hand panel against the sliding cut. For the $d_{\perp} > 0.55$ cut, we should expect 7$\times 10^5$ galaxies per steradian per mag change in $d_{\perp}$, and for the sliding cut we should expect an extra 5$\times 10^5$ change in the number of galaxies per steradian per mag change in $d_{\perp}$. Thus, we should expect a change of roughly 1.2$\times 10^6$ targeted galaxies per steradian per mag offset in $d_{\perp}$ (the two cuts are not strongly covariant). Therefore, given the 0.0064 mag offset in $d_{\perp}$ between the NGC and SGC, we should expect 7700 galaxies per steradian in the SGC. This is 2.7\% of the number density of the CMASS sample. When we apply this offset to the target selection, we find there only remains a 0.2\% excess in the number density of objects in the SGC compared to the NGC. 
\begin{figure*}
\begin{minipage}{7in}
\centering
 \includegraphics[width=7in]{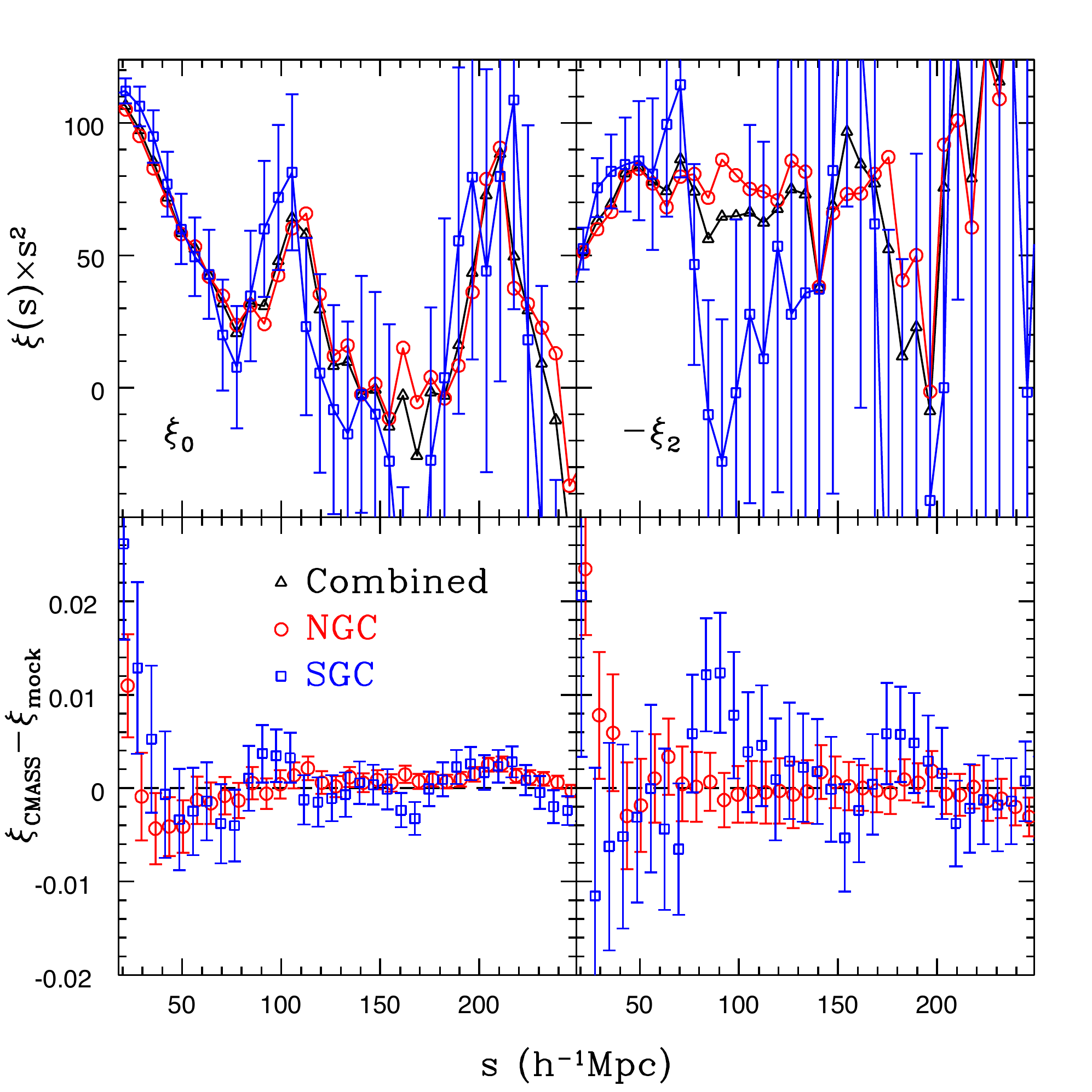}
  \caption{Top panels: The measured redshift space correlation functions, $\xi_{0}$ and $\xi_{2}$, of CMASS data in the Northern (NGC; red) and Southern Galactic Caps (SGC; blue), and their pair-weighted average (black triangles), using FKP weights and the $w_{star}$ weights. The error-bars are the standard deviations of the $\xi_{\ell}$ in the mocks drawn from the SGC footprint. For both the NGC and SGC measurements, the BAO feature is apparent at $s\sim100h^{-1}$Mpc. Bottom panels: The difference between the measured $\xi_{0,2}$ of the NGC (red) and SGC (south) CMASS samples and the mean of their respective mocks, after scaling the mocks for a best-fit bias. The error-bars are the standard deviations of the $\xi_{\ell}$ in the mocks drawn from the respective SGC and NGC footprints.}
  \label{fig:xiNScomb}
 \end{minipage}
\end{figure*}

Fig. \ref{fig:nzNSwmerr} displays the number density as a function of redshift, $n(z)$, for CMASS galaxies in the NGC (black) and SGC (blue), using bins of width $\Delta z = 0.01$. The number density is 10\% smaller around the peak of the distribution in the South, and this becomes more dramatic when we apply the \cite{Sch10b} offsets to the target selection. However, the number density is 20\% larger at $z=0.6$. The error bars are determined from the variations we find in the $n(z)$ of mocks cut to the same angular footprint as our Southern data set (the $n(z)$ for each individual mock varies due to the cosmic variance inherent in large-scale-structure). Using these mocks, we can also determine the covariance between the $n(z)$ bins, thus allowing us to calculate the $\chi^2$ between the SGC and the NGC. When the offsets are applied, we find $\chi^2 =  36$ (for 27 bins). The $\chi^2$ is higher (39) when the offsets are not applied to the target selection. We find that 55 of the 600 Southern mock $n(z)$ have a $\chi^2$ that is greater than 36 (when compared to the average of the 600 mocks) and 34 have a $\chi^2$ that is greater than 39 (this is roughly in line with the probabilities of 11\% and 6\% one obtains for these $\chi^2$ values and 27 degrees of freedom). Thus, applying the offsets makes the redshift distributions of the NGC and SGC samples more consistent, but the differences between them slightly unusual.

In summary, for both the CMASS and LOWZ samples, we find that the difference in their number densities is consistent with the level of color offset between the NGC and SGC as determined by \cite{Sch10b} using their spectrum method. Further, the offset is understood --- it is within the expected rms of DR8 calibration errors and found between the two regions that have the least available data for relative calibration. We can apply these color offsets to the selection of galaxies in the South in an attempt to make a homogenous sample, as doing so only makes the cuts more restrictive. However, there is some uncertainty inherent in the level of the offset, and, based on the mocks, the expected variance in the number density between the NGC and SGC is 2\%. Further, the $n(z)$ distributions remain slightly inconsistent. We therefore believe the most conservative approach is to treat the two samples as having separate selection functions, due (at least in part) to the fact that there are offsets in the photometry between the two regions. Thus, we analyze all galaxies observed in the South separately, accepting they comprise a denser sample than those in the NGC.

\begin{figure*}
\begin{minipage}{7in}
\centering
 \includegraphics[width=7in]{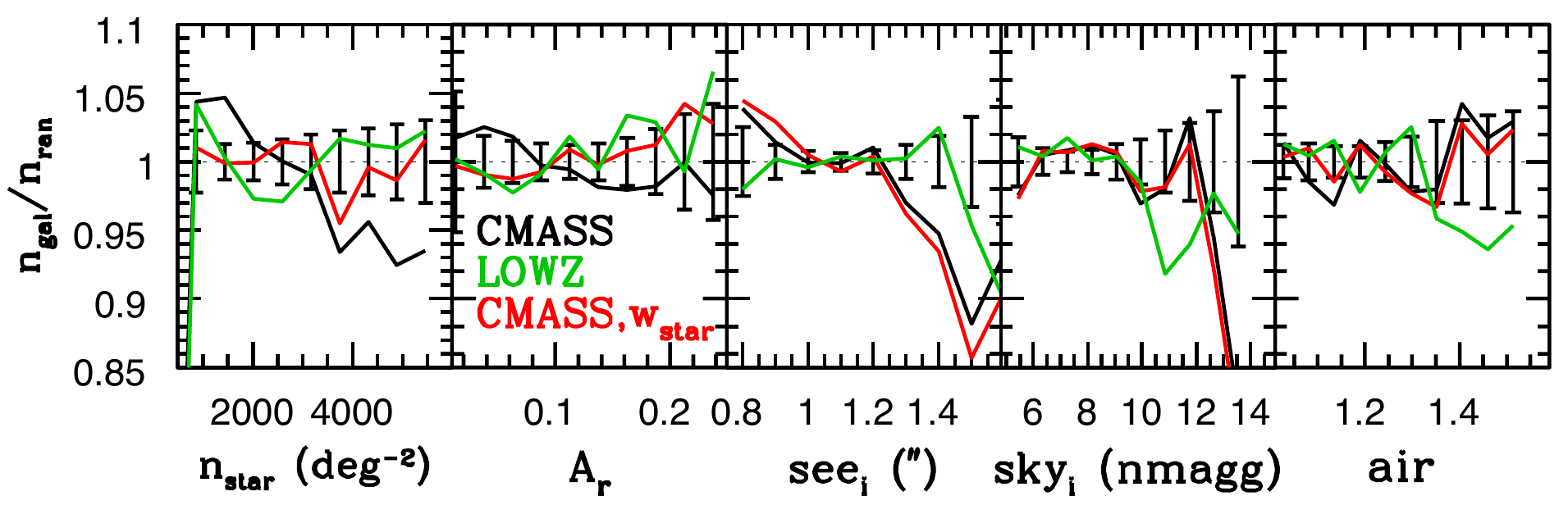}
  \caption{The projected number density of CMASS galaxies as a function of the potential systematics: stellar density ($n_{star}$), Galactic extinction in the $r$-band ($A_r$), the $i$-band seeing (see$_i$), $i$-band background sky flux in nanomaggies/arcsec$^2$, denoted $nmagg$, (`sky$_i$';  $nmagg$ are related to the magnitude $m$, via $m = $22.5 -2.5log$(nmagg)$), and the airmass (air). The black lines display the results for the CMASS sample, and the green the results for the LOWZ sample. The red lines display the residual CMASS relationships after applying weights that account for the linear relationships between galaxy density, stellar density, and fiber magnitude ($w_{star}$). The expected errors are determined by finding the standard deviation of the relationships measured from individual mock CMASS catalogs.}
  \label{fig:ngvnsys}
\end{minipage}
\end{figure*}

\subsection{Measurements of Clustering}
\label{sec:nsxi}
Fig. \ref{fig:xiNScomb} displays the measured $\xi(s)$ in the Northern (NGC; red) and Southern Galactic Caps (SGC; blue). These measurements include the weights that correct for stars, fiber collisions, and redshift failures, defined by Eq. \ref{eq:wtot} and also the FKP weights. The area covered by the SGC data is only one quarter that of the NGC, and therefore the uncertainty in the SGC $\xi(s)$ is about twice as large as the NGC. At almost all scales, the $\xi_0$ measurements appear consistent with each other; the only notable exception is a significant dip in the measurements at 170 $h^{-1}$Mpc. Both of the $\xi_0(s)$ measurements display a prominent increase in clustering at around 100 $h^{-1}$Mpc, suggesting significant BAO peaks, though the peak does appear at a smaller scale in the Southern measurement. Interestingly, both the NGC and SGC measurements appear to also have a peak in $\xi_0(s)$ at around $s = 215 h^{-1}$Mpc, but we note that the uncertainty on the measurements at these scales is much larger than around the BAO scale. The $\xi_2$ measurements appear slightly less consistent, especially between 75 and 95 $h^{-1}$Mpc, where the measurements are clearly inconsistent within the 1$\sigma$ error-bars. 

The black points in Fig. \ref{fig:xiNScomb} display combined NGC and SGC measurements, which are produced by summing the DD, DR, and RR pairs, which is appropriate when FKP weights are used. The number of randoms in each region use the same normalization with respect to the number density in each region (we use just over 15 times the number of galaxies). Thus, the relative normalization of randoms to galaxies between the two hemispheres is matched as if the samples were treated individually, and the results are optimally combined.

We test the consistency of the measurements by summing the covariance matrices of the NGC and SGC and determining $\chi^2$ in the standard fashion. For $s < 250h^{-1}$Mpc (35 data points), we find $\chi^2=45.4$ for $\xi_0$ and 32.0 for $\xi_2$, so despite the apparent differences, $\xi_2$ is actually more consistent than $\xi_0$. For $\xi_0$, the $\chi^2$ is slightly high --- 11\% of consistent samples drawn from a Gaussian distribution will have a higher $\chi^2$. Reducing the range of the fit to $25 < s < 150h^{-1}$Mpc (the primary range we study), the $\chi^2$ is 23.2 (18 data points), which shows consistency (18\% of consistent samples will have a larger $\chi^2$). \cite{SanchezCos12} find the same $\chi^2/$dof (1.3) when they fit their $\xi(s)$ measurements, which use an alternative binning, between $40 < s < 200h^{-1}$Mpc. Scaling to the NGC sample, we find that the best-fit relative bias, $b_{rel}$, of the SGC sample is $1.057\pm0.038$ ($\chi^2_{min} = 20.9$), when fit to $25 < s < 150h^{-1}$Mpc. Relaxing the minimum bound to $s=10h^{-1}$ (adding two data points), we find $b_{rel} = 0.983\pm0.015$ with $\chi^2_{min} = 20.5$, $b_{rel} = 1$ is just outside the 1$\sigma$ bounds, suggesting that the clustering in the two regions is consistent to within 1.1$\sigma$.

We further test the consistency of the NGC and SGC measurements by finding the best-fit bias for the NGC and SGC samples by scaling the mocks, as described in Section \ref{sec:mocks}, in the range $25 < s < 150h^{-1}$Mpc. We find the best-fit $\tilde{b} = 1.904\pm$0.039, with $\chi^2_{min} = 24.3 $ (18 measurements) for the NGC data and 2.06$\pm$0.07, with $\chi^2_{min} = 18.8$ for the SGC. The difference is nearly 2$\sigma$. The best-fit bias of the combined sample, $\tilde{b}=1.936\pm0.035$, is very close to the weighted average of the two samples ($\tilde{b} = 1.943\pm0.035$).  Increasing the minimum scale to $s=30h^{-1}$Mpc, we find a significant change in the best-fit bias for the NGC ($\tilde{b} = 1.87\pm0.05$), but we find negligible change for the SGC ($\tilde{b} = 2.08\pm0.11$). For the combined sample, the bias decrease is even larger, as $\tilde{b} = 1.886\pm0.048$ when we fit $30 < s < 150h^{-1}$Mpc. The values of $\tilde{b}$ (and other parameters we measure throughout this section) are summarized in Table \ref{tab:allpar} (found in Section \ref{sec:con}).

Fixing the bias at the best-fit value from $\xi_0$, we find the best-fit value of $\tilde{f}$ from the $\xi_2$ measurements by scaling the mock $\xi_2$. For the NGC, we find $\tilde{f}= 0.691\pm0.052$ (with $\chi^2_{min} = 10.9$) and for the SGC data $\tilde{f}= 0.79\pm0.09$ (with $\chi^2_{min} = 11.8$). For the combined sample, $\tilde{f}=0.711\pm0.044$ ($\chi^2_{min} = 10.2$) --- very similar to the weighted average of $0.716\pm0.045$. This suggests that the information content in $\xi_2(s)$ measurements related to the velocity field is consistent between the two regions.

As described in Section \ref{sec:mod}, we can use the mocks to fit for a bias and stretch parameter, $\tilde{\alpha}$. We stress that these $\tilde{\alpha}$ values should reflect both changes we expect in the best-fit $\Omega_{m}h^{2}$ and distance constraints one may obtain from the BAO feature, i.e., it only reflects the level of disagreement we should expect in derived cosmological parameters using the NGC/SGC footprint, and whether the level of disagreement is consistent with what we expect to find given cosmic variance, but it contains no information on specific parameters. For the NGC footprint, we find $\chi^2_{min} = 22.1$ at $\tilde{\alpha}=0.990, \tilde{b} = 1.888$; marginalizing over the bias, $\tilde{\alpha}=0.994\pm0.023$. For the SGC footprint, $\chi^2_{min} = 12.3$ at $\tilde{\alpha}=1.090, \tilde{b} = 2.319$; $\tilde{\alpha} = 1.083 \pm 0.029$ when we marginalize over bias. For the combined sample, we find $\chi^2_{min} = 21.5$ at $\tilde{\alpha} = 1.019$, $\tilde{b} = 1.982$; marginalizing over the bias, $\tilde{\alpha} = 1.020\pm0.019$. This is smaller than the weighted average, $1.028\pm0.018$, of the two $\tilde{\alpha}$ measurements, but it is still greater than a 1$\sigma$ shift from the result obtained using only the NGC data. This implies that one may find differences of 1$\sigma$ on recovered cosmological parameters when comparing results from only the NGC data to the combined sample. Indeed, \cite{SanchezCos12} find this level of variation. 

The difference between the NGC and SGC $\tilde{\alpha}$ values we measure is $2.5\sigma$. We find negligible changes in the values of $\tilde{\alpha}$ we obtain from the measured SGC $\xi_0$ if we apply the \cite{Sch10b} offset to the selection of SGC CMASS galaxies, use any of the separate weighting schemes described in the appendix, or neglect to apply any weight at all; that is, we have not been able to identify any systematic that may cause the differences we observe. Any true difference would represent a violation of isotropy. \cite{alph} find that the tension between their BAO scale measurements is reduced to 1.4$\sigma$ when reconstruction is applied to the CMASS galaxy density field (and is 2.5$\sigma$ without reconstruction). As reconstruction generally improves the signal-to-noise in BAO scale estimation, this reduction in the tension between the two measurements implies the difference is indeed driven by noise.

In general, the level of disagreement between the NGC and SGC correlation functions is between 1 and 2$\sigma$. The differences in the bias when scaling the mocks ($1.9\sigma$) and when we fit for a relative bias ($b_{rel} = 1$ is 1.5$\sigma$ from the best-fit) are both less than $2\sigma$. The $n(z)$ distributions disagree at a similar level of significance. The $n(z)$ discrepancy is likely related to the disagreement in the clustering. Finally, when fixing $\tilde{\alpha} = 1.017$ (the upper 1$\sigma$ bound on $\tilde{\alpha}$ from the NGC sample), $\chi^2_{min} = 16.4$ (meaning the $\chi^2$/dof is less than one) when testing the $\xi_0$ of the SGC sample (at $\tilde{b}=2.11$). The SGC footprint is currently only 705deg$^2$ --- 28\% of its final (planned) size (2500 deg$^2$). If the differences are indeed in the noise, we expect that all results between the NGC and SGC will become more consistent as the BOSS survey continues and the sample grows.

\section{Angular Variations in Target Catalog}
\label{sec:angvar}

\cite{imsys} found significant correlations between the number density of galaxies in the SDSS imaging data with a photometric selection similar to that of the CMASS sample, and various parameters. In particular, the number density of observed galaxies decreased significantly as a function of the stellar density. We repeat the tests performed by Ross et al. (2011b), now using spectroscopically-confirmed galaxies. We are only using data within the DR9 mask (which is about 1/3 of the imaging area used in \citealt{imsys}) and we now have access to mocks that allow us to quantify the statistical variations we should expect to find.

\subsection{Galaxy density vs. potential systematics}
We determine the number density of the DR9 spectroscopically-identified galaxies as a function of stellar density, seeing, Galactic extinction, airmass, and sky background (all during the imaging observations). To perform these tests, we make {\rm HEALPix} \citep{Gor05} maps of DR9 galaxies and compare them to maps of the number of stars with $17.5 < i < 19.9$ or maps of the mean values of the potential observational systematic based on data from the SDSS DR8 Catalog Archive Server (CAS)\footnote{http://skyserver.sdss3.org/dr8/en/} within pixels at N$_{\rm side}$ = 256, which splits the sky into equal area pixels of 0.0525 deg$^{2}$. Rather than pixelate the mask, in each pixel we determine the number of galaxies, $n_{gal}$ and the number of randoms, $n_{ran}$, (multiplied by the factor $n_{gal,tot}/n_{ran,tot}$), and therefore map $n_{gal}/n_{ran}$.

Fig. \ref{fig:ngvnsys} displays the relationships between the number of galaxies observed and potential observational systematics. We weight each galaxy for redshift completeness and close pairs as described in Section \ref{sec:weights} and we combine the NGC and SGC data by using the same normalization of randoms to galaxies in each respective region. We find no significant differences in our analysis if we analyze the two regions separately. We apply the same tests on the 600 mocks (described in Section \ref{sec:mocks}) and use the standard deviation as the errors in each bin displayed in Fig. \ref{fig:ngvnsys}. The relationships for CMASS galaxies are displayed in black. As in \cite{imsys}, we find a 10\% decrease in the number density of galaxies between areas with high and low stellar density. 

As quantified in Section 4.1 of \cite{imsys}, 3\% of the decrease in galaxy density results from the fact that, within 10$^{\prime\prime}$ of stars, seeing reduces the ability to detect galaxies (with little dependence on the magnitude of the star between $17.5 < i_{mod} < 19.9$). The relationship is not found in DR7 data (Bauer, A., private communication). The most significant change between the DR7 and DR8 photometric pipelines was a refinement in the sky background subtraction algorithm, in order to improve the photometry of bright galaxies \citep{DR8}. One effect of this change is to increase the low-surface brightness extent of both galaxies and stars, causing more objects to be linked together. In regions of higher stellar density, this means that the deblender more often has to deal with complicated superpositions of many objects. To control processing time, the deblending code will separate out up to 25 overlapping objects in one parent, but no more. This happens more often with the new code, meaning that there are more missing galaxies in regions of high stellar density than before. This may explain the remaining 7\% effect. 

We also find an anti-correlation with Galactic extinction --- this is at least partly due to the fact that the Galactic extinction and stellar density are correlated. Indeed, we find that the correlation with extinction becomes insignificant once weights are applied (see Section \ref{sec:angweight}) to correct for the relationship with stellar density. See \cite{Yah07} for a more detailed study on the ways in which the Galactic extinction, as determined by the \cite{SFD} dust maps, correlates with the observed density of galaxies. 

We find a sharp decrease in the number density of galaxies in areas with poor seeing; this effect was explained in \cite{imsys} as being due to the fact that the star/galaxy separation criteria defined by eqs. \ref{eq:sgsep1} and \ref{eq:sgsep2} remove more true galaxies in areas where the seeing is poor. This systematic relationship had little effect on the measured clustering in \cite{imsys}, as the pattern of seeing in the DR8 imaging is essentially random on large scales. For sky background and airmass, the level of fluctuations are close to what we should expect due to cosmic variance (as represented by the error bars).

\begin{figure}
\includegraphics[width=84mm]{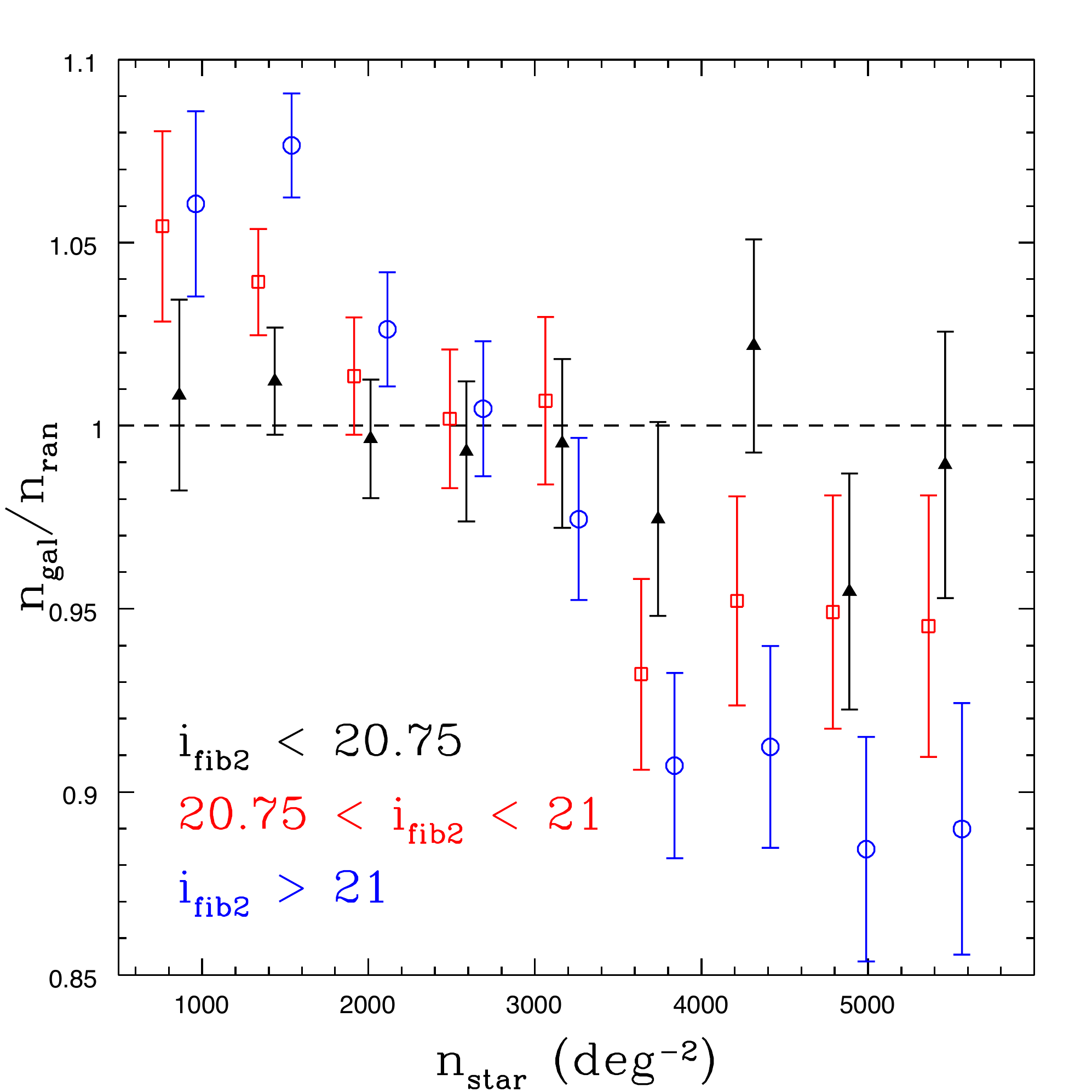}
  \caption{Same as the black points in the left panel of Fig \ref{fig:ngvnsys}, except that we have broken the CMASS sample into three subsamples based on the labeled fiber magnitudes, $i_{fib2}$: $i_{fib2} < 20.75$ has 78,065 good redshifts, $20.75 < i_{fib2} < 21$ has 85,284 good redshifts, and $i_{fib2} > 21$ has 112304 good redshifts.}
  \label{fig:ngfib}
\end{figure}

\begin{figure}
  \includegraphics[width=84mm]{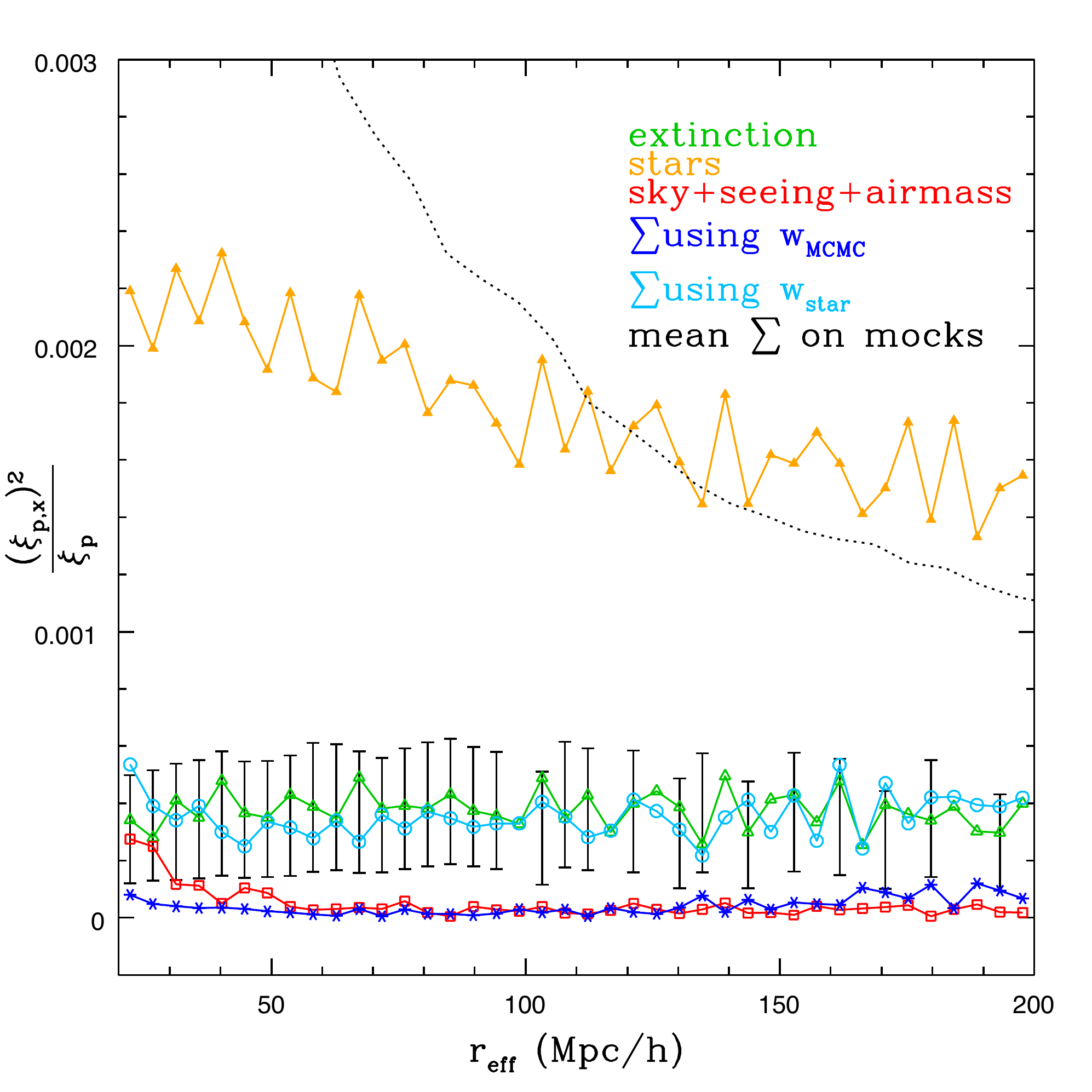}
  \caption{The cross-correlation with the full CMASS sample, $\xi_{p,x}(r_{eff})$, squared, divided by the auto-correlation, $\xi_p$, for stars (orange), Galactic extinction (green), the sum of seeing, sky background, and airmass (red), the sum of all five when applying the linear-fit weights to all five potential systematics ($w_{MCMC}$, blue, which we consider in the appendix), the sum of all five when applying the linear-fit weights for only stellar density as a function of $i_{fib2}$ ($w_{star}$, light blue), and the mean sum of all five and its standard deviation on the mocks (black error-bars). The dotted black line displays the expected statistical uncertainty, determined from the variance of the mock $\xi_0(s)$ measurements.}
  \label{fig:wsyscomb}
\end{figure}
\begin{figure*}
\begin{minipage}{7in}
\centering
 \includegraphics[width=5.5in]{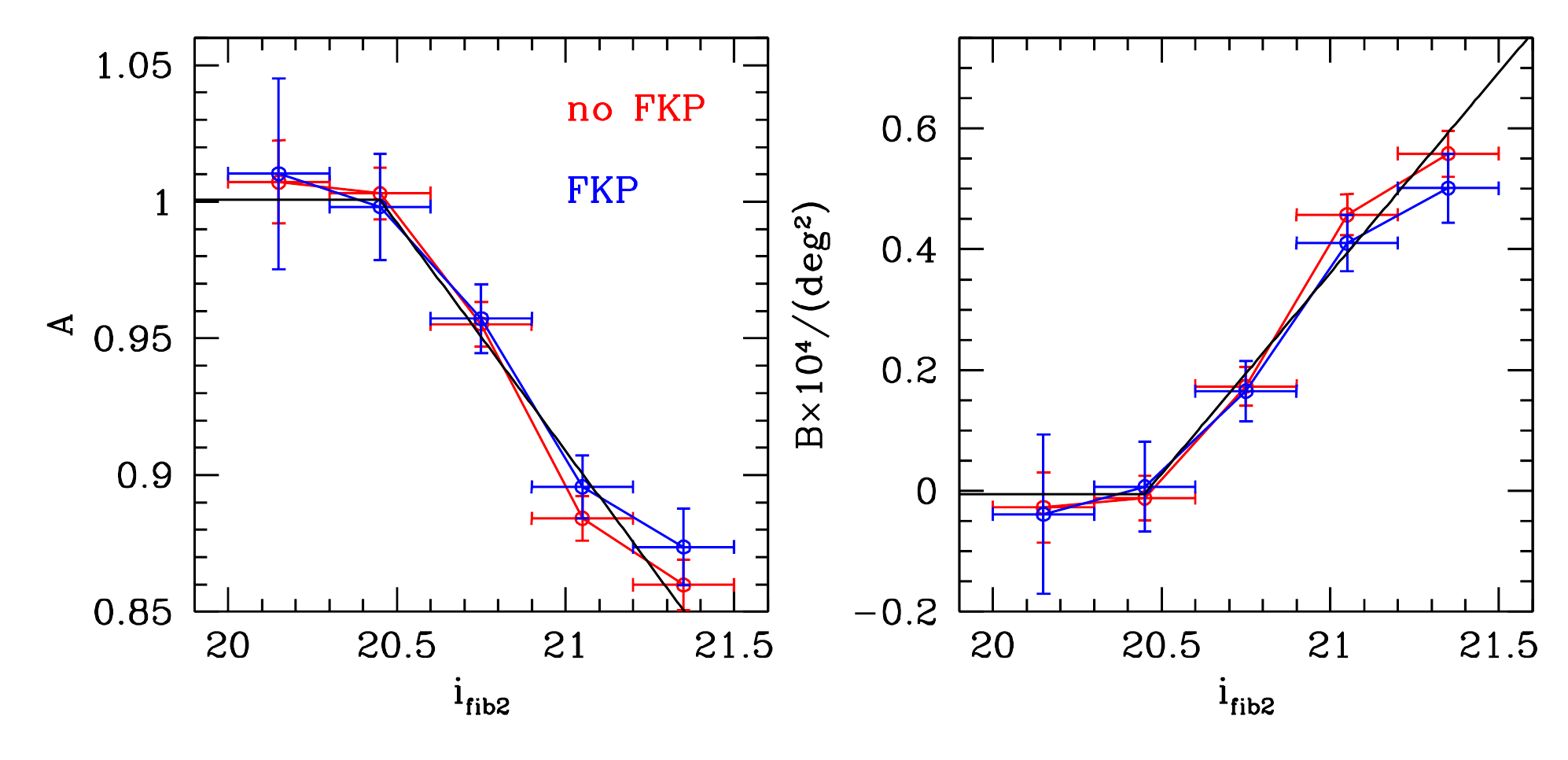}
 \caption{The best-fit coefficients to the relationship $n_{gal}/n_{ran} = A+Bn_{star}$ as a function of the $i_{fib2}$ magnitude of the galaxies. The blue points display the results when we use FKP weights  (see Eq. \ref{eq:fkp}) and the red points show the results when we do not use this weighting. The black lines display our fit to these coefficients, which we use to determine weights as a function of $i_{fib2}$ and $n_{star}$.}
 \label{fig:wstlinAB}
 \end{minipage}
 \end{figure*}

The relationships for the LOWZ sample are displayed in Fig. \ref{fig:ngvnsys} with green lines. In contrast to the CMASS sample, we do not find any systematic dependency with stellar density, Galactic extinction, or seeing. This is likely due to the fact that LOWZ galaxies are, on average, considerably brighter than CMASS galaxies, and their detection should therefore be less affected by imaging systematics. Given that the volume of the LOWZ sample is considerably smaller than the CMASS sample, we should expect larger cosmic variance. Indeed, it appears that all of the variance in number density we find for the LOWZ sample can be attributed to cosmic variance.

In \cite{imsys}, the relationship between the number density of galaxies and stellar density was found to depend on the surface brightness of the galaxy. Given that the mean surface brightness of CMASS galaxies is lower at higher redshift, the relationship between the galaxy density and stellar density may depend on redshift. In Fig. \ref{fig:ngfib}, we show the relationship between galaxy density and stellar density when splitting the sample into three sets based on the $i_{fib2}$ magnitude, as this magnitude uses a fixed aperture and is thus essentially a surface brightness measurement. The slope in the relationship clearly grows more negative at fainter $i_{fib2}$. The fact that the effect correlates strongly with surface brightness is further evidence that it is related to a systematic in the DR8 imaging, likely related to the sky subtraction routine, as opposed to a real large-scale density fluctuation perfectly aligned with the Galaxy.

For each of the potential systematics displayed in Fig. \ref{fig:ngvnsys}, we determine the auto-correlation, $\xi_p$, and cross-correlation, $\xi_{p,x}$ with the CMASS sample as a function of the effective scale, $r_{eff}$ (all of which are defined by Eq. \ref{eq:xipix} and the surrounding text). As described in \cite{HoLG} and \cite{imsys}, the effect of any potential systematic on the measured correlation function can be estimated as $\xi_{p,x}(r_{eff})^2/\xi_{p}(r_{eff})$. Fig. \ref{fig:wsyscomb} displays this ratio for the five potential systematics displayed in Fig. \ref{fig:ngvnsys}. We confirm with the spectroscopic sample the result found in the angular clustering analysis \citep{imsys}: the presence of stars has the greatest systematic effect. The effect of Galactic extinction is second largest, but \cite{imsys} found it to be almost entirely degenerate with the effect of stars. The sky background, airmass, and seeing all have negligible effects. However, we have found more significant correlations with sky background when the sample is split further by color. Overall, we should expect a difference of $\sim$ 0.002 between fiducial $\xi(s)$ measurements and those with corrections for systematics. 

The mean of the sum and standard deviation of $\xi_{p,x}(r_{eff})^2/\xi_{p}(r_{eff})$ for all five potential systematics we consider overf the mock catalogs, are displayed with black error-bars in Fig. \ref{fig:wsyscomb}. This is non-zero on average because the auto-correlation of each systematic is positive and the cross-correlations, which have 0 mean, are squared. Thus, we should expect a non-zero mean on the sum of these contributions, even when there is no real systematic effect.

\subsection{Angular Weights}
\label{sec:angweight}

As shown in Fig. \ref{fig:wsyscomb}, the primary source of systematic error is due to the relationship with stellar density. To account for this systematic effect, we apply weights that counteract the systematic relationship. Fig. \ref{fig:ngfib} suggests that the systematic relationships depend on the surface brightness of the galaxy. We thus use this information in order to determine `linear-fit stellar density weights', which we denote $w_{star}$. We split the sample by $i_{fib2}$ and assume $n_{gal}/n_{ran} = A+B n_{star}$ for each sub-sample, now using N$_{\rm side}=128$ for the resolution of the maps. The result of this approach is shown in Fig. \ref{fig:wstlinAB}. We find that for $i_{fib2} < 20.45$, the relationship is consistent with being constant. At fainter $i_{fib2}$, we find a linear relationship with $A(i_{fib2})$ and $B(i_{fib2})$, and we thus use this linear fit to determine the $w_{star}$ weights (which ignores the rest of the potential systematics). The linear fit is given by
\begin{eqnarray}
A = A_0+A_1i_{fib2}\\
B/(deg^2) = B_0+B_1i_{fib2}
\end{eqnarray}
where $A_0 = 3.96, A_1= -0.14, B_0 = -1.18\times10^{-3}$, $B_1 = 5.76\times10^{-5}$ (for the case where the FKP weights are applied). For $i_{fib2} < 20.45$, $A$ and $B$ are set to the $A(20.45)$, $B(20.45)$ given by the above equations.

The residual relationships after applying the $w_{star}$ weights are displayed in red in Fig. \ref{fig:ngvnsys} --- the relationship with Galactic extinction changes from having a slightly negative to a slightly positive slope and the seeing, sky background, and airmass relationship remain similar to the unweighted case. For all but seeing, the relationship appear consistent with the variations we expect due to cosmic variance (as shown by the error-bars in Fig. \ref{fig:ngvnsys}). Further, the sum of the five potential systematic contributions is consistent with the mean mock sum when the $w_{star}$ weights (as shown in Fig. \ref{fig:wsyscomb}) are applied to the CMASS data. We therefore believe the $w_{star}$ weights are  appropriate to apply to the CMASS sample. Additionally, we believe the $i_{fib2}$ dependence of the $w_{star}$ weights should (mostly) account for changes in redshift. In the appendix, we compare the $w_{star}$ weights to two other weighting schemes and find that the application of the $w_{star}$ weights has the least potential to remove true fluctuations from the density field and does not bias the clustering measurements of our mock galaxy samples.

\begin{figure*}
\begin{minipage}{7in}
\centering
 \includegraphics[width=7in]{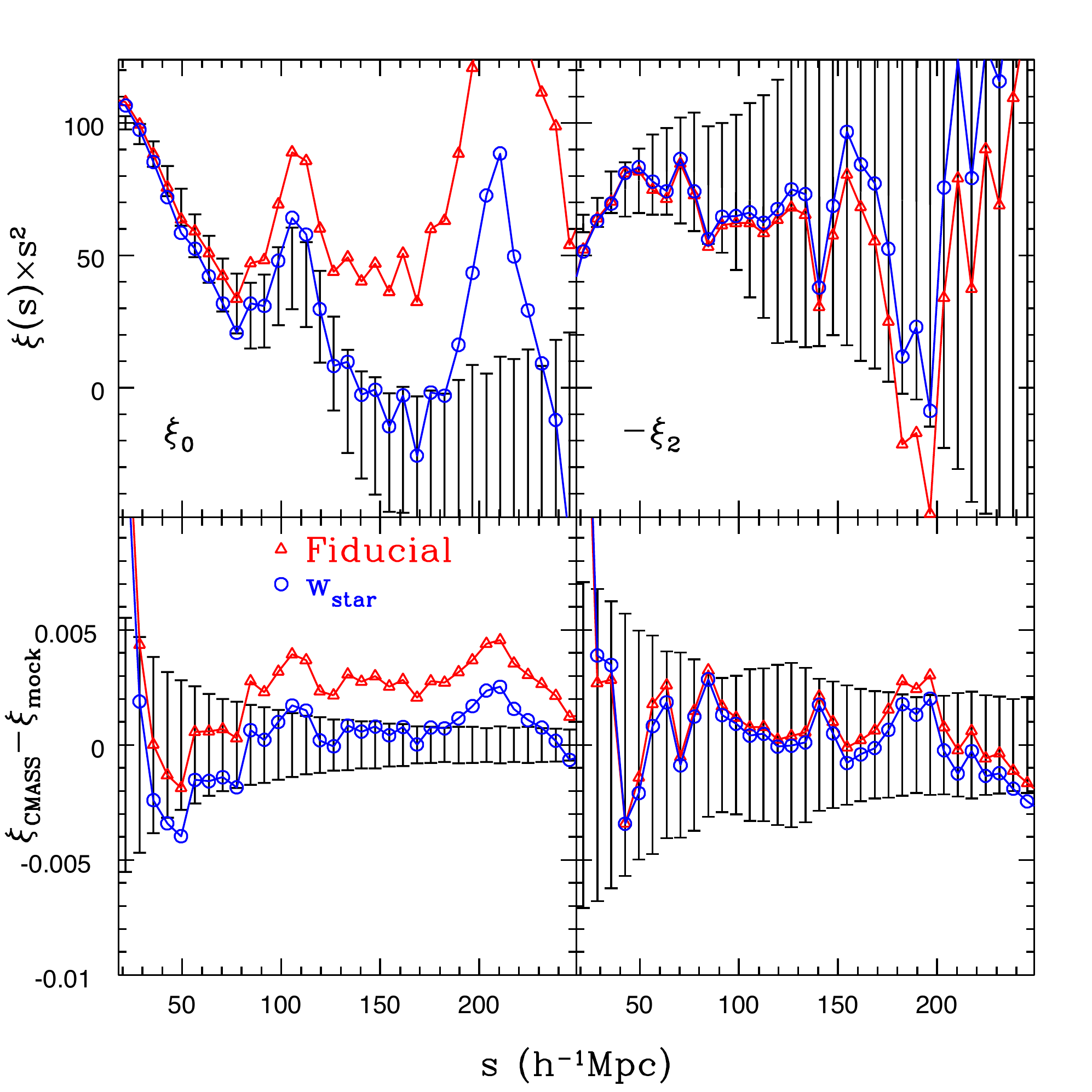}
  \caption{The measured redshift space correlation functions of CMASS galaxies using the fiducial catalog (red triangles), and applying weights that correct for the linear relationships between galaxy density and stellar density and $i_{fib2}$ magnitude (blue circles). The bottom panels display the difference between the measured $\xi_{0}$ and $\xi_{2}$ and that mean calculated from the mock $\xi_{0}$ and $\xi_{2}$. Black error bars represent the standard deviation of the mock $\xi_{0}$ and $\xi_{2}$. We analyze the apparent feature at $s\sim200h^{-1}$Mpc in Section \ref{sec:remain}.}
  \label{fig:xiwts}
 \end{minipage}
\end{figure*}

\subsection{Effect of Angular Weights on CMASS Clustering}
\label{sec:wxi}

\begin{figure*}
\begin{minipage}{7in}
\centering
 \includegraphics[width=5.5in]{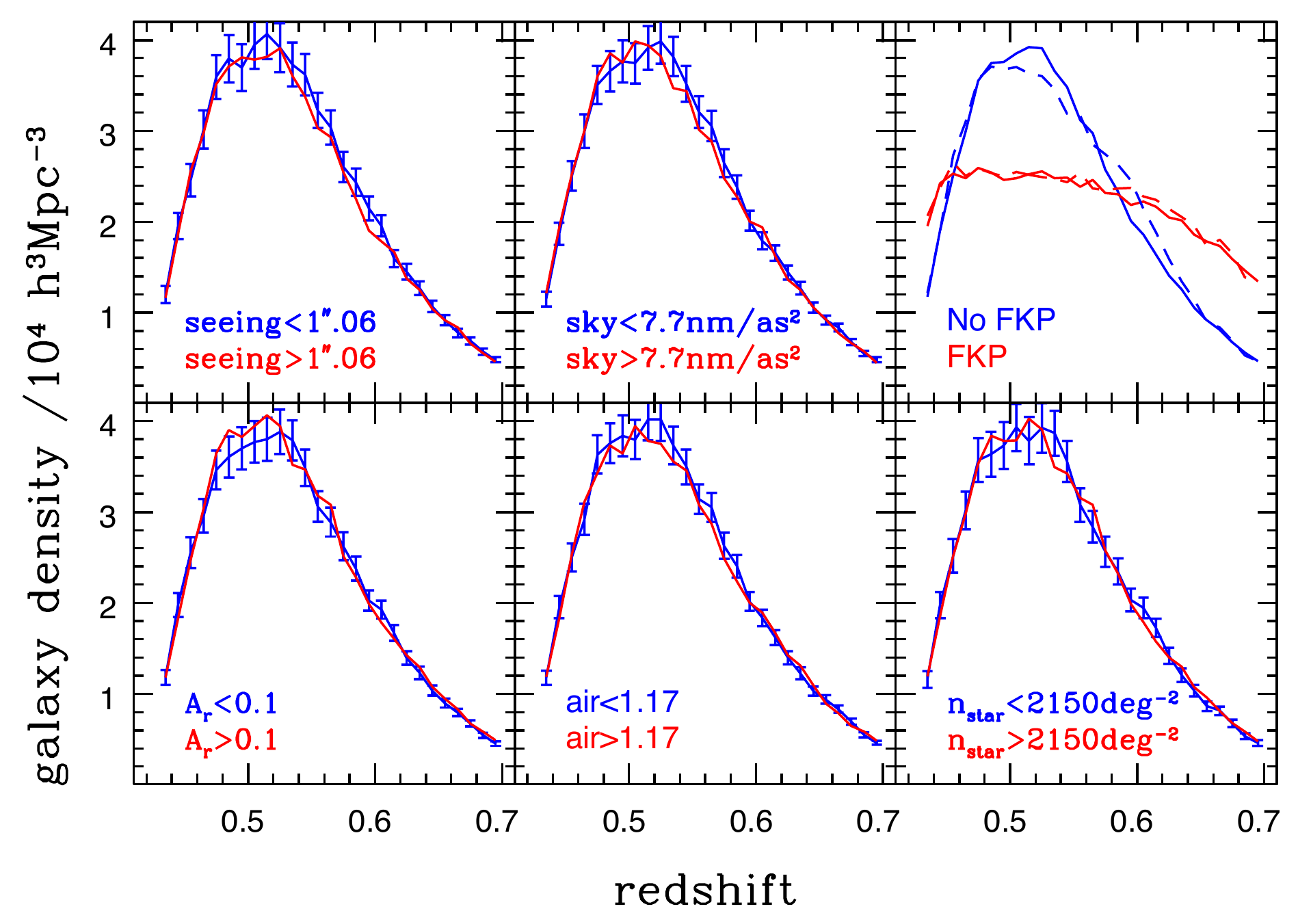}
  \caption{The redshift distributions of CMASS objects in the Northern Galactic Cap (NGC), splitting the area in half on seeing, $i$-band sky background in nanomaggies per square arcsecond, the $r$-band Galactic extinction ($A_r$), airmass, and the number density of stars ($n_{star}$). The errors are determined by finding the standard deviation of the mock $n(z)$. The upper-right panel displays the result with (red) and without (blue) FKP weights, using solid lines for the NGC sample and dashed lines for the SGC sample.}
  \label{fig:nzsys}
\end{minipage}
\end{figure*}

The top panels of Fig. \ref{fig:xiwts} display the resulting $\xi_{0}(s)$ and $\xi_{2}(s)$ measurements for CMASS galaxies for the fiducial sample (which include FKP weights and weights for close-pairs and redshift failures, see Fig. \ref{fig:cpzf}), and when we include the $w_{star}$ weights. As expected (Fig. \ref{fig:wsyscomb}), the weights for stellar density cause a nearly constant decrease of 0.002 in the measured $\xi_0$. This change is greater than the statistical uncertainty at scales greater than 110 $h^{-1}$Mpc. The $w_{star}$ weights only have a slight effect on the $\xi_2$ measurements. The difference is always smaller than the statistical uncertainty, reaching $\sim0.5\sigma$ at the largest scales. 

As described in Section \ref{sec:mocks}, we can scale the mocks to determine a best-fit bias, $\tilde{b}$. The consistency of the different weighting schemes can be further tested by comparing the best-fit bias and associated $\chi^2_{min}$, which we determine in all cases using the covariance matrices calculated using the mocks. Applying the $w_{star}$ weights to the CMASS sample when calculating $\xi(s)$ and fitting between $25 < s < 150 h^{-1}$Mpc, the best-fit $\tilde{b}=1.936\pm0.035$, with $\chi^2_{min} = 22.7$ (18 measurements). When we do not apply the $w_{star}$ weights to the CMASS sample, the best-fit bias increases by $\sim0.5\sigma$ to $\tilde{b}=1.949^{+0.034}_{-0.035}$, and the $\chi^2_{min}$ increases to 34.2. Assuming Gaussian statistics, only 1.2\% of consistent samples would have a $\chi^2 > 34.2$, while 20.2\% would have $\chi^2 > 22.7$. The values of $\tilde{b}$ (and other parameters we measure throughout this section) are summarized in Table \ref{tab:allpar} (found in Section \ref{sec:con}).

We find similar results when we fit the weighted and un-weighted $\xi_0$ measurements for both $\tilde{b}$ and $\tilde{\alpha}$ (again in the range $25 < s < 150 h^{-1}$Mpc). The $\chi^2_{min} = 33.9$ at $\tilde{\alpha} = 1.009$, $b = 1.971$ when the weights are not applied; marginalizing over the bias, we find $\tilde{\alpha} = 1.007\pm0.019$. When applying the $w_{star}$ weights, we find $\chi^2_{min} = 21.5$ is at $\tilde{\alpha} = 1.019$, $b = 1.982$; marginalizing over the bias, $\tilde{\alpha} = 1.020\pm0.019$. We note that this test only reflects the level of systematic change we should expect in derived cosmological parameters. The change in the $\tilde{\alpha}$ value therefore suggests that the application of the weights could cause a shift in the best-fit cosmology of close to two-thirds $\sigma$ when constraints are derived from the full shape of $\xi_0$. \cite{alph} measure the same BAO peak position, to within 0.1\%, whether or not the $w_{star}$ weights are applied, suggesting that the change in our $\tilde{\alpha}$ measurement reflects the change in the shape of $\xi_0(s)$. 

Fig. \ref{fig:xiwts} suggests that the weights have little affect on $\xi_2$. This is confirmed by fixing $\tilde{b}$ at the best-fit value from the $\xi_0$ measurements and finding the best-fit $\tilde{f}$ value, by scaling the mocks. When the $w_{star}$ weights are applied, we find $\tilde{f}=0.711\pm0.044$ with $\chi^2_{min}=11.8$ (18 data points); when the weights are not applied, we find $\tilde{f}=0.710\pm0.044$ with $\chi^2_{min}=12.7$.  

\section{Radial selection function}
\label{sec:zsys}
We may be concerned that any parameter that causes a systematic effect in the angular distribution of galaxies may also cause change in the redshift distribution. To test this possibility, we split the sample in half, based on each of the same five potential systematics in turn, and determine the redshift distributions. The results are shown in Fig. \ref{fig:nzsys} when using the weights fit to the linear relationships between galaxy density, stellar density and $i_{fib2}$ ($w_{star}$). None of the distributions are significantly outside of the errors we determine based on the standard deviation in the distributions of mock samples within the NGC or SGC footprints (though we only plot the results for the NGC; see Section \ref{sec:mocks}). 

We use FKP weights \citep{FKP}, defined by Eq. \ref{eq:fkp}, to optimally weight the data as a function of redshift. This changes the $n(z)$ from the blue curves to the red in the upper-right panel of Fig. \ref{fig:nzsys}, with solid lines representing data from the NGC and dashed lines for the SGC. Including the FKP weights effectively equalizes the contribution of every redshift interval we consider in the $\xi_{\ell}$ calculation. This is illustrated by the fact that the mean redshift changes from $z=0.55$ to $z=0.57$. The FKP weights also make the NGC and SGC selection functions more similar to one another. 

\begin{figure}
  \includegraphics[width=84mm]{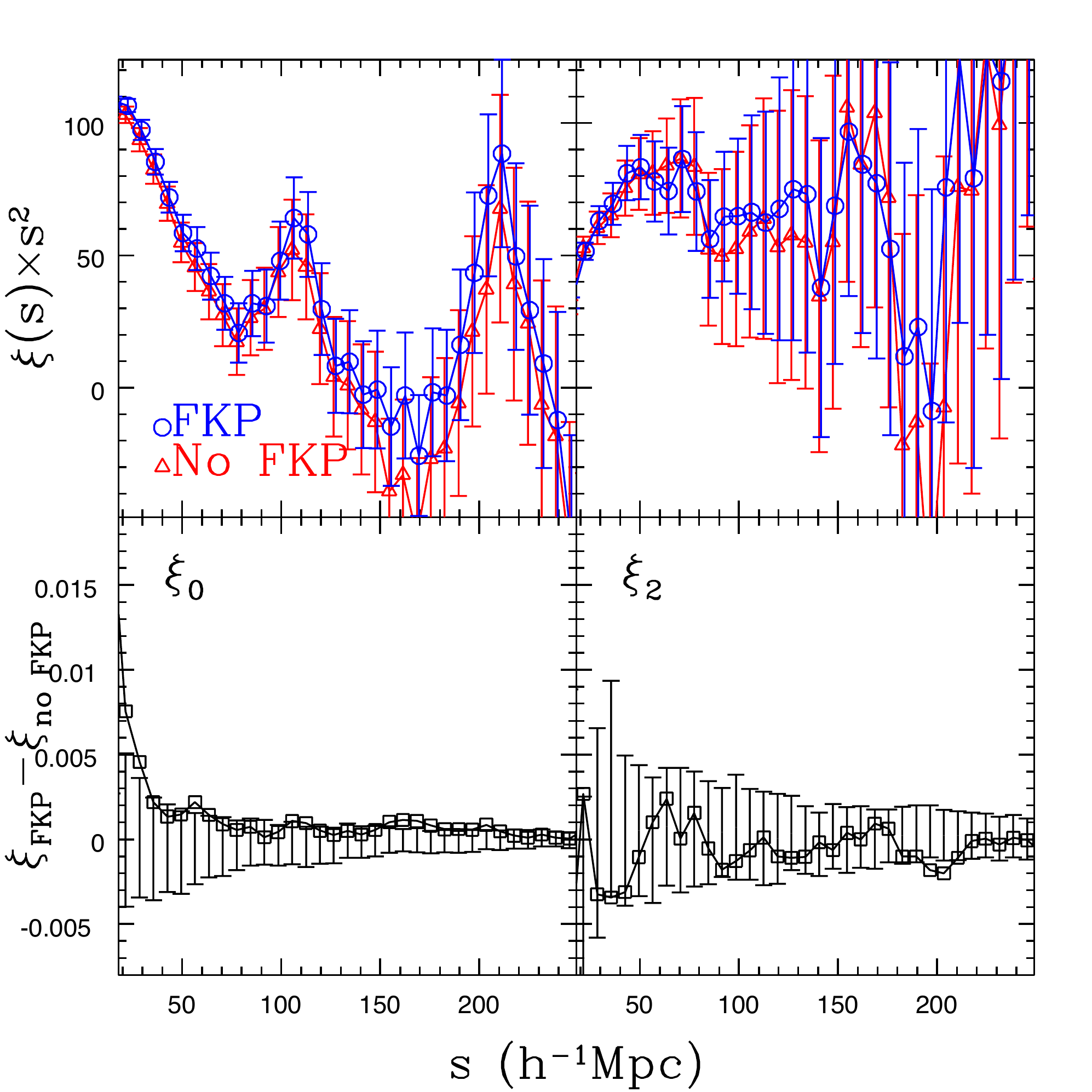}
  \caption{Top panels: The measured redshift space auto-correlation functions of the combined CMASS sample measured with (blue) and without (red) FKP weights. Points represent measurements; error-bars are determined using from the variance of the mock calculations. Bottom panels: Points display the difference between the measurements using and not using the FKP weights. Error-bars represent the mean and the variance of the difference using and not using FKP weights for the mock calculations.}
  \label{fig:xifkp}
\end{figure}

We display $\xi_{\ell}$ measurements, with and without FKP weights, in Fig. \ref{fig:xifkp}. This serves as a check that these weights have not imparted any systematic errors and illustrates the advantage of applying FKP weights.  For $\xi_0$, the amplitudes are marginally higher at all scales when the FKP weights are applied. This result is due in part to the fact that the FKP procedure assigns larger weights to the higher redshift data, which is more likely to include more luminous galaxies which thus have higher bias. We find, as expected, that the application of the FKP weights reduces the uncertainty on derived parameters by at least 10\% and that the derived $\tilde{f}$ and $\tilde{\alpha}$ are consistent to within 0.5$\sigma$ whether or not the FKP weights are applied (without FKP weights, we find $\tilde{\alpha} = 1.033\pm0.025$ and $\tilde{f}=0.711\pm0.044$).

\subsection{Testing models of the radial selection function}
\label{sec:ztest}
To model the expected galaxy distribution we must assign redshifts to the random catalogs we used to calculate $\xi_{\ell}$. This is difficult to achieve without using the data itself, as we would need a full theoretical model for the galaxy population targeted. Without such a model, we are limited by the fact that we can only estimate the true $n(z)$ empirically. However, we can test the effects of this dependence using the mocks. The mocks were constructed assuming a fiducial $n(z)$ (which is the $n(z)$ we measure from the data) and the $n(z)$ of each individual mock will scatter around this input $n(z)$ due to cosmic variance. For each mock, we consider three methods to determine the $n(z)$ applied to the random catalog: 
\begin{enumerate} 
{\bf \item `spline'}, where a spline fit to the observed redshift distribution of galaxies, using bins of width $\Delta z = 0.01$, is used to determine the $n(z)$, we sample from this to construct a random catalog and 
{\bf \item `shuffled'}, where for each point in the random sample we assign the redshift of a randomly-selected redshift from the galaxy sample.
{\bf \item `true'}, where we use the input $n(z)$. The true $n(z)$ is, of course, not available for any observed sample and we test the difference between clustering measured using the true $n(z)$ and either the spline of shuffled $n(z)$ in this section.
 \end{enumerate}

We create a random catalog for each of the 600 individual mocks using both the spline method and the shuffled method and compare the results to those derived using the true underlying $n(z)$. For the spline method, we use an $N$-node spline. We examine the cases where $N=10,20,30,50$. We expect the results using the $N$-node spline to approach the limit of the results of the shuffled catalog when $N$ is very large. Importantly, we can compare all results to the true case, thereby quantifying the bias and additional uncertainty imparted by the need to self-average as a function of redshift, and thus addressing the concerns raised in \cite{Labini}.

\begin{figure}
  \includegraphics[width=84mm]{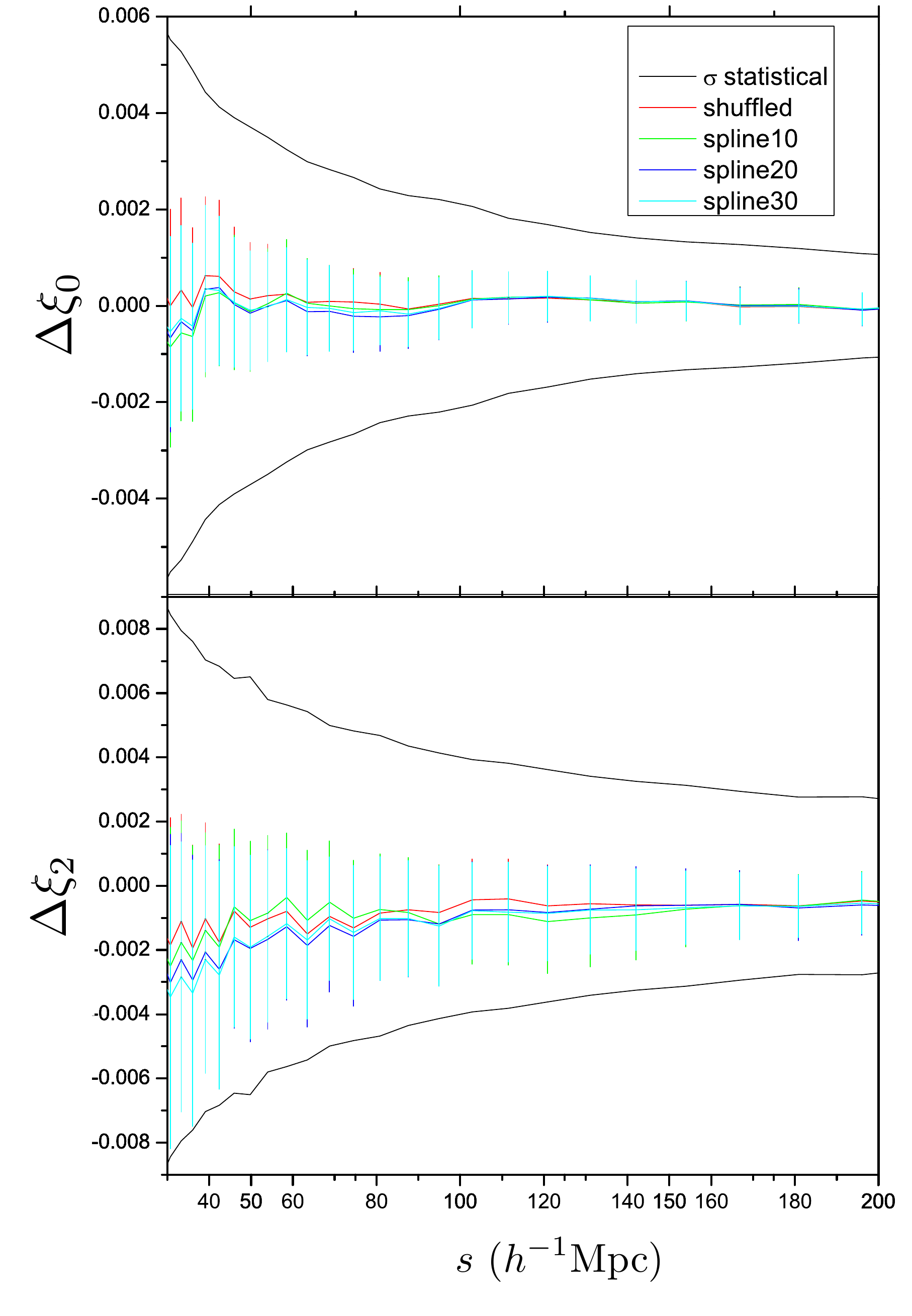}
  \caption{Top panel: Average offsets and 1$\sigma$ deviations from a true $\xi_0$ for different methods of assigning redshifts to random catalogs, determined using the 600 mock galaxy catalogs. The solid black line corresponds to the statistical errorbars. Bottom panel: As above, for $\xi_2$. }
  \label{fig:x0sys}
\end{figure}

The top panel of Fig.~\ref{fig:x0sys} shows the average bias of $\xi_0$ measurements and its standard deviation, determined using 600 realizations of $\xi_0$ computed from mocks using different random catalogs. For the measurements of the monopole the average bias for all methods of constructing a random catalog is a small fraction of the statistical errors and the standard deviation of the bias is about a third of the statistical errors. The bias is smallest when using the shuffled random catalog and appears negligible for $\xi_0$.

The bottom panel of Fig.~\ref{fig:x0sys} shows the results of the same test on the measurements of $\xi_2$. For the second Legendre moment of the correlation function the systematic offset is larger than for $\xi_0$, but is still small compared to the statistical errors. The standard deviation of the bias is $\sim$50\% of the statistical errors. For both monopole and quadrupole of the correlation function, the shuffled catalog performs the best; however, the average bias for both $\xi_{0}$ and $\xi_{2}$ is
also small for the $N$-spline methods.

\begin{figure}
  \includegraphics[width=84mm]{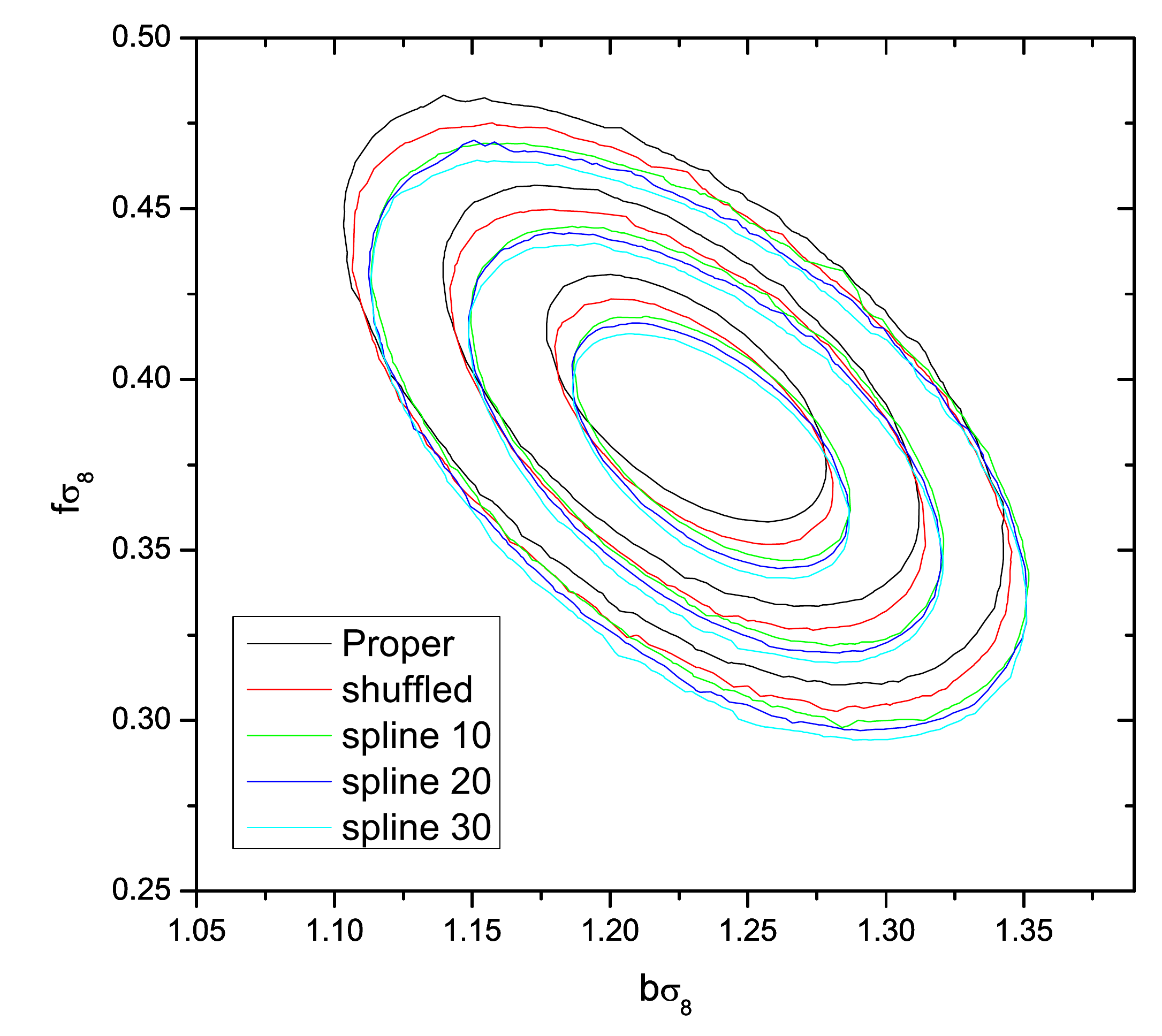}
  \caption{Contours showing the 68\%, 95\%, and 99.7\% confidence level contours on bias and growth factor estimated from $\xi_0$ and $\xi_2$ using different methods to determine the radial distribution used for the random catalogs.}
  \label{fig:bfsys}
\end{figure}

The potential systematic error induced by the treatment of the random catalog appears largest for $\xi_2$. To see the effect of $n(z)$ systematics, we find the best-fit values of bias $b$ and growth rate $f$ when performing a joint fit to $\xi_0$ and $\xi_2$ for each mock
catalog, first by using the random catalog with the true $n(z)$ and then repeating the same analysis using $N$-splined and shuffled random catalogs. Figure~\ref{fig:bfsys} shows the 1, 2, and 3$\sigma$ contours on the joint measurement of $b$ and $f$ when using the true $n(z)$ (black), a shuffled $n(z)$, and a 10- (green), 20- (blue), 30- (cyan) node spline. The $n(z)$ systematics push the measured bias towards slightly higher values and the measured growth rate towards lower values but all contours are consistent within 1$\sigma$. The results derived with the shuffled catalog are on average closer to the results derived with a true catalog than results obtained using a spline fit. Similar fits are performed to the CMASS $\xi_{\ell}$ in \cite{ReidRSD12}.

\begin{figure}
  \includegraphics[width=84mm]{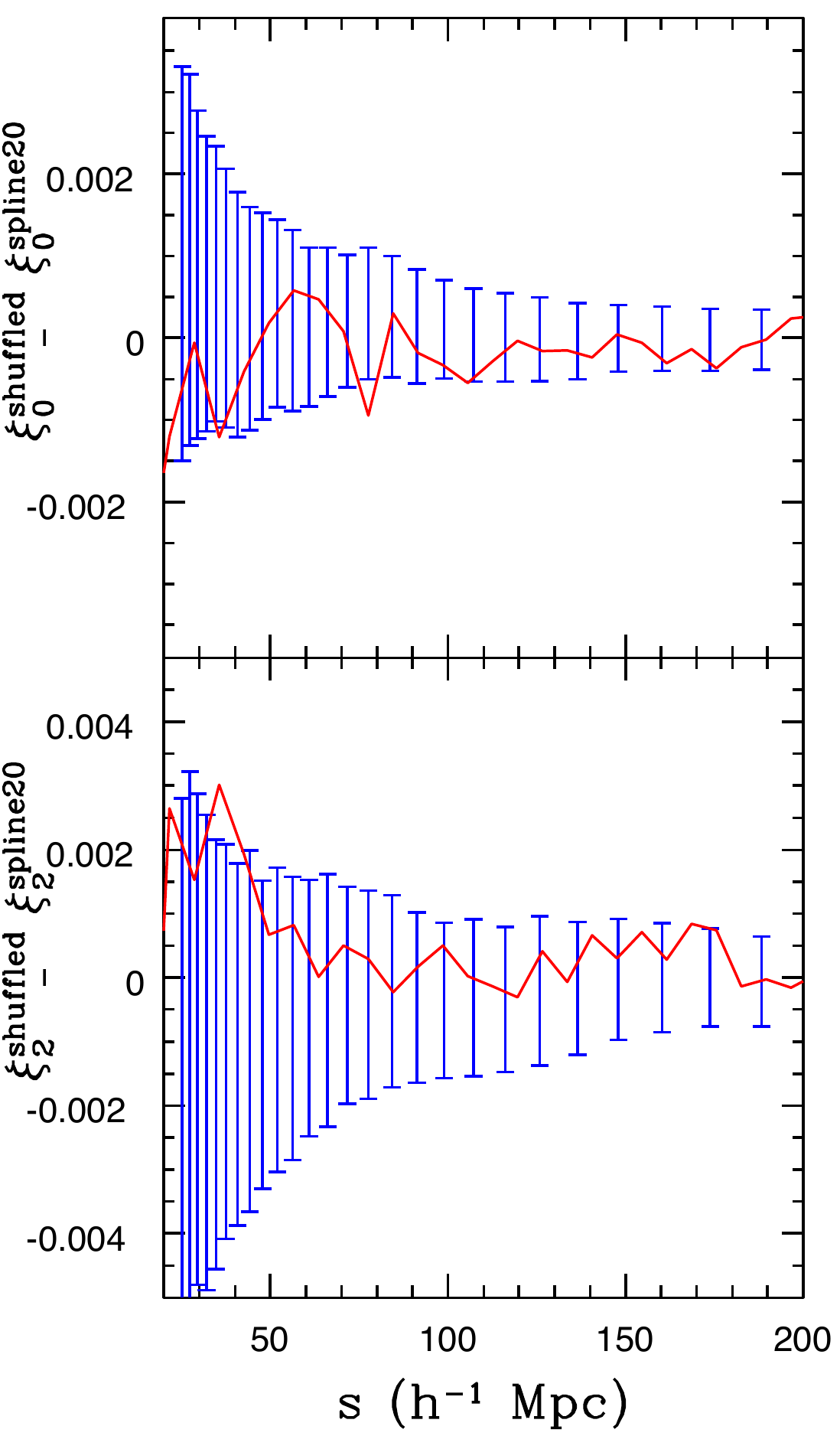}
  \caption{Top panel: The red line displays difference between $\xi_0$ measurements made using the Northern Galactic Cap (NGC) CMASS sample when using a 20-node spline and when randomly selecting redshifts from the galaxy sample (`shuffled') to assign redshifts to the random catalogs. The error bars represent the mean and standard deviation of this difference in the 600 mock galaxy catalogs occupying the NGC footprint. Bottom panel: As above, for $\xi_2$. }
  \label{fig:xi02dcmass}
\end{figure}

The tests outlined in this section suggest that there is some systematic uncertainty introduced by the method in which random points are assigned redshifts. In general, this causes both an increase in the statistical uncertainty and a systematic bias. The added statistical uncertainty is at most 5\% (as given by $\sqrt{1+0.33^2}$) of the fiducial uncertainty for $\xi_{0}(s)$ and 12\% of the fiducial uncertainty for $\xi_2(s)$. This added statistical uncertainty is accounted for by measuring the mock correlation functions used for the covariance matrix using the same method as we employ on the data. Fig. \ref{fig:xi02dcmass} shows that difference between the CMASS $\xi_{\ell}$ measurements made using a 20-node spline and shuffled random catalog is indeed at the level we expect from the mocks.

The systematic bias induced by the treatment of the randoms is negligible for $\xi_0(s)$, but is larger for $\xi_2(s)$. In both cases, using a shuffled random catalog, on average, produces the least biased result. Therefore, we use the shuffled method to obtain redshifts for the random catalogs we use in $\xi(s)$ calculations.

\subsection{Clustering Split at $z = 0.52$}
\label{sec:xizspl}
One may worry that the clustering at higher redshift may be more prone to systematic errors, given that, all else being equal, higher redshift objects should be fainter. Therefore, we split our CMASS data into two samples, one with $z < 0.52$ and the other with $z > 0.52$. This split is close to the peak of the redshift distribution and represents the redshift at which the CMASS sample transitions from being approximately volume-limited to magnitude-limited (as can be inferred by inspecting the $n(z)$ in, e.g., Fig. \ref{fig:number_density_vs_redshift}). We also find that the $w_{star}$ weights become much more important at $z > 0.52$, as the mean $i_{fib2}$ magnitude is 21.01 above $z=0.52$ and 20.74 below. Thus, we expect differences in $\xi(s)$ measurements to be greatest when split at $z = 0.52$ (and indeed, differences in the measured $\xi_{\ell}$ are smaller when we split at, e.g., $z = 0.55$).

\begin{figure}
  \includegraphics[width=84mm]{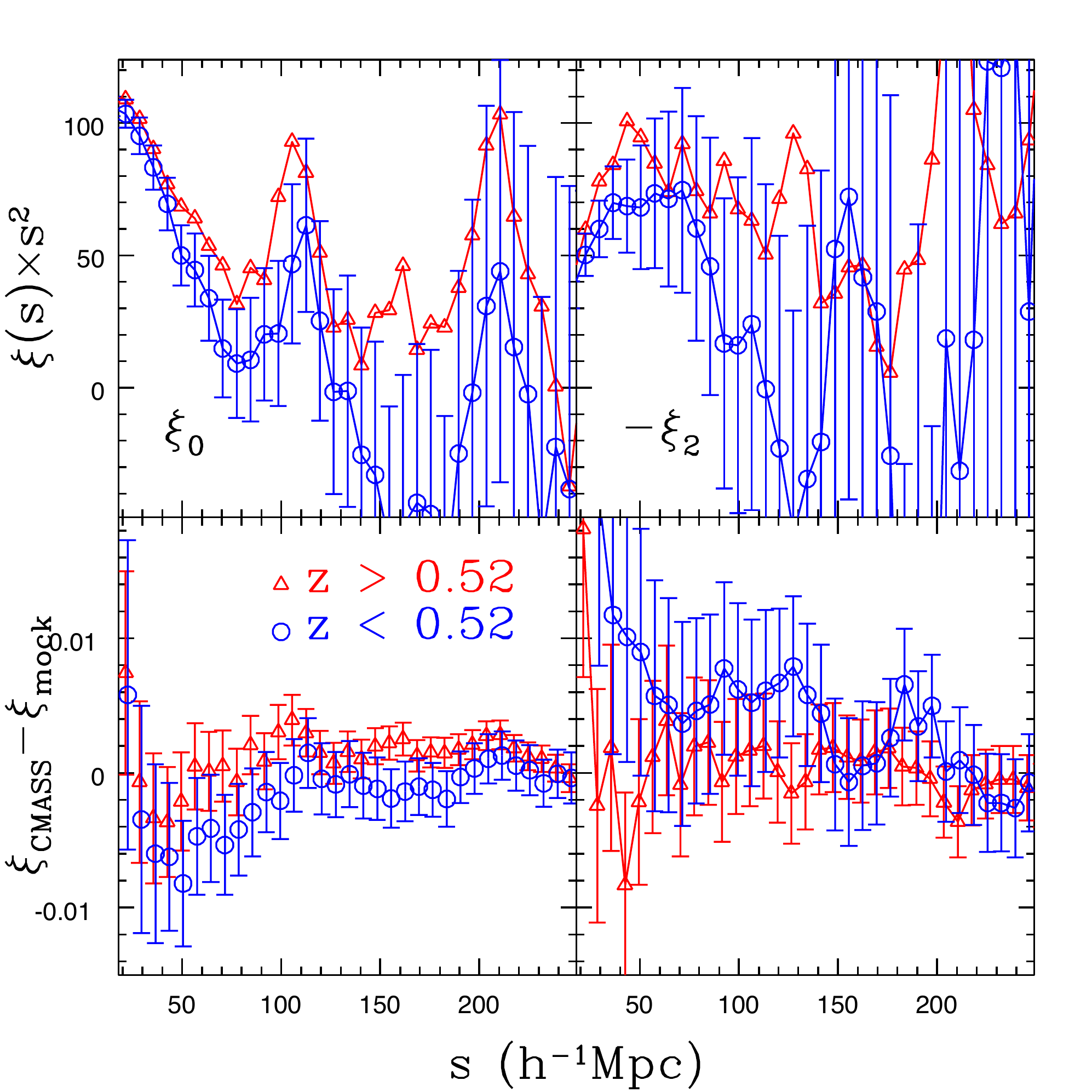}
  \caption{Top panels: The measured redshift space auto-correlation functions of the combined CMASS data split by redshift into samples with $z > 0.52$ (red) and $z < 0.52$ (blue). Points represent measurements; error-bars are determined by from the mocks split at the same redshifts. Bottom panels: The difference between the CMASS measurement and the mean of the mock measurements scaled to the best-fit bias of the respective CMASS samples. The values of $s$ for the $z < 0.52$ sample have been shifted horizontally by 1 $h^{-1}$Mpc for clarity. }
  \label{fig:xizspl}
\end{figure}

The resulting $\xi(s)$ are displayed in the top panels of Fig. \ref{fig:xizspl}, with open blue circles representing $z < 0.52$ and red triangles representing $z> 0.52$. The amplitudes of the $z > 0.52$ measurements are significantly larger at all scales than the lower redshift ones. This may partially be due to the fact that galaxies at $z > 0.52$ are more luminous, and thus we may expect them to have a higher bias. We fit these data to our mocks between $25 < s < 150h^{-1}$Mpc and account for the 6.4\% change in the linear growth factor between $z= 0.61$ and 0.48 (the mean redshifts of the respective samples). We indeed find a higher bias for the $z>0.52$ sample, as $\tilde{b}=2.02\pm0.04$ for $z > 0.52$ ($\chi^2_{min} = 22.9$) and $\tilde{b}= 1.85\pm0.06$ for $z<0.52$ ($\chi^2_{min} = 15.1$). The values of $\tilde{b}$ (and other parameters we measure throughout this section) are summarized in Table \ref{tab:allpar} (found in Section \ref{sec:con}). 

The bottom-left panel of Fig. \ref{fig:xizspl} displays the difference between the measurement and the mean of mocks (600 each for $z< 0.52$ and $z>0.52$), after scaling the mocks to the best-fit bias. Even after this is done, the amplitudes of the $z > 0.52$ $\xi_0$ measurements are larger across all scales than the $z< 0.52$ counterparts. This illustrates the level of covariance between $s$ bins in the $\xi_{\ell}$ measurements (which allows best-fit solutions where most of the data is either above or below the model). Both measurements are consistent with the mocks at scales $150 < s < 250h^{-1}$Mpc (14 data points), as $\chi^2=7.6$ for $z < 0.52$ and 17.7 for $z> 0.52$ (91\% of consistent samples will have $\chi^2>7.6$ and 22\% will have $\chi^2>17.7$).

The right-hand panels of Fig. \ref{fig:xizspl} display the same content as the left, but for $\xi_2$. The values of $|\xi_2|$ are consistently smaller for $z< 0.52$. We test the significance of this result by fixing the bias at the value determined from $\xi_0$ and scaling the mean of the $\xi_2$ mocks to find a best-fit $\tilde{f}$, via Eq. \ref{eq:mod2}. For $z > 0.52$, $\tilde{f} = 0.75\pm0.05$ ($\chi^2=14.1$); for $z < 0.52$, $\tilde{f}=0.59\pm0.08$ ($\chi^2 = 6.9$). Accounting for the fact that we expect a 6\% decrease in $f$ between $z=0.61$ and 0.48, this is a 1.3$\sigma$ discrepancy and is not surprising given we intentionally split the sample at a redshift where we expected to find the largest differences.

We find that splitting the sample at $z=0.52$ yields consistent $\alpha$ values when marginalizing over the bias: for $z<0.52$ we find $\tilde{\alpha} = 1.016\pm0.038$ and for $z > 0.52$ we find $\tilde{\alpha} = 1.013\pm0.021$. Both best-fit $\tilde{\alpha}$ values are smaller than the best-fit for the combined sample ($\tilde{\alpha} = 1.021\pm0.019$), implying that the cross-correlation between the two slices contains significant information relevant to $\tilde{\alpha}$. The consistency of the results further implies that fits to standard $\Lambda$CDM cosmological parameters will yield consistent results. We find similar levels of consistency when splitting the individual NGC and SGC regions at $z=0.52$ and performing the same tests. 

\section{Clustering at the Largest Scales}
\label{sec:remain}

\begin{figure}
  \includegraphics[width=84mm]{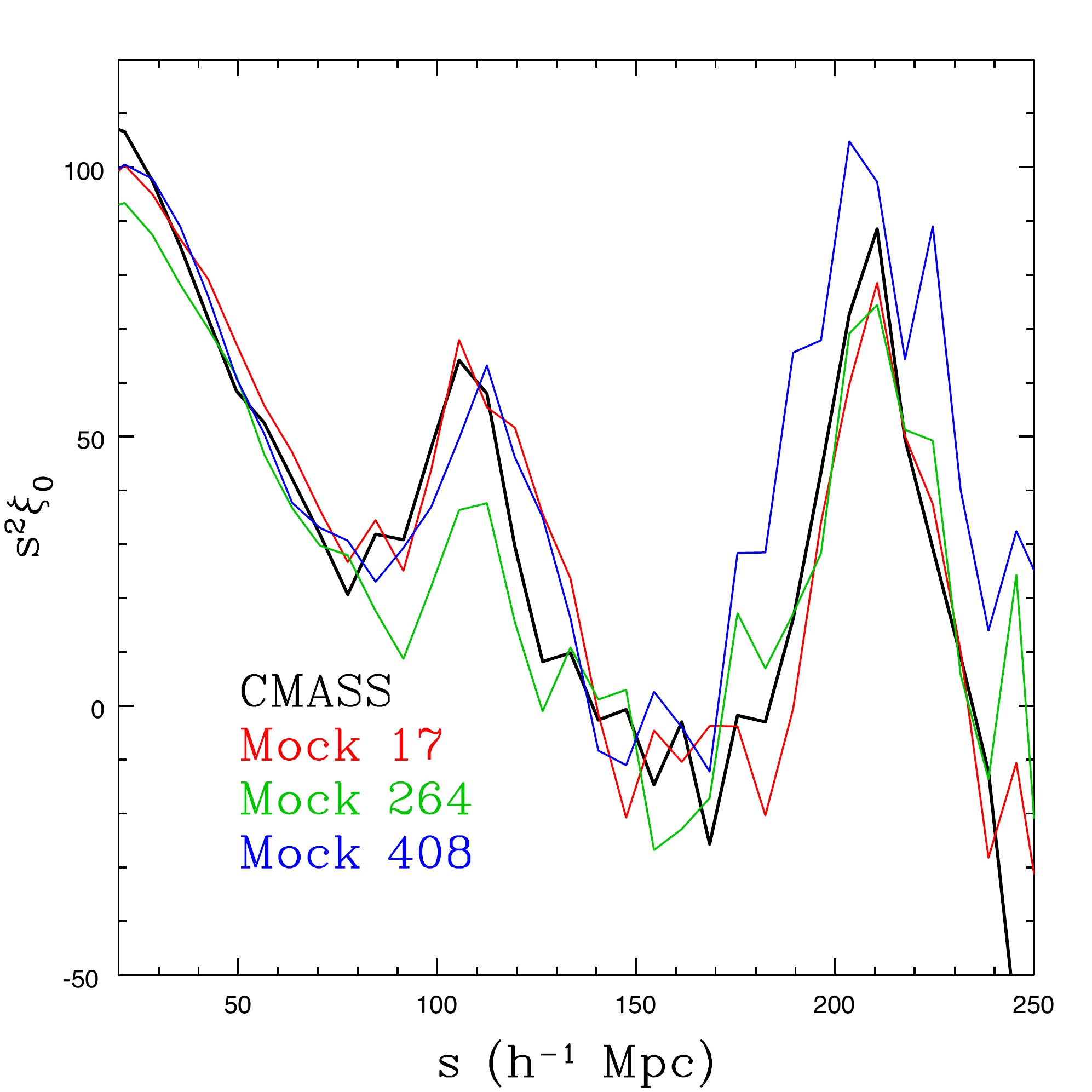}
  \caption{The measured monopole of the correlation function, $\xi_0$, multiplied by $s^2$ for the combined CMASS samples and three mock catalogs with similar levels of clustering at $s\sim200h^{-1}$.}
  \label{fig:com200}
\end{figure}

\begin{figure*}
\begin{minipage}{7in}
\centering
 \includegraphics[width=7in]{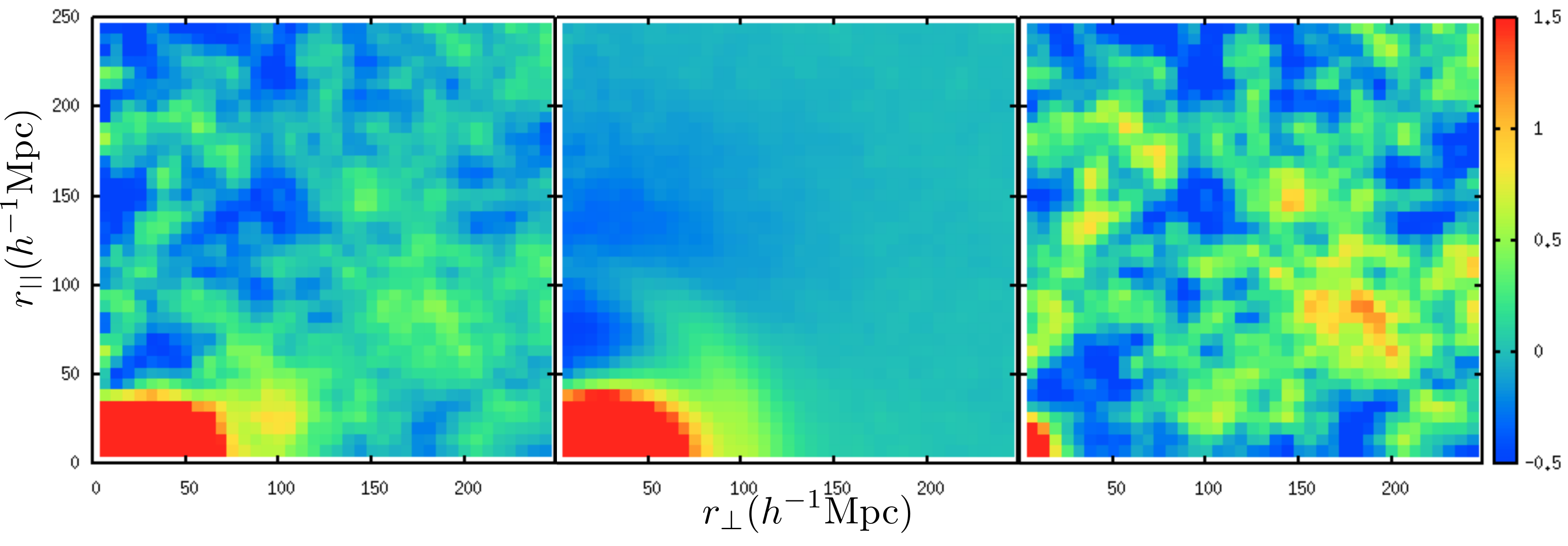}
  \caption{Left panel: The measured correlation function, $\xi(r_{\perp},r_{||})$, plotted a function of the radial, $r_{||}$, and transverse, $r_{\perp}$, distance for the CMASS sample. To compare all scales, we plot the hyperbolic sine of $(75\xi[r_{\perp},r_{||}])$. Middle panel: The mean $\rm{sinh}(75\xi[r_{\perp},r_{||}])$ of 600 mocks masked to simulate the Northern Galactic Cap (NGC) footprint of the CMASS data set. Right panel: $(\xi_{CMASS}[r_{\perp},r_{||}]-\xi_{mock}[r_{\perp},r_{||}])/2\sigma$, where $\sigma$ is the standard deviation of the $\xi_{mock}[r_{\perp},r_{||}]$ (and sinh scaling is no longer used; we divide by 2$\sigma$, rather than 1, for clarity). }
  \label{fig:rppi}
  \end{minipage}
\end{figure*}

To this point have focused on scales $s < 150 h^{-1}$Mpc. In this section, we focus on larger scales. Models of the galaxy correlation function in a Universe dominated by dark energy and cold dark matter cross zero at a scale just beyond the BAO peak and asymptote towards zero. This is true even for models with a high level of primordial non-Gaussianity (where the amplitudes around the BAO scale and zero-crossing scale increase). Thus, $\xi_0(s)$ measurements that differ from this behaviour indicate the presence of systematic effects in the galaxy density field, effects not accounted for in the standard paradigm, or correlated noise.

Comparing our measured $\xi_0$ (using the $w_{star}$ and FKP weights) to the mean of that of the mocks between $150 < s < 250 h^{-1}$ (14 measurements), we find $\chi^2 = 27.3$. This value is rather large, as only 2\% of consistent samples have a greater $\chi^2$ value. We note that if the $w_{star}$ weights are not applied, $\chi^2=57.3$ (only $3\times10^{-5}$\% of consistent samples would have such a large $\chi^2$). If we do not use the FKP weights, but still use the FKP covariance matrix (to make sure the measurement and not the covariance is driving the $\chi^2$), the $\chi^2$ decreases to 23.2. This is still large enough that only 5.7\% of consistent samples would have a larger $\chi^2$. This discrepancy is a product of the full sample, over all redshifts (and the associated lower variance), as neither sample when split at $z=0.52$ returned an abnormally large $\chi^2$ value. The best-fit bias of the CMASS sample (when using FKP weights) is 2\% higher than that of the mocks (1.938 compared to 1.9) and, naively, $\chi^2 \propto b^4$. Performing such a scaling reduces the $\chi^2$ to 25.2 --- still large enough that we should expect a larger $\chi^2$ value for only 3.3\% of consistent samples.

The effect of any systematic associated with the angular mask, e.g., variations in stellar density or errors in the normalization of pair counts, is to add roughly a constant amplitude to $\xi_0$. Therefore, we add a constant, $A$, to the $\xi_0$ of the mocks and find the best-fit value, fitting between $150 < s < 250 h^{-1}$Mpc. The $\chi^2$ is minimized at $A = 0.0006$, but is only reduced to 25.8 (23.8 if scaling by the bias). Given that we have reduced the number of degrees of freedom to 13, the probabilities of being consistent remain the same (to within the quoted number of significant digits). That is, $A$ (and, thus, any remaining purely angular systematic) is not detected with any significance.

\subsection{Anisotropic Clustering and Feature at 200 $h^{-1}$ Mpc}

\begin{figure}
  \includegraphics[width=84mm]{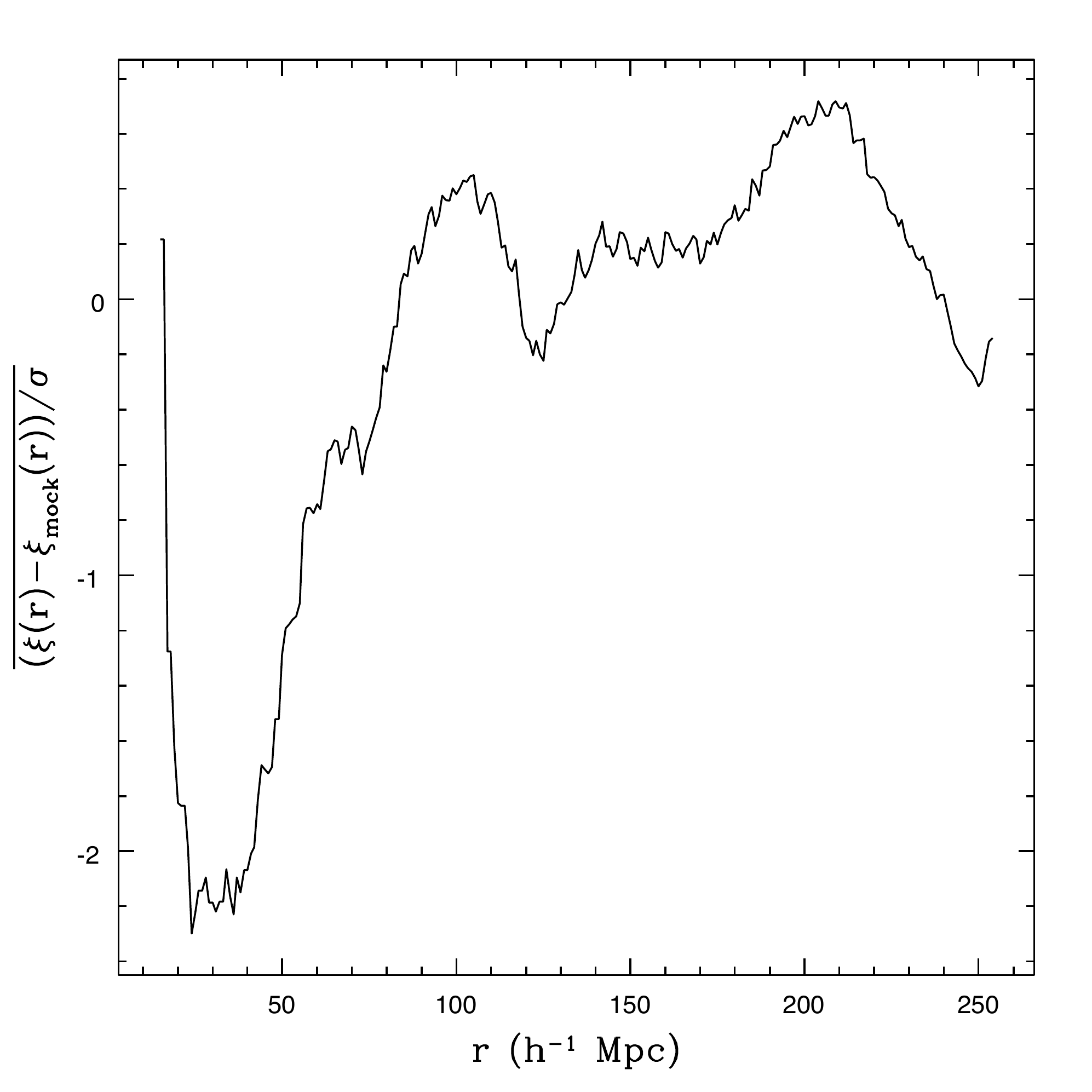}
  \caption{The average value of  $(\xi[r_{||},r_{\perp}]-\bar{\xi}_{mock}[r_{||},r_{\perp}])/\sigma(r_{||},r_{\perp})$ calculated in rings with $\Delta r = 10h^{-1}$ and $(r-\Delta r)^2 < r^2_{||} +r^2_{\perp} < (r+\Delta r)^2$.}
  \label{fig:ringsig}
\end{figure}

The inconsistency we find between the measured clustering and the mean of mocks appears driven by an excess at $s\sim200h^{-1}$Mpc. No matter how we split the CMASS sample or change our analysis, the bump-like feature around 200 $h^{-1}$Mpc remains. It is strange that this feature is nearly as robust in the NGC and SGC samples alone and also in both the samples split at $z=0.52$. However, we can find mocks with similar large-scale $\xi_0$ to the CMASS sample. The $\xi_0$ of three such realizations are plotted along with the $\xi_0$ of the combined CMASS sample in Fig. \ref{fig:com200}. Clearly, cosmic variance allows the possibility of obtaining peaks in $\xi_0$ at around 200 $h^{-1}$Mpc that are qualitatively similar to that we observe. 

We investigate further by examining the clustering of BOSS galaxies as a function of radial, $r_{||}$, and transverse, $r_{\perp}$, distances, $\xi(r_{\perp},r_{||})$. Given the redshift-space separation $s$ and the cosine of the angle to the line-of-sight $\mu$, $r_{||} = \mu s$ and $r_{\perp} = \sqrt{s^2-r^2_{||}}$. The left-hand panel of Fig. \ref{fig:rppi} displays the hyperbolic sine, sinh, of 75 times $\xi(r_{\perp},r_{||})$. This transformation allows information on all scales to be displayed on a single figure. The effect of redshift space distortions is apparent, as amplitudes along the line-of-sight are clearly decreased relative to those at the same transverse separations. 

\begin{figure*}
\begin{minipage}{5in}
\centering
 \includegraphics[width=5in]{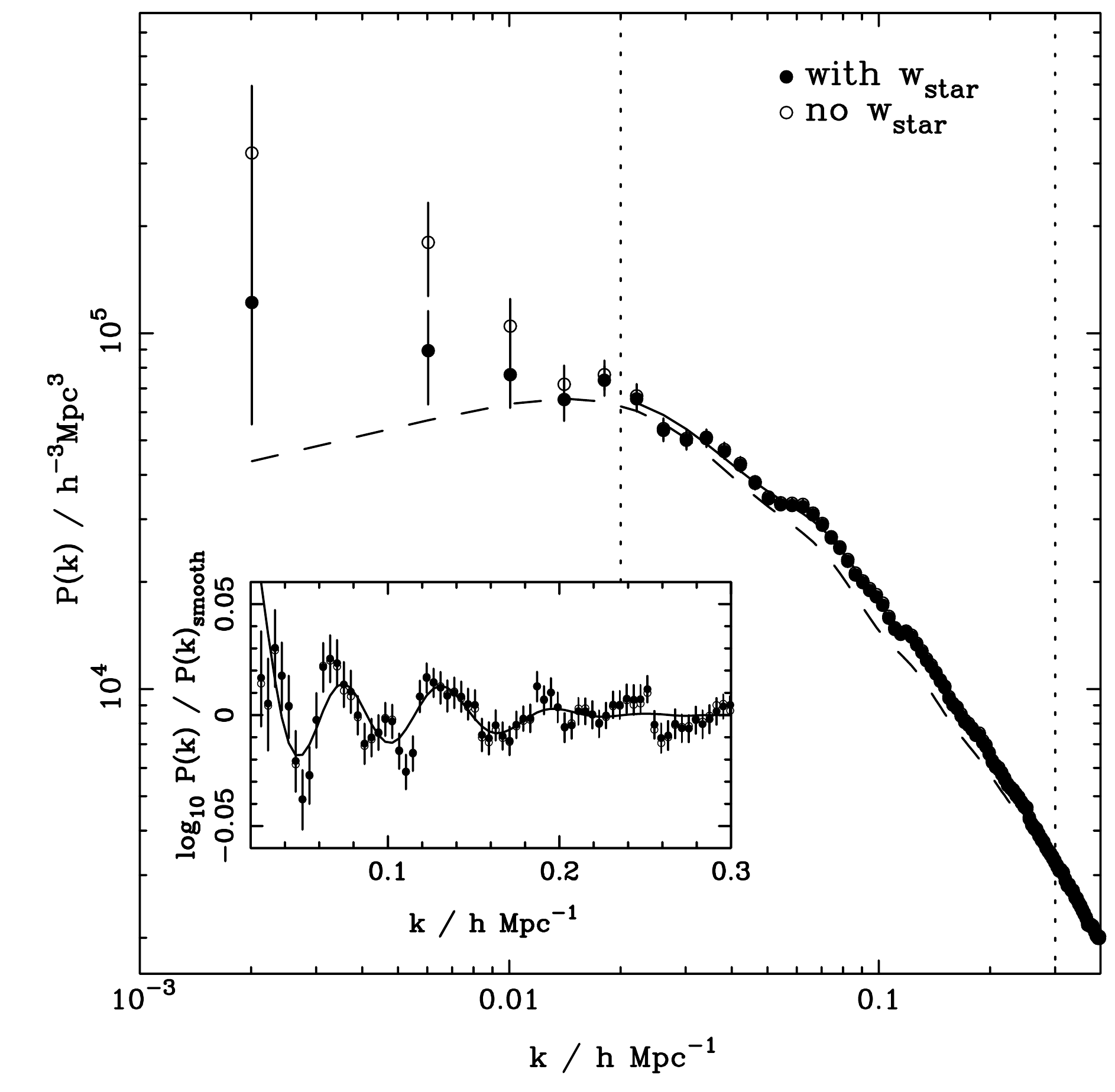}
  \caption{The measured spherically-averaged (in redshift-space) power spectrum, $P(k)$, of the full DR9 CMASS sample with only the fiducial angular weight applied (open circles) and with the linear-fit weights for stellar density, $w_{star}$, applied (solid circles). The average of the mock $P(k)$ is displayed with a dashed line. The best-fit smooth model $P(k)$ determined by Blanton et al. 2012 is plotted with a solid line. The inset panel displays the same information, divided by the smooth-fit, with the solid line displaying the best-fit model including BAO. }
  \label{fig:Pk}
 \end{minipage}
\end{figure*}

There is a ring at around 100$h^{-1}$Mpc, as expected for the BAO feature. The extra information in the $r_{\perp},r_{||}$ dependence of the BAO feature is examined in Blanton et al. (in prep.). There also appears to be an excess in a ring around 200$h^{-1}$Mpc. The middle panel displays the mean sinh(75$\xi[r_{\perp},r_{||}]$) in the mocks. At scales less than 100$h^{-1}$Mpc, this has a similar appearance to the measurements. The right-hand panel displays the difference between the measurements and the mocks, divided by twice the standard deviation of $\xi_{mock}[r_{\perp},r_{||}]$. In general, there is excess at the largest scales, and is most pronounced at scales $\sim200h^{-1}$Mpc. The excess at $r\sim200h^{-1}$Mpc is largest for pairs between 10$^{\rm o}$ and 50$^{\rm o}$ from the line-of-sight. The fact that the excess appears at approximately constant $r$, rather than $r_{\perp}$ or $r_{||}$, suggests that the feature is not due to a systematic strictly related to the angular mask (as this would show up at constant $r_{\perp}$) or the process of obtaining spectroscopic redshifts (as this would appear at constant $r_{||}$). Finally, the feature shows up at nearly identical physical scales when the sample is split at $z=0.52$ (see Fig. \ref{fig:xizspl}), further implying the feature is not associated with a fixed angular scale.  

To further assess the significance of the feature at $\sim$200$h^{-1}$Mpc, we design a statistic that reflects the degree to which there is an excess (or decrement) of signal at constant separation. Thus, we determine the difference between the measured correlation function and the mean of the mock correlation function divided by the standard deviation determined from the mocks, $\sigma(r_{||},r_{\perp})$,
averaged over measurements within a constant separation bin. Thus, we measure
\begin{equation}
t(r,\Delta r) = \left|\frac{\sum \Theta(r_{||},r_{\perp})(\xi[r_{||},r_{\perp}]-\bar{\xi}_{mock}[r_{||},r_{\perp}])/\sigma(r_{||},r_{\perp})}{\sum \Theta(r_{||},r_{\perp})}\right|
\label{eq:sigt}
\end{equation}
where $\Theta(r_{||},r_{\perp}) = 1$ if $(r-\Delta r)^2 < r^2_{||} +r^2_{\perp} < (r+\Delta r)^2$ and 0 otherwise. In essence, this statistic is just a binned version of the information plotted in the right panel of Fig. \ref{fig:rppi}.

Fig. \ref{fig:ringsig} displays the results of performing this test on the measurements, setting $\Delta r = 10 h^{-1}$Mpc. The most discrepant results are at $r=30h^{-1}$Mpc, although this is likely suggestive of a different preferred cosmology from the one used by the mocks. At $r = 204 h^{-1}$Mpc, we find a peak with amplitude 0.72. If we perform the same test on each of the 600 mocks, we find a peak with amplitude greater than 0.72 at $r>120h^{-1}$Mpc in 84 of the mocks, which suggests that the observed size of the peak is common in the mocks. If we demand that the peak be at least as wide as we find in the CMASS data by searching for features where $t > 0.5$ over 25$h^{-1}$Mpc and $t_{max} > 0.72$ (as is the case for the CMASS data), only 29 mocks (4.8\%) are selected. This implies that the combination of the size and width of the feature centered at 200$h^{-1}$Mpc is driving the large $\chi^2$ values of the $\xi_0$ measurements with $s > 150h^{-1}$Mpc. The local maximum at $r=100h^{-1}$Mpc, $t_{max} = 0.45$, and $t > 0.3$ over 25$h^{-1}$Mpc is not significant, as 45\% (267) of the mocks have a feature at least as large and wide centred at $r > 80h^{-1}$Mpc.

Fig. \ref{fig:com200} displays the $\xi_0$ of three realizations where $t_{max} > 0.65$ and is close to 200$h^{-1}$Mpc, showing that these cases are qualitatively similar to the CMASS $\xi_0$. The SDSS I/II LRG $\xi_0(s)$ measurements have larger than expected amplitudes at large scales.  However, a systematic study of the SDSS I/II LRG sample, similar to our own, has not been published and \cite{xu12} still suggest its large-scale clustering amplitudes are within 2$\sigma$ of those expected. The final BOSS data set will have three times more data than the DR9 sample, and thus future data releases will confirm if the feature at 200$h^{-1}$ is simply noise, a yet-to-be-determined systematic, or a real feature in the clustering of galaxies. We do not know of any model that predicts such a feature, as, e.g., the predicted $\xi(s)$ given a strongly non-Gaussian primordial power-spectrum display much smoother variation. This is studied further in Ross et al. (in prep.).

\subsection{Power Spectrum Measurements}
We investigate results using the spherically-averaged power spectrum, $P(k)$, in order to isolate the large scale density modes (i.e., low $k$; the $P(k)$ measurements are less covariant between $k$ bins) and as a consistency check on results derived from $\xi_0$ measurements (which should contain the same information). In Fig. \ref{fig:Pk}, we display the measured $P(k)$, calculated as described in Section \ref{sec:est}. The open circles show the measurement without using the weights accounting for stellar density, while the solid circles display the measurement when these $w_{star}$ weights are applied. These weights only cause a significant difference for the smallest $k$ (largest scales), unlike the situation for $\xi_0(s)$, where the difference was nearly independent of scale. The inset panel displays the same information, divided by the best-fit smooth model found in \cite{alph}. The open and solid circles are indistinguishable from each other. Clearly, the $w_{star}$ weights do not affect the $P(k)$ measurements at the scales related to the BAO feature.

We scale the mock $P(k)$ to find a best-fit $\tilde{b}$ values in the same manner as performed throughout for $\xi(s)$. We determine the covariance matrix of {\rm ln}$[P(k)k^3$], from the 600 mock catalogs and minimize the $\chi^2$ of {\rm ln}$[P(k)k^3$]. We use the logarithmic scaling to account for the fact that we expect the cosmic-variance contribution to the $P(k)$ uncertainty to be proportional to $P^2(k)$ (see, e.g., \citealt{FKP}). This scaling does not significantly alter the best-fit values we determine, but it does result in significantly smaller $\chi^2_{min}$ values. We find $\tilde{b}=1.983\pm0.035$ fitting $k < 0.05h$Mpc$^{-1}$ with $\chi^2_{min} = 19.2$ (11 degrees of freedom) when the $w_{star}$ weights are applied and $\tilde{b}=2.001\pm0.035$ with $\chi^2_{min} = 38.8$ when no weights are applied. The $\chi^2_{min}$ increases dramatically, by a factor greater than 2, without the weights. This shows how dramatic an effect the weights have --- only 0.006\% of consistent samples would have a $\chi^2 > 38.8$, while 5.8\% would have $\chi^2>19.2$.

The effect of the $w_{star}$ weights on the $P(k)$ measurement shows significant scale dependence, unlike for $\xi(s)$, where the change was nearly constant. We can assume that any unaccounted-for systematic will have the same $k$ dependence as the $w_{star}$ weights and determine if adding a factor $A[P(k)-P_{weight}(k)]$ improves our $\tilde{b}$ fit. We find that the $\chi^2_{min} = 13.96$ at $\tilde{b}=1.974\pm0.035$ and $A=-0.41$. That is, a 41\% stronger systematic correction decreases the $\chi^2$ by 5.2. These results strongly suggest that proper treatment of the weights are vital in any attempt to obtain robust measurements that use $P(k)$ measurements at low $k$, e.g., measurements of primordial non-Gaussianity or the scale of matter radiation equality (from the overall peak in $P(k)$). The degeneracy between the systematic correction and the constraints on primordial non-Gaussianity one can obtain is studied further in Ross et al. (in prep.).

The best-fit bias obtained from the $P(k)$ measurement is nearly 1$\sigma$ larger than what we obtain when $\xi(s)$ is fit at scales $>25h^{-1}$Mpc. Measurements of absolute bias values are notoriously difficult, and the recovered value is often driven by the minimum scale that is fit (since this measurement has the least uncertainty; see, e.g., \citealt{ARoss11a}). Indeed, it is difficult even using dark matter simulations, as \cite{Manera10} found systematic differences (at a level similar to what we find here) in large scale bias measurements when comparing results obtained from matter-halo cross-power spectrum and halo auto correlation measurements. Thus, obtaining robust absolute bias measurements is a generic problem rather than an issue specific to the BOSS DR9 galaxy sample.

\begin{figure}
 \includegraphics[width=84mm]{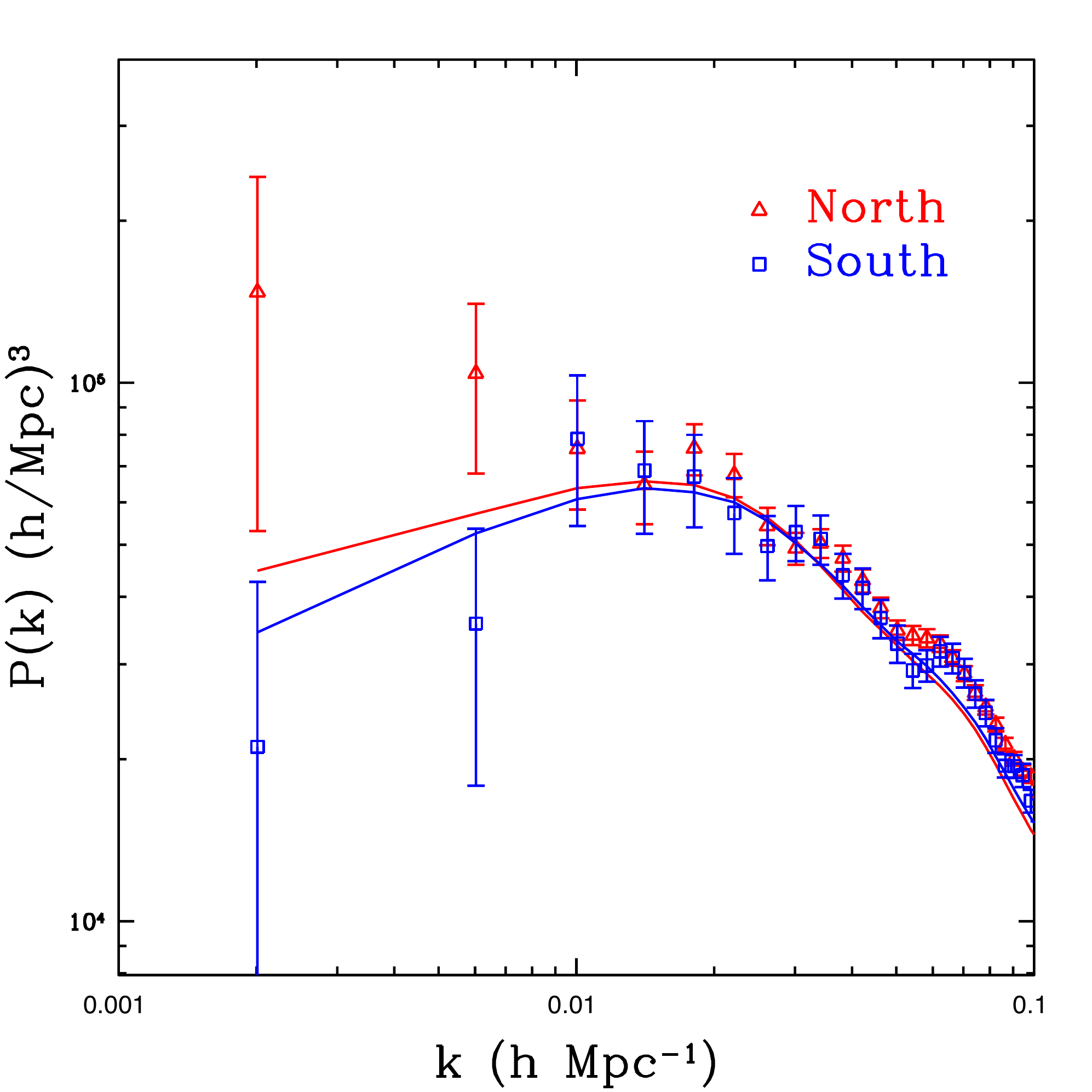}
  \caption{The $P(k)$ measurements for the Northern (NGC; open red triangles) and Southern Galactic Cap samples (SGC; open blue squares), with the mean of the respective mock samples displayed with a solid line. The difference between the two lines illustrates the effect of the different windows of the NGC and SGC on the expected $P(k)$. }
  \label{fig:PkNS}
\end{figure}

\begin{figure}
 \includegraphics[width=84mm]{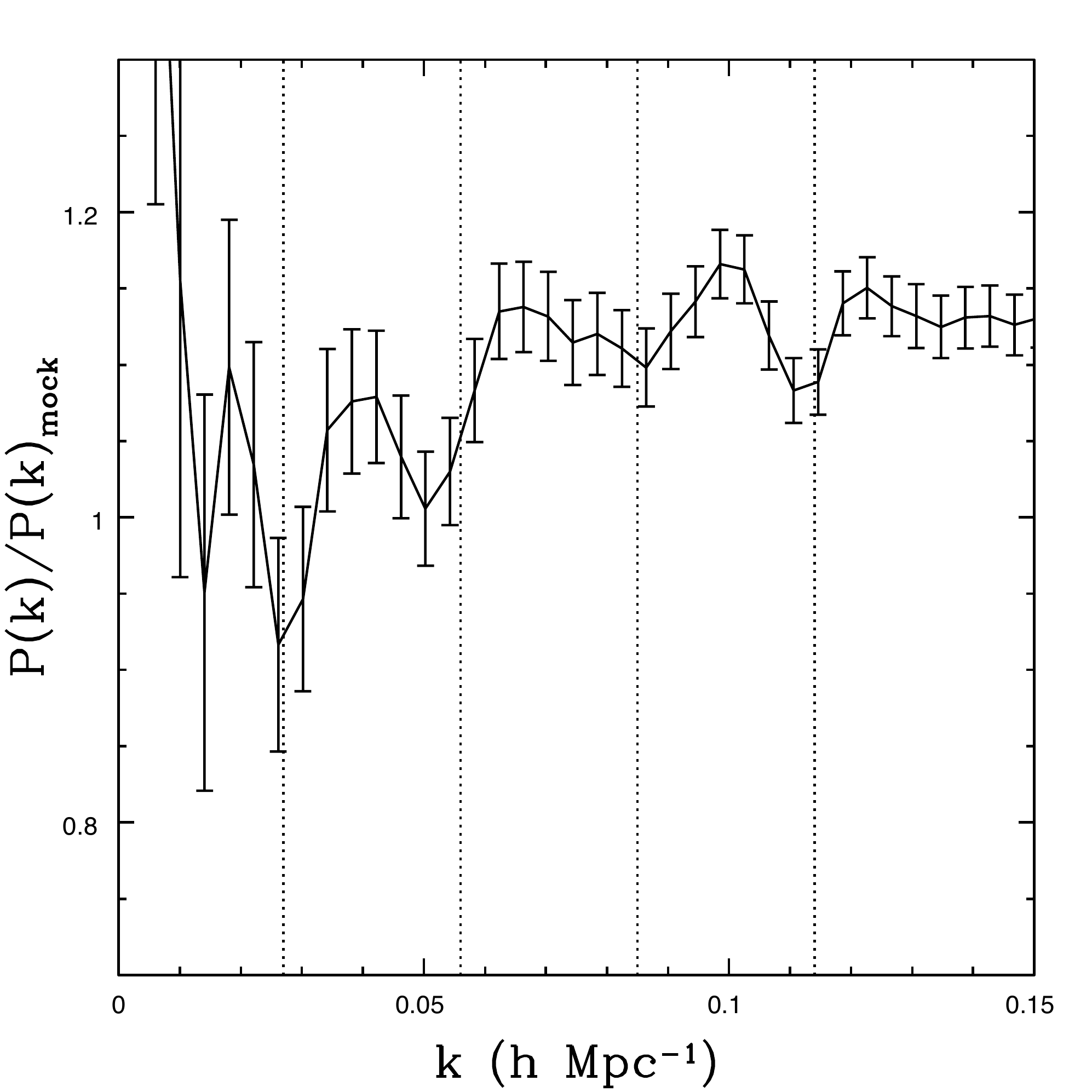}
  \caption{Solid lines display the ratio between the CMASS $P(k)$ measurement to the average of the mock $P(k)$. The vertical dotted lines denote a spacing of 0.029 $h$Mpc$^{-1}$ in $k$, with the first line at $k=0.027 h$Mpc$^{-1}$. }
  \label{fig:Pkwig}
\end{figure}

Fig. \ref{fig:PkNS} displays the $P(k)$ measured for the NGC (open triangles) and SGC (open squares) samples and the average of the mock $P(k)$ for each region separately. The windows of the respective regions create a large difference in the expected $P(k)$, as the shape of the mock $P(k)$ are considerably different at small $k$. Clearly, a direct comparison of the respective $P(k)$ is not appropriate. Scaling the mock $P(k)$ for the respective regions, we find $\tilde{b} = 1.991\pm0.039$ for the NGC with $\chi^2_{min}=18.5$, and $\tilde{b} = 1.93\pm0.07$ for the SGC ($\chi^2_{min} = 11.9$). The difference between the respective $\tilde{b}$ is opposite what we found from the $\xi_0(s)$ measurements, where the bias of the SGC sample was significantly larger. This suggests that smaller scale modes ($k>0.05$) have significant influence on correlation function measurements of the Southern sample at $s > 25h^{-1}$Mpc. Unlike for $\xi(s)$, the bias values determined using $P(k)$ are consistent between the two regions. 

Fig. \ref{fig:Pkwig} displays the ratio between the measured $P(k)$ and the mean mock $P(k)$ when scaling $k$ linearly. The measured $\xi_0$ has a peak at 204$h^{-1}$Mpc, which suggests we should find a periodic feature in the $P(k)$ with wavelength $\sim$0.03 $h$Mpc$^{-1}$ in $k$. We have plotted dotted lines with a spacing of 0.029 $h$Mpc$^{-1}$, with the first at $k=0.027h$Mpc$^{-1}$. There appears to be local minima at $k$ values near each dotted line. This appears most significant at $k\sim0.03h$Mpc$^{-1}$. We caution that the possibility this feature is due to an undiscovered systematic may need to be considered when performing analysis of the shape of the CMASS power spectrum.

\section{Conclusions}
\label{sec:con}

\begin{table*}
\begin{minipage}{7in}
\caption{The parameters and $\chi^2_{min}$ we derive from the clustering of BOSS DR9 CMASS galaxies, for different treatments and subsamples of the data: $\tilde{b}$ is a measure of the amplitude of the measured clustering, $\tilde{f}$ is a measure of the amplitude of $\xi_2$, and $\tilde{\alpha}$ measures the preferred dilation in scale, relative to the average of the mock $\xi_0$.}
\begin{tabular}{lcccccc}
\hline
\hline
Estimator & Hemisphere &  $z$ range &  $w_{star}$ weights? & $\tilde{b}$, $\chi^2$/dof & $\tilde{f}$, $\chi^2$/dof & $\tilde{\alpha}$ \\
\hline
$\xi(s)$ & Both & $0.43 < z < 0.7$ & yes & $1.936\pm0.035$, 22.7/17 & $0.711\pm0.044$, 11.8/17 & $1.020\pm0.019$\\
$P(k)$ & Both & $0.43 < z < 0.7$ & yes & $1.983\pm0.035$, 19.2/11 & -, - & -\\
$\xi(s)$ & NGC & $0.43 < z < 0.7$ & yes & $1.904\pm0.039$, 24.3/17 & $0.691\pm0.052$, 10.9/17 & $0.994\pm0.023$\\
$P(k)$ & NGC & $0.43 < z < 0.7$ & yes & $1.991\pm0.039$, 18.5/11 & -, - & -\\
$\xi(s)$ & SGC & $0.43 < z < 0.7$ & yes & $2.06\pm0.07$, 18.8/17 & $0.79\pm0.09$, 11.8/17 & $1.083\pm0.029$\\
$P(k)$ & SGC & $0.43 < z < 0.7$ & yes & $1.93\pm0.07$, 11.9/11 & -, - & -\\
$\xi(s)$ & Both & $0.43 < z < 0.52$ & yes & $1.85\pm0.06$, 15.1/17 & $0.59\pm0.08$, 6.9/17 & $1.016\pm0.038$\\
$\xi(s)$ & Both & $0.52 < z < 0.7$ & yes & $2.02\pm0.04$, 22.9/17 & $0.75\pm0.05$, 14.1/17 & $1.013\pm0.021$\\
$\xi(s)$ & Both & $0.43 < z < 0.7$ & no & $1.949^{+0.034}_{-0.035}$, 34.2/17 & $0.710\pm0.044$, 12.7/17 & $1.007\pm0.019$\\
$P(k)$ & Both & $0.43 < z < 0.7$ & no & $2.001\pm0.035$, 38.8/11 & -, - & -\\

\hline
\label{tab:allpar}
\end{tabular}
\end{minipage}
\end{table*}

We have investigated potential systematic effects on the three-dimensional clustering of the DR9 sample of BOSS galaxies and tested the robustness of the results when we split the data into the regions we expect may be the most different for observational (Northern/Southern Galactic Caps) and physical (split at $z = 0.52$) reasons. Our main findings are summarized as follows:

\noindent $\bullet$ Redshift failures occur at preferred locations on spectroscopic tiles (see Fig. 3), but the nearest redshifts within the same sector (which therefore share the same observing conditions) have an $n(z)$ like that of the overall sample (see Fig. 4). We therefore account for redshift failures by up-weighting the nearest targeted object within the same sector. We find this approach has a minor effect on the measured $\xi(s)$ (see the red lines in Fig. 5).

\noindent $\bullet$ We account for target objects that lack spectra due to fiber collisions by up-weighting the nearest targeted object. At large scales ($s > 10h^{-1}$Mpc), this should be equivalent to the more sophisticated method proposed in \cite{HGuo}. This scheme increases the $\xi(s)$ measurements in a manner consistent with a small increase in the galaxy bias (see the blue lines in Fig. 5).

\noindent $\bullet$ The overall number densities of observed galaxies in the Southern Galactic hemisphere are higher for both the LOWZ (8\%) and CMASS (3.2\%) samples. If we apply the offsets in color found by \cite{Sch10b} to the selection of BOSS galaxies, the number densities become consistent within 2\% for LOWZ and 0.2\% for CMASS. After applying the \cite{Sch10b} offsets, the $n(z)$ are discrepant at a level that we find in 10\% of mock samples; that is, the difference is not significant.

\noindent $\bullet$ The measured clustering in the NGC and SGC generally agree to within 2$\sigma$, depending on the specific test that is performed. For $\xi(s)$, the bias disagrees at 1.9$\sigma$, but for $P(k)$, the discrepancy is only 0.3$\sigma$. Measurements of the amplitude of $\xi_2$, $\tilde{f}$, are consistent to within 1$\sigma$.

\noindent $\bullet$ The measured clustering in the NGC and SGC disagrees most at around the BAO scale; we find this causes a difference in stretch parameter, $\tilde{\alpha}$, (which in this study encodes information on both the BAO scale and the shape of $\xi(s)$) of 2.5$\sigma$. We find no treatment of the data (e.g., alternative weighting) that significantly alters the level of discrepancy. 

\noindent $\bullet$ We weight CMASS galaxies based on linear relationships between the expected number density of galaxies as a function of their $i_{fib2}$ magnitude and the local stellar density ($w_{star}$ weights). We find no evidence that similar weights are necessary for (the brighter) LOWZ galaxies. By applying the method used to determine the weights to mocks (which have no need for correction) we find that these weights produce no bias on the mean measured $\xi(s)$, whereas more aggressive weighting schemes may (see the appendix). Application of these weights reduces the measured cross-correlations of the CMASS galaxies with stellar density, Galactic extinction, seeing, sky background, and airmass to the level we expect to find randomly (see Fig. 13). These weights account for redshift dependence through the $i_{fib2}$ relationship, but may no longer be sufficient when the sample is split by color. 

\noindent $\bullet$ Applying the $w_{star}$ weights produces a 0.7$\sigma$ shift in the measured value of $\tilde{\alpha}$, most of which we believe is due to the change in shape of the correlation function. Applying the weights reduces the $\chi^2$ from 34.2 to 22.7 when $\xi_0(s)$ measurements are fit in the range $25 < s < 150h^{-1}$Mpc (18 data points). The $w_{star}$ weights have little effect on $\xi_2$ measurements, as the best-fit $\tilde{f}$ does not change and the $\chi^2$ is reduced only from 12.7 to 11.8 when fitting scales $25 < s < 150h^{-1}$Mpc.

\noindent $\bullet$ We use the mocks to determine the least-biased way to simulate the radial selection function of CMASS galaxies, in order to produce a random catalog. Randomly selecting the redshift of a galaxy in the sample (`shuffled') produces a smaller bias than performing a spline fit to the measured $n(z)$ (`spline'). In all cases, the bias is negligibly small for $\xi_0$, but is as high as 50\% of the statistical uncertainty for $\xi_2$. We therefore advocate using shuffled random catalogs and note that any constraints obtained using $\xi_2$ measurements should take this bias into account. We find that the differences between $\xi_0$ and $\xi_2$ measurements we obtain using CMASS galaxies using the spline and shuffled methods are consistent with the level found in the mocks.

\noindent $\bullet$ The $\xi_0(s)$ measurements, when split at $z=0.52$, yield bias and $\alpha$ values that are consistent within 1$\sigma$, when fit in the range $25 < s < 150h^{-1}$Mpc. The $\xi_2$ measurements are somewhat discrepant, as the best-fit $\tilde{f}$ values differ by 1.7$\sigma$.

\noindent $\bullet$ The $\xi_0(s)$ measurements, at scales $s > 150h^{-1}$Mpc, are inconsistent at a greater than 94\% level with the expected clustering. Allowing for a constant offset in large-scale clustering (as angular systematics tend to contribute) produces no improvement.

\noindent $\bullet$ The inconsistency at large scales is dominated by a peak in $\xi_0(s)$ at $s\sim200h^{-1}$Mpc. This feature appears in a ring when we measure $\xi(r_{\perp},r_{||})$, implying that it is not due to a systematic solely to either the mask or target catalog (as would show up in $r_{\perp}$) or the process of obtaining spectroscopic redshifts (as would show up in $r_{||}$). Similar features are found that are at least as significant in 4.8\% of the mocks that we test.

\noindent $\bullet$ The $w_{star}$ weights have a dramatic effect on $P(k)$ measurements and the effect depends strongly on $k$. The $\chi^2$ decreases from 38.8 to 19.2 before and after $w_{star}$ weights are applied to $P(k)$ measurements fit at $k < 0.05$ (12 data points). We find the $\chi^2$ reduces to 14.0 when allowing a further correction of the form $A(P[k]_{noweight}-P[k]_{weight})$. This implies that one must be careful when obtaining any information that depends on measurements of $P(k)$ at $k < 0.01$ and this issue is studied further in Ross et al (in prep.).

The fundamental conclusions of this work are that, for BOSS DR9 CMASS galaxies, we recommend the application of weights to account for fiber collisions, redshift failures, and the systematic effect of stars and we find no further systematics dependencies. We therefore expect \cite{alph}, \cite{SanchezCos12}, \cite{ReidRSD12}, and \cite{SamRSD12} (all of which apply the same weights) to obtain robust cosmological constraints using the clustering of BOSS DR9 galaxies. 

\cite{alph} find the measured BAO position does not depend on the application of the $w_{star}$ weights --- the position changes by less than 0.1$\sigma$. Therefore, our results do not suggest there is a significant unaccounted-for systematic error in previous BAO measurements. However, our study does suggest that any previous finding of a large-scale excess in clustering measurements may have been due to systematic errors, similar to the ones we correct for, and requires extra scrutiny. Indeed, \cite{imsys} show that much of the excess in large scale power presented in \cite{Thomas11} (who used SDSS-II imaging data) is removed when the systematic effect of stars on the projected density field is accounted for.

The SDSS-III BOSS DR9 sample of galaxies represent only one third of the final BOSS CMASS sample and one quarter of the final LOWZ sample. Further data sets will allow potential systematics to be tested to an even greater extent and reveal whether the feature at 200$h^{-1}$Mpc is noise.

\section*{Acknowledgements}
AJR is grateful to the UK Science and Technology Facilities Council for financial support through the grant ST/I001204/1.
WJP is grateful for support from the the UK Science and Technology
Facilities Research Council, the Leverhulme Trust, and the European
Research Council.
MAS acknowledges the support of NSF grant AST-0707266.
MECS was supported by the NSF under Award No. AST-0901965

We thank the anonymous referee for comments that helped improve the quality of this work.

Funding for SDSS-III has been provided by the Alfred P. Sloan
Foundation, the Participating Institutions, the National Science
Foundation, and the U.S. Department of Energy Office of Science.
The SDSS-III web site is http://www.sdss3.org/.

SDSS-III is managed by the Astrophysical Research Consortium for the
Participating Institutions of the SDSS-III Collaboration including the
University of Arizona,
the Brazilian Participation Group,
Brookhaven National Laboratory,
University of Cambridge,
Carnegie Mellon University,
University of Florida,
the French Participation Group,
the German Participation Group,
Harvard University,
the Instituto de Astrofisica de Canarias,
the Michigan State/Notre Dame/JINA Participation Group,
Johns Hopkins University,
Lawrence Berkeley National Laboratory,
Max Planck Institute for Astrophysics,
Max Planck Institute for Extraterrestrial Physics,
New Mexico State University,
New York University,
Ohio State University,
Pennsylvania State University,
University of Portsmouth,
Princeton University,
the Spanish Participation Group,
University of Tokyo,
University of Utah,
Vanderbilt University,
University of Virginia,
University of Washington,
and Yale University.

\begin{appendix}
\section{Angular Weighting Schemes}
\begin{figure*}
\begin{minipage}{7in}
\centering
 \includegraphics[width=7in]{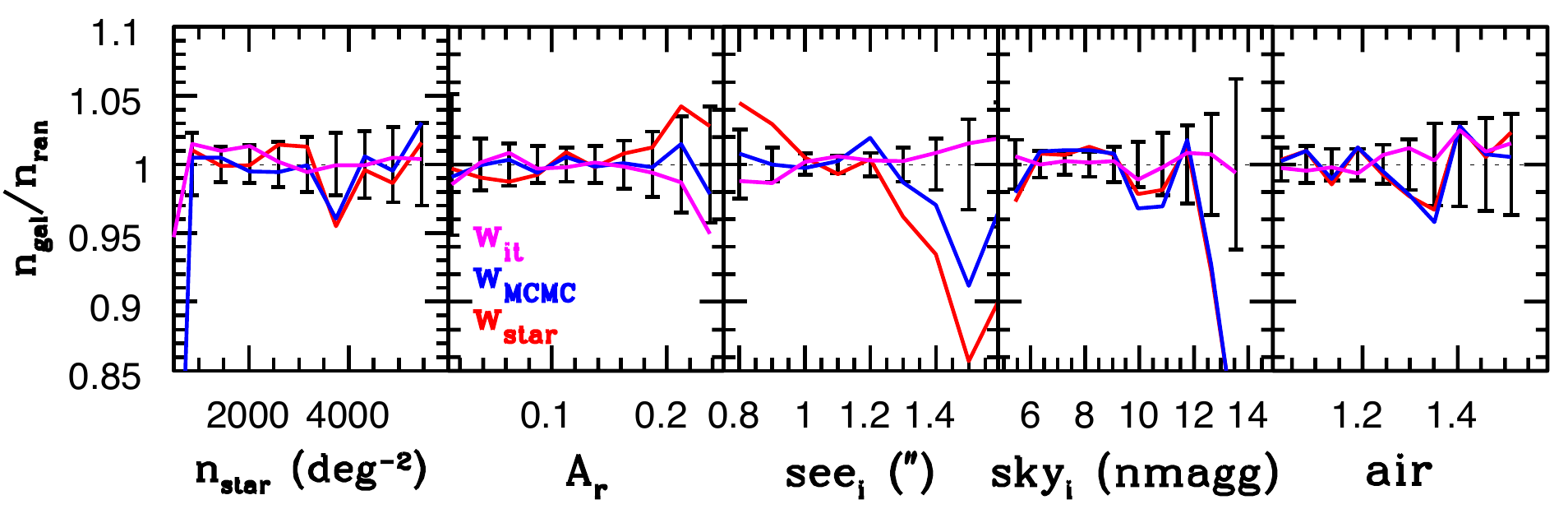}
  \caption{Same as Fig. \ref{fig:ngvnsys}, except we now plot the residual relationships after applying iterative weights (magenta; $w_{it}$), the residual relationships after using a Monte-Carlo Markov Chain to simultaneously fit linear relationships in order to determine the weights (blue; $w_{MCMC}$) and the residual when the weights are split as a function of the fiber magnitude, but calculated only based on stellar density (red; $w_{star}$).}
  \label{fig:ngvnsysw}
\end{minipage}
\end{figure*}

We considered three different weighting schemes in order to account for the systematic correlations found in Section \ref{sec:angvar}. These included: 
\begin{enumerate}
{\bf \item `iterative weights'} which we denote $w_{it}$: This technique was applied in \cite{imsys}. It assumes that the effects of each systematic are separable, and proceeds by starting with one systematic and setting the weight in every Healpix pixel equal to the inverse of the quantity plotted in black in Fig. \ref{fig:ngvnsys}. One then moves on to the next systematic and re-calculates the relationship between the number density of galaxies and the systematic, and then multiplies the weights by the inverse of the relationship. If the effects are indeed separable, the $n_{gal}(sys)$ relationships should all remain consistent with unity after all of the weights have been calculated. 

To determine $w_{it}$, we proceed in the order stellar density, Galactic extinction, airmass, seeing, and sky background. If each is truly separable, the order should not matter, and we do find negligible differences for any permutation of the order we have tested. The residual relationships between the galaxy number density and the potential systematic, when weighting by the full $w_{it}$, are displayed with magenta lines in Fig. \ref{fig:ngvnsysw}. In every case, the relationship is almost fully removed. This implies the weighting is too aggressive, as we should actually expect variations consistent with the size of the error-bars in Fig. \ref{fig:ngvnsysw}. 

We can test the extent to which the $w_{it}$ weights may remove true power from clustering measurements by applying weights to each mock sample (which of course contain no imaging systematics) following the methods we apply to the data. The black triangles in Fig. \ref{fig:mockwbias} display the average difference between the fiducial $\xi$ measurements and when the full iterative weights, $w_{it}$, are calculated and applied to each mock. For the monopole, this decreases the expected result by about half the statistical uncertainty (displayed with the black dotted lines). There is also a non-zero bias for the $\xi_2$ measurements (top panel), but the difference is insignificant compared to the statistical uncertainty.
\begin{figure}
  \includegraphics[width=84mm]{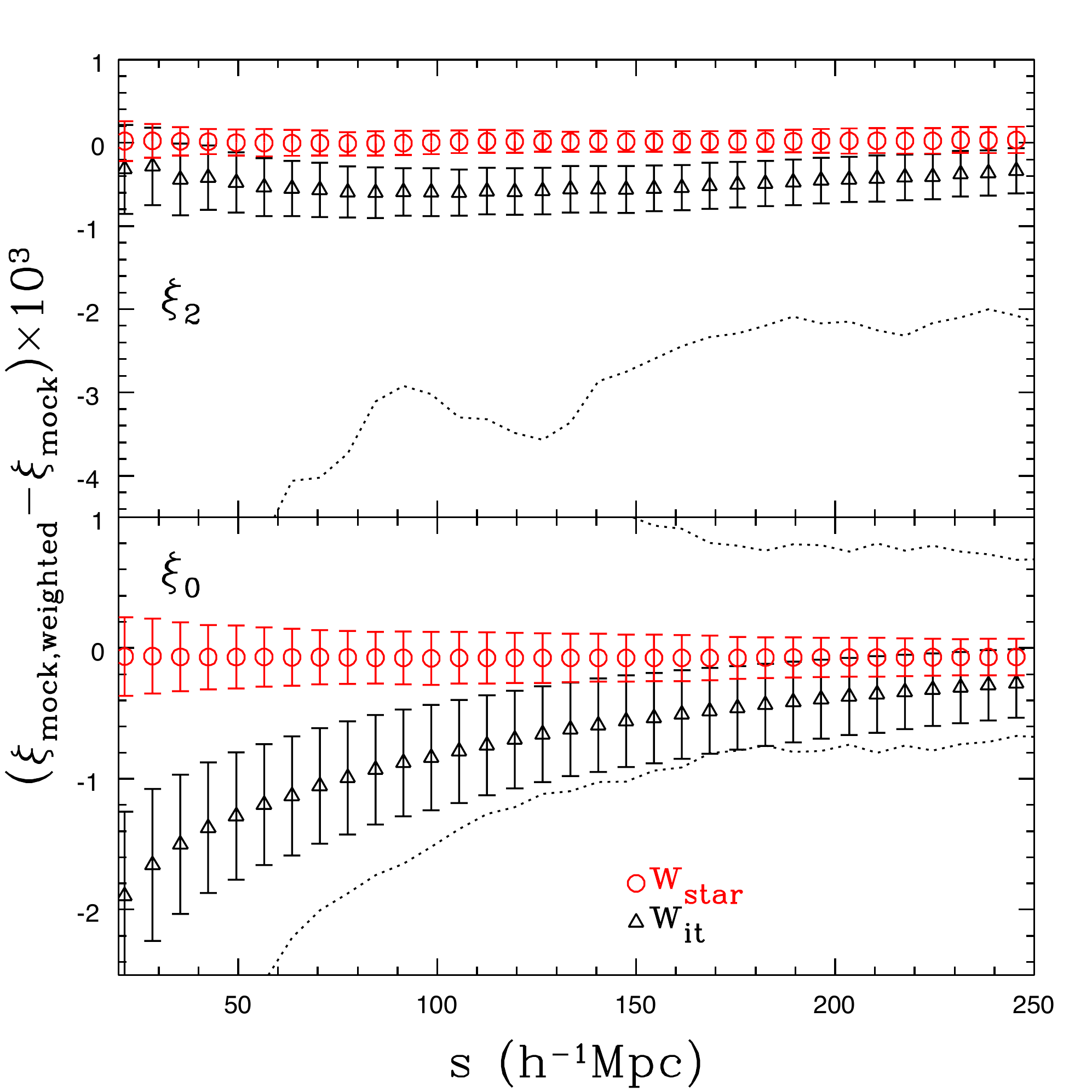}
  \caption{The average difference between the fiducial redshift-space correlation function of the mocks, $\xi$, and that using weights for each mock using the full iterative method, (black triangles $w_{it}$), and that using weights for each mock using only a linear fit to the relationship with stellar density (red circles, $w_{star}$). Error-bars represent the standard deviation of the difference. The black dotted lines display the variance on $\xi$ found in the mocks. We note that the mocks do not require weights --- a deviation from zero implies that a bias is imparted by the weight scheme.}
  \label{fig:mockwbias}
\end{figure}

{\bf \item `linear-fit MCMC weights'}, which we denote $w_{MCMC}$: These weights are calculated by using a Monte-Carlo Markov chain (MCMC) to simultaneously find the linear coefficients that best describe the total $n_g(n_{sys})$ relationships. 

The $w_{MCMC}$ weights are determined by finding the best-fit solution to
\begin{equation}
n_{gal}/n_{ran} = K+A n_{star}+B A_r+C see_i + D sky_i + E air
\label{eq:mcmcw}
\end{equation}
where $K,A,B,C,D,E$ are the coefficients we fit for and $see$ is the Seeing, $sky$ is the sky background, and $air$ is the airmass. This is solved efficiently using a MCMC, as coefficients can be applied to the healpix map simultaneously (thereby accounting for any covariances between the potential systematics). The value of $w_{MCMC}$ is then the inverse of the best-fit relationship. The residual relationships after applying the $w_{MCMC}$ weights are displayed in blue in Fig. \ref{fig:ngvnsysw}. These weights allow more variation than the $w_{it}$ weights. However, the sum of $(\xi_{p,x}(r_{eff})^2/\xi_{p}(r_{eff})$ over all five potential systematics for CMASS galaxies with the $w_{MCMC}$ weights, displayed in blue in Fig. \ref{fig:wsyscomb}, is substantially smaller than we expect from the mocks (black error bars in Fig. \ref{fig:wsyscomb}). This result implies that the $w_{MCMC}$ weights may also over-correct the CMASS galaxy density field and remove true fluctuations.

{\bf\item `linear-fit stellar density weights'}, which we denote $w_{star}$: These weights are calculated by performing linear fits to the dependence of galaxy density with stellar density and $i_{fib2}$ magnitude, as described in Section \ref{sec:angweight} and adopted for the analysis we performed throughout this paper.
\end{enumerate}

\begin{figure}
  \includegraphics[width=84mm]{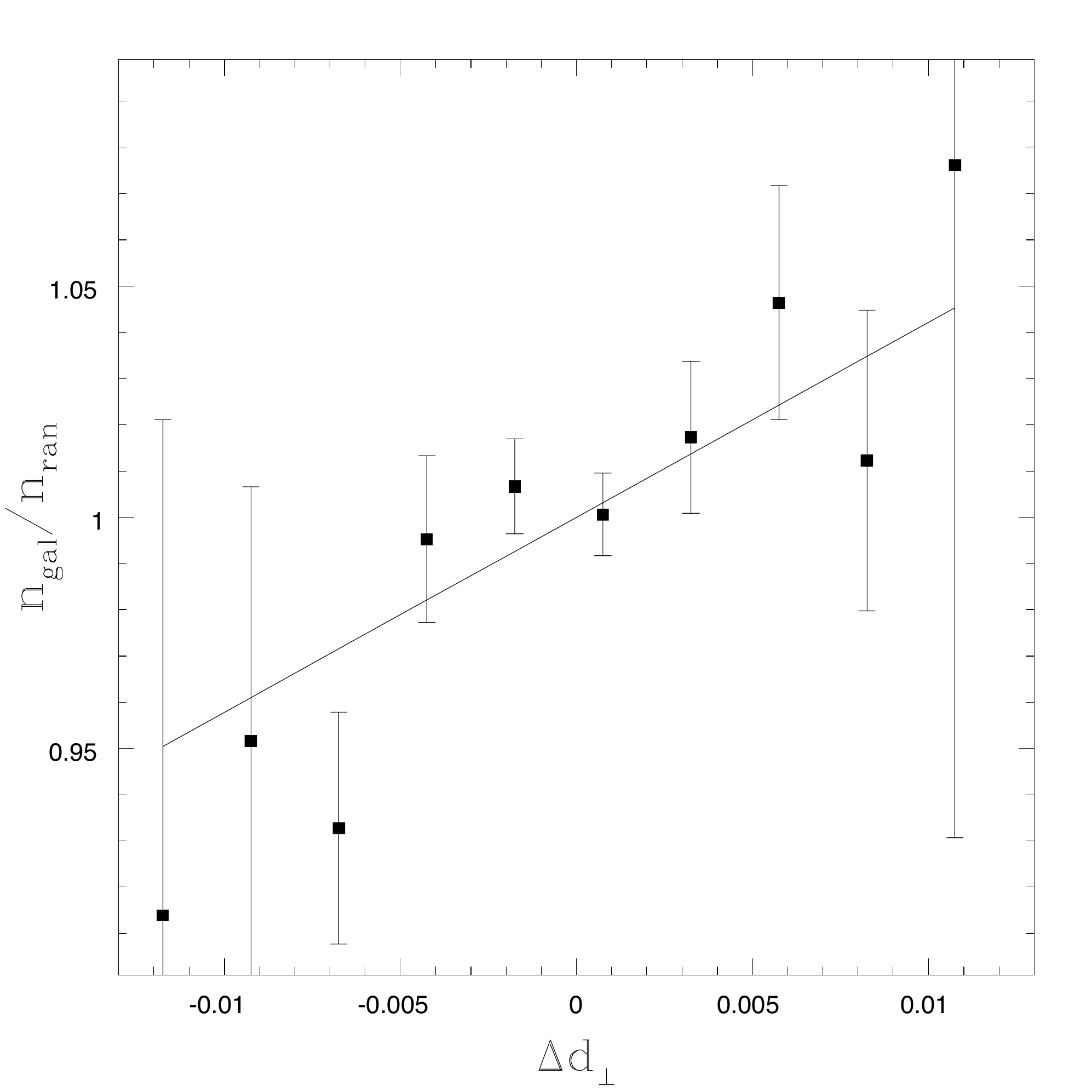}
  \caption{The relationship between the projected number density of CMASS galaxies versus the offset in $\Delta d_{\perp}$, as determined from the Schlafly et al. (2011) offsets determined for SDSS imaging runs. The error-bars reflect the variation found in mock galaxy catalogs occupying the same area. The solid line is the expected relationship, based on the scalings found in Section \ref{sec:NS}.}
  \label{fig:ngvdpF}
\end{figure}

The effect of applying only weights for stars, fit to the linear relationship between $n_{gal}$ and $n_{star}$, on the mocks is displayed in red circles in Fig. \ref{fig:mockwbias}. The difference is consistent with zero for both $\xi_{0}$ and  $\xi_{2}$. This suggests that the $w_{star}$ weights do not over-smooth the galaxy density field. Further, the sum of the five potential systematic contributions is consistent with the mean mock sum when the $w_{star}$ weights (as shown in Fig. \ref{fig:wsyscomb}) are applied to the CMASS data. These results are in contrast to our previous tests that suggested both the $w_{it}$ and $w_{MCMC}$ remove true power. Therefore, we believe the $w_{star}$ weights are the most appropriate to apply to the CMASS sample. 

Finally, we considered the fact that \cite{Sch10b} determined color offsets for every SDSS imaging run available at the time of their study. This included 95\% of the CMASS data in the NGC and 63\% in the SGC. Restricting our analysis to these regions, we can measure the relationship between the projected number density of CMASS galaxies and the offset in $d_{\perp}$, as given by \cite{Sch10b}. This is displayed in Fig. \ref{fig:ngvdpF}, where we have applied the $w_{star}$ weights to the CMASS sample and the error bars reflect the variation found in mock samples over the same footprint. The solid line displays the relationship expected from scalings found in Section \ref{sec:NS}, determined to be $n_{gal}/n_{ran} = 1 + 4.217\Delta d_{\perp}$, and appears consistent with what we measure. We measured $\xi_0(s)$ in this region, applying a weight to account for the predicted scaling with $\Delta d_{\perp}$, and found negligible differences between the recovered measurements and those using the full footprint, with separate NGC and SGC selection functions, and the $w_{star}$ weights (our recommended procedure). Within a single hemisphere, the effect of the offsets is similar to that of seeing --- there is a systematic relationship, but the pattern of the imaging runs is effectively random and therefore the relationship does not impart spurious power at scales relevant to our analysis. \cite{HoLG} reached a similar conclusion when analyzing angular power spectra of the BOSS imaging data. The exception is when one considers the NGC and SGC together, as the mean offset between the two regions is large enough to produce a significant offset in the number densities of the two regions, and thus imparts a systematic error in the large-scale clustering if separate NGC and SGC selection functions are not applied.

\end{appendix}

\label{lastpage}

\end{document}